\documentclass[amsmath,amssymb,aps,prd,10pt,floatfix,preprintnumbers,superscriptaddress,nofootinbib,twocolumn]{revtex4-1}

\pdfoutput=1 
\usepackage[utf8]{inputenc} 
\usepackage{amsmath}
\usepackage{amssymb}
\usepackage{citesort}
\usepackage{xcolor}
\usepackage{lineno}
\usepackage{graphicx}
\usepackage{url}
\usepackage[colorlinks, citecolor=blue,anchorcolor=blue,menucolor=blue, 
linkcolor=blue,filecolor=blue,runcolor=blue,urlcolor=blue,frenchlinks=true,breaklinks=false]{hyperref}

\newcommand{\head}[1]{\vspace{6pt} \noindent \textbf{\textcolor{blue}{ #1}}\ }

\long\def\junk#1{\null}
\newcommand{\orcid}[1]{\,\href{https://orcid.org/#1}{\includegraphics[width=9pt]{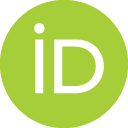}}\,}
\newcommand{\orcidHA}{0000-0002-2017-7706} %
\newcommand{\orcidSA}{0000-0001-5450-0447} %
\newcommand{\orcidFG}{0000-0002-8506-274X} %
\newcommand{\orcidAG}{0000-0002-8553-7338} %
\newcommand{\orcidAL}{0000-0001-7968-5388} %
\newcommand{\orcidFO}{0000-0001-6799-2436} %
\newcommand{\orcidOZ}{0000-0003-3783-6330} %
\newcommand{\orcidRS}{0000-0002-9157-6819} %
\newcommand{\orcidDB}{0000-0002-9246-7366} %
\newcommand{\orcidMB}{0000-0003-3938-1544} %
\newcommand{\orcidCG}{0000-0002-3518-0617} %
\newcommand{\orcidFH}{0000-0001-7563-687X} %

\newcommand{\orcidIN}{0000-0003-0087-6355} %

\newcommand{\orcidVB}{0000-0003-0148-0272} %
\newcommand{\orcidSC}{0000-0003-0479-7689} %
\newcommand{\orcidACS}{0000-0002-7107-5902} %
\newcommand{\orcidJF}{0000-0003-0860-9569} %
\newcommand{\orcidHJ}{0000-0002-2964-9845} %
\newcommand{\orcidAK}{0000-0002-4090-0084} %
\newcommand{\orcidTM}{0000-0002-1723-4028} %
\newcommand{\orcidPS}{0000-0003-1990-0992} %
\newcommand{\orcidMS}{0000-0003-4893-8041} %
\renewcommand{\affiliation}[1]{}
\newcommand{\ipm}{\affiliation{School of Particles and Accelerators, Institute for Research in Fundamental     Sciences (IPM), P.O.Box 19395-5531, Tehran, Iran}}

\newcommand{\smu}{\affiliation{Department of Physics, Southern Methodist University,
    Dallas, TX 75275-0175, U.S.A.}}
\newcommand{\desy}{\affiliation{Deutsches Elektronen-Synchrotron (DESY), Notkestrasse 85, D-22607 Hamburg, Germany}}
\newcommand{\dubna}{\affiliation{Joint Institute for Nuclear Research, Joliot-Curie 6, Dubna, Moscow region, Russia, 141980}}
\newcommand{\maxPlanck}{\affiliation{Max-Planck-Institut f\"ur Physik, F\"ohringer Ring 6, D-80805 M\"unchen, Germany}}{}
\newcommand{\oxford}{\affiliation{Particle Physics, Denys Wilkinson Bdg, Keble Road, University of Oxford, OX1 3RH Oxford, UK}}

\newcommand{\cracow}{\affiliation{T. Kosciuszko Cracow University of Technology, PL-30-084, Cracow, Poland}}
\newcommand{\krakow}{\affiliation{Institute of Nuclear Physics Polish Academy of Sciences, PL-31342 Krakow, Poland}}

\newcommand{\cern}{\affiliation{CERN, CH-1211 Geneva 23, Switzerland}}

\newcommand{\liverpool}{\affiliation{Department of Mathematical Sciences, University of Liverpool, Liverpool L69 3BX, United Kingdom}}
\newcommand{\cea}{\affiliation{IRFU, CEA, Universit\'e Paris-Saclay, F-91191
  Gif-sur-Yvette, France}}

\begin{document}
\modulolinenumbers[5]
\setlength\linenumbersep{2pt}

\title{\Large \bf xFitter: An  Open Source QCD Analysis Framework 
\mbox{\textcolor{blue}{\small A resource and reference document for the Snowmass study}
}%
}

\collaboration{xFitter Collaboration}
\thanks{%
xFitter Contact: Sasha Glazov: alexandre.glazov@desy.de
\\
Webpage: \href{https://www.xfitter.org/xFitter/xFitter/}{www.xFitter.org}
}

\author{The xFitter Developers' Team:}\noaffiliation{}

\author{H.~Abdolmaleki\orcid{\orcidHA}}\ipm{}
\author{S.~Amoroso\orcid{\orcidSA}}\desy{}
\author{V.~Bertone\orcid{\orcidVB}}\cea{}
\author{M.~Botje\orcid{\orcidMB}}
\author{D.~Britzger\orcid{\orcidDB}}\maxPlanck{}
\author{S.~Camarda\orcid{\orcidSC}}\cern{}
\author{A.~Cooper-Sarkar\orcid{\orcidACS}}\oxford{}
\author{J.~Fiaschi\orcid{\orcidJF}}\liverpool{}
\author{F.~Giuli\orcid{\orcidFG}}\cern{}
\author{A.~Glazov\orcid{\orcidAG}}\desy{}
\author{C.~Gwenlan\orcid{\orcidCG}}\oxford{}
\author{F.~Hautmann\orcid{\orcidFH}}
\author{H.~Jung\orcid{\orcidHJ}}\desy{} 
\author{A.~Kusina\orcid{\orcidAK}}\krakow{}
\author{A.~Luszczak\orcid{\orcidAL}}\cracow{}
\author{T.~M{\"a}kel{\"a}\orcid{\orcidTM}}\desy{}
\author{I.~Novikov\orcid{\orcidIN}}\desy{}
\author{F.~Olness\orcid{\orcidFO}}\smu{}
\author{R.~Sadykov\orcid{\orcidRS}}\dubna{}
\author{P.~Starovoitov\orcid{\orcidPS}}
\author{M.~Sutton\orcid{\orcidMS}}
\author{O.~Zenaiev\orcid{\orcidOZ}\vspace{0.5cm}}

\begin{abstract}
\vspace{1cm}
We provide an overview of the xFitter open-source software package, 
review the general capabilities of the program, 
and highlight applications relevant to the Snowmass study. 
An updated version of the program (2.2.0) is available on CERN GitLab,\footnote{xFitter version 2.2.0: \mbox{\url{https://gitlab.cern.ch/fitters/xfitter}}}
and this has been updated to a C++ codebase with enhanced and extended features. 
We also discuss some of the ongoing and future code developments 
that may be useful for precision studies. 
We  survey   recent analyses performed by the xFitter developers' team
including:
W and Z production, 
photon PDFs, 
Drell-Yan forward-backward asymmetry studies, 
resummation of small-$x$ contributions, 
heavy quark production, 
constraints on the strange PDF, 
determination of the pion PDF,
and
determination of the pion Fragmentation Functions.
Finally, we briefly summarize selected applications of xFitter in the literature.
The xFitter program is a versatile, flexible, modular,
and comprehensive tool that
can provide impact studies for possible future facilities.
We encourage the use of xFitter,
and welcome new contributions from the community.
\vspace{2cm}
\end{abstract}

\date{ 15 March 2022}

\maketitle
\tableofcontents{}

\newpage
\section{Introduction to \texorpdfstring{\lowercase{x}F\lowercase{itter}}{xFitter}}

\subsection{\texorpdfstring{\lowercase{x}F\lowercase{itter}}{xFitter} Overview}

{\bf xFitter}~\cite{Alekhin:2014irh} 
(formerly {\bf HERAfitter}) is an open-source software package that provides a framework for the
determination of the {\bf parton distribution functions} (PDFs) of the
proton and related subjects.\footnote{%
The xFitter program is available on the web at: 
\href{https://www.xfitter.org/xFitter/xFitter/}{www.xFitter.org}
} %
xFitter version 2.0.1 is currently available, and 
version 2.2.0 will be released imminently and  offers an expanded set of tools and options. 
It incorporates experimental data from a wide range of experiments 
including fixed-target, Tevatron, HERA, and LHC data sets. 
xFitter can analyze this data using predictions up to 
next-to-next-to-leading-order (NNLO)  in perturbation theory with a variety of 
theoretical calculations including   numerous
methodological options for carrying out PDF fits and plotting tools which
help visualize the results. 
While primarily based on the collinear factorization foundation, 
xFitter also provides facilities for fits of dipole models
and {\bf transverse-momentum dependent} (TMD) distributions. 
The package can be used to study the impact of new precise measurements from
hadron colliders, and  also assess the  impact of future  colliders.  
This paper provides a brief overview of xFitter 
with emphasis of the features relevant for the Snowmass2021 study.

\head{The need for precision PDFs:}
The PDFs are the essential components
that allow us to make theoretical predictions for experimental
measurements of collider experiments with initial state protons and hadrons.
Despite the recent progress of PDF analyses (including NLO and
NNLO calculations), the uncertainty for many precision measurements at the LHC
stems nowadays primarily from the PDFs~\cite{HERAFitterdevelopersTeam:2015cre,Accomando:2019vqt,Amoroso:2020fjw}.
Hence, our ability to fully
characterize the Higgs boson and constrain 
Beyond-Standard-Model (BSM) physics     
signatures ultimately
comes down to how accurately we determine the underlying PDFs; this is
the focus of the xFitter project.

\head{Open Source Code:}
The xFitter package is provided at {\tt www.xFitter.org},
and a write-up of
the program can be found in Ref.~\cite{Alekhin:2014irh},
and an overview of available tutorials in Ref.~\cite{Bertone:2017tig}
including some presented at MCnet-CTEQ schools. 
The xFitter framework has already been used for more than 100+ analyses
including many LHC studies.
The code structure of the xFitter package is modular, and it allows for various theoretical and
methodological options.
Currently it contains interfaces to
 QCDNUM~\cite{Botje:2010ay},
 APFEL~\cite{Bertone:2013vaa,Bertone:2017gds},
 LHAPDF~\cite{Buckley:2014ana},
 APPLGRID~\cite{Carli:2010rw},
 APFELGRID~\cite{Bertone:2016lga},
 FastNLO~\cite{Stober:2015nlg},
 HATHOR~\cite{Aliev:2010zk}, %
among other packages. 

xFitter also has a large number of data sets available, 
including a variety of fixed target experiments, 
HERA, Tevatron, and  LHC. 
It is also possible to add new custom data sets such as 
LHeC~\cite{LHeC:2020van} and EIC~\cite{Accardi:2012qut,Proceedings:2020eah})
pseudo-data.

\subsection{\texorpdfstring{\lowercase{x}F\lowercase{itter}}{xFitter}  Capabilities}

\head{PDF Fits \& Analysis:}
First and foremost, xFitter provides a flexible open-source framework 
for performing PDF fits to data. 
The PDFs are the fundamental object that xFitter works with, and it
has a variety of utilities to read, write, and manipulate the standardized PDF file format and
associated uncertainties.
For example, xFitter is able to read and write PDFs in the LHAPDF6
format~\cite{Buckley:2014ana}.

\head{xFitter-draw:}
xFitter can also automatically generate comparison plots of data vs.\ theory.
 There are a variety of options
for the definition of the $\chi^2$ function and the treatment of
experimental uncertainties.
Examples are presented in Ref.~\cite{Alekhin:2014irh}.

\head{Nuclear PDFs:}
xFitter has also been extended to produce nuclear PDFs;
this was used to produce the TUJU19 nPDF set
of Ref.~\cite{Walt:2019slu}.

\head{Pion PDFs:}
xFitter can also produce meson PDFs, and
Ref.~\cite{Novikov:2020snp} illustrates this
for the case of pion PDFs.

\head{Pseudo-Data:}
An important application of xFitter is to understand how a particular
data set or experiment will impact the PDFs.
A typical study might be to  use  pseudo-data from a proposed
experiment (e.g.  LHeC or EIC) to constrain the relative uncertainty on the
underlying PDFs. 
For example, Ref.~\cite{Abdolmaleki:2019acd} used LHeC pseudo-data  to
constrain the strange PDF with  charged-current DIS charm production data.
Additionally, forward-backward Drell-Yan asymmetry pseudo-data  
were prepared to simulate the end of Run-II LHC (i.e. $300~fb^{-1}$), and also the HL-LHC; these pseudo-data have been used for PDF profiling in 
Refs.~\cite{Accomando:2019vqt} and~\cite{Abdolmaleki:2019ubu}.

\head{Profiling \& Reweighting:}
xFitter is able to perform PDF profiling and reweighting studies.
The reweighting method allows xFitter to update the probability
distribution of a PDF uncertainty set (such as a set of NNPDF
replicas) to reflect the influence of new data.
For the PDF profiling, xFitter compares data and MC predictions 
based on the $\chi^2$-minimization, and then constrains the
individual PDF eigenvector sets taking into account the data
uncertainties.
For example, Ref.~\cite{HERAFitterdevelopersTeam:2015cre} used the Tevatron $W$-boson charge asymmetry
and of the $Z$-boson production cross sections data to study the impact
on the PDFs using Hessian profiling and Bayesian reweighting
techniques.

In a separate study~\cite{Accomando:2019vqt}, the forward-backward asymmetry in neutral current Drell-Yan production provides powerful constraints on 
the valence quark PDFs, and this in turn can impact  both 
SM~\cite{Accomando:2018nig,Accomando:2017scx} 
and BSM~\cite{Accomando:2016tah,Accomando:2016ehi}  
physics.

\head{NNLO \& QED PDFs:}
As many PDF analyses are now extended out to NNLO, the NLO QED effects
can also become important.
For example, including QED processes in the parton evolution will
break the isospin symmetry as the up and down quarks have different
couplings to the photon.
xFitter is able to include NLO QED effects, and this is illustrated in
Ref.~\cite{xFitterDevelopersTeam:2017fxf}
which computes the photon PDF as determined
using a NNLO QCD and NLO QED analysis.

\head{Transverse-momentum-dependent distributions:}  
Transverse-momentum-dependent (TMD) parton distribution functions~\cite{Angeles-Martinez:2015sea} 
encode nonperturbative information on hadron structure, extending to the transverse plane the one-dimensional 
picture given by collinear PDFs, and providing a 3D imaging of hadron structure. Similarly to collinear PDFs, 
TMDs can be parameterised and fitted to experimental data. Within the xFitter framework, the extraction of 
TMDs from fits to experimental data has been carried out in the cases of CCFM 
evolution~\cite{Hautmann:2013tba,Dooling:2014kia,Hautmann:2014uua} 
and Parton Branching evolution~\cite{Martinez:2018jxt,Hautmann:2017fcj,Hautmann:2017xtx}.  
xFitter is able to write and manipulate TMDs in the TMDlib format~\cite{Hautmann:2014kza,Abdulov:2021ivr}.

\head{Small-$x$ resummation:}
xFitter can also study the impact of the $\ln(1/x)$-resummation corrections to the DGLAP splitting functions and  DIS coefficient functions. 
The resummation formalism for both the splitting functions and the coefficient functions  is developed in~\cite{Catani:1994sq}. 
The Leading-Log $\ln(1/x)$ (LLx) \cite{Kuraev:1976ge,Kuraev:1977fs,Balitsky:1978ic,Jaroszewicz:1982gr} and 
Next-to-Leading-Log $\ln(1/x)$ (NLLx) \cite{Catani:1994sq,Catani:1993rn,Fadin:1998py,Ciafaloni:1998gs} 
resummed calculations are implemented in 
  the public code HELL~\cite{Bonvini:2016wki,Bonvini:2017ogt}. The phenomenological effects of the  $\ln(1/x)$  resummation
  are investigated in Ref.~\cite{xFitterDevelopersTeam:2018hym}. 
In a related study~\cite{Bonvini:2019wxf}, a more flexible PDF parameterisation 
is used with xFitter which provides a better description of the combined inclusive HERA I+II data, expecially at low $x$.

\head{Pole \& $\overline{MS}$ running masses:}
Another  feature of xFitter is the ability to handle both pole
masses and $\overline{MS}$ running masses. 
While the pole mass is more closely connected to what is measured in
experiments, the $\overline{MS}$ mass has advantages on the
theoretical side of improved perturbative convergence.
xFitter was used to perform a high precision determination of the
$\overline{MS}$ charm mass in this new
framework~\cite{Bertone:2016ywq}.

\head{Dipole models:}
We have several dipole models ~{\cite{GolecBiernat:1998js,Bartels:2002cj,Iancu:2003ge}} implemented in xFitter which describe HERA inclusive and diffractive DIS cross sections very well and are a natural description of QCD reaction in the low x and low $Q^2$ region, fits to HERA data are shown in~{\cite{Luszczak:2013rxa,Luszczak:2016bxd}}.
The gluon distribution ~\cite{Luszczak:2016bxd} determined from dipole model in xFitter can be applied to the description of selected LHC processes such as gamma-proton and nucleus-nucleus collisions ~{\cite{Luszczak:2017dwf,Luszczak:2019vdc}}.

\subsection{\texorpdfstring{\lowercase{x}F\lowercase{itter and}}{xFitter and}  Snowmass2021}

We briefly discuss how xFitter might contribute to some of the future projects
being studied in the Snowmass2021 planning process.

\head{LHC \& HL-LHC:}
The xFitter package has been used for more than 100 analyses
including many LHC studies; a more complete list is available at {\tt www.xFitter.org}.
Applying this work to data taken at HL-LHC is a natural extension. 

To highlight just one LHC example, the strange quark PDF has
generated considerable attention in the recent literature.
There is a comprehensive study~\cite{Cooper-Sarkar:2018ufj}
that examines the compatibility of 
both the ATLAS~\cite{Aaboud:2016btc,Aad:2014xca}
and CMS~\cite{Chatrchyan:2013uja}
data in a uniform framework using the xFitter program.

\head{EIC \& LHeC:}
The EIC~\cite{AbdulKhalek:2021gbh} and LHeC~\cite{LHeC:2020van} facilities will provide lepton--nucleon scattering in
a collider configuration with a variety of beams.

In addition to exploring the proton PDFs, these colliders
can also study nuclear PDFs with nuclear beams, and also
meson (pion \& kaon) structure via leading neutron production.
xFitter is capable of studying both  nuclear PDFs~\cite{Walt:2019slu}
and meson PDFs~\cite{Novikov:2020snp}.
Additionally, xFitter can also compute the transverse momentum dependent (TMD)
distributions~\cite{Alekhin:2014irh}.

\head{DUNE:}
The Deep Underground Neutrino Experiment (DUNE) will use an intense neutrino beam generated
at Fermilab to study open questions about neutrino oscillations.  The massive DUNE detectors will also contribute to the study of proton decay and Grand Unified Theories, as well as observe neutrino signals from supernova core-collapse~\cite{Arguelles:2019xgp}. 

In particular, the NuSTEC white paper~\cite{Alvarez-Ruso:2017oui} outlines the status and challenges of neutrino-nucleus interactions, with special attention to DUNE. 
Improvements in PDF nuclear correction factors and the generation of nuclear PDFs fit, specifically to neutrino--nucleus interactions in the relevant energy range, can help minimize systematic errors for the +30\% fraction of DUNE events coming from the DIS region.  
This would enhance analyses in both the near and far detectors. 
Thus, improvements by xFitter on both proton and nuclear PDFs~\cite{Walt:2019slu}  can contribute 
to the DUNE project.

\head{UHE Cosmic Rays}
Recent advances in neutrino astronomy have enabled us to study
ultra-high energy cosmic rays (UHECR) by studying atmospheric
neutrinos.  For example, IceCube~\cite{Aartsen:2016xlq,Aartsen:2013jdh}
has isolated more than 100 high-energy cosmic neutrinos, with energies
between 100~TeV and 10~PeV.

Interpretation of these measurements would benefit from accurate PDFs
in the low-$x$ region.  An example application is the evaluation of
the prompt flux of atmospheric neutrinos originating from the
semileptonic decays of heavy-flavored hadrons produced in the
interactions of UHECR with nuclei in the
atmosphere~\cite{Bhattacharya:2016jce,Zenaiev:2019ktw}.  The prompt
atmospheric-neutrino flux represents a relevant background for
searches of highly energetic cosmic neutrinos.
Thus, increased precision of both PDFs and nuclear corrections in the very
low-$x$ region would improve theoretical predictions in this extreme kinematic region.

\section{Future Developments with \texorpdfstring{\lowercase{x}F\lowercase{itter}}{xFitter}}

The latest xFitter release, 2.2.0, represents a significant restructuring of the code. The new version provides significantly improved modularity for PDF parameterization, evolution,
theory predictions and minimization. This additional flexibility simplifies further developments of the code.

The planned developments contain improvements of the existing functionality and expansion of the code capabilities. Both goals will be achieved in a modular manner, without increase of the requirements for the core package. An emphasis is given on usage of modern, industry-standard libraries.  

There are several ideas on improvement of the execution time required for minimization. This is important in particular in view of large amount of data samples which will come from LHC, EIC, and other experiments. xFitter already contains an interface to the FK-tables (``apfelgrid'') via interface to APFEL. These tables are based on APPLGRID predictions and combine evolution and convolution in one step. The interface to modern minimization packages, such as CERES~\cite{ceres-solver}, which is included in version 2.2.0, provides additional opportunity to employ automatic differentiation.
Introduction of this capability is planned for the next release. Combined with improved minimization code, this should improve convergence time for specific problems by
more that order of magnitude.

xFitter already contains highly efficient code for likelihood function calculation. It will be improved further, by more extensive use of BLAS and EIGEN libraries.

The state-of-the-art PDF fits include NNLO QCD plus NLO electroweak corrections. However, in many cases these corrections are provided as fixed k-factor tables. It is planned to include them using grid-based methods or with semi-analytic routines~\cite{H1:2018mkk,Britzger:2020kgg}.

xFitter plans further developments beyond 
conventional PDFs. In particular, 
the TMDs which have already been introduced 
through the parton branching 
method~\cite{Hautmann:2017fcj} will be developed 
by including branching scale-dependent resolution parameters~\cite{Hautmann:2019biw} as 
a new functionality.  Furthermore, 
TMDs will also be introduced 
independently of the parton branching method 
using the existing interface to the DYTURBO~\cite{Camarda:2019zyx} package. 
It is also planned to introduce functionality for simultaneous PDF and fragmentation function fits. 

The inclusion of theoretical uncertainties in PDFs 
is currently one of the main targets of 
state-of-the-art PDF global fits.   xFitter plans 
to provide PDF sets 
incorporating theoretical uncertainties by applying  
the resummation-scale  
technique~\cite{Bertone:2022sso,Bertone:2022ope}.  

NLO EW corrections are important at the precision level of the Drell-Yan measurements at the LHC and the Tevatron.
Complete one-loop EW corrections for NC and CC Drell-Yan processes are realized in ReneSANCe Monte Carlo event generator
that uses the SANC modules for differential cross-sections. The predictions obtained by ReneSANCe were thoroughly cross-checked
with the results of MCSANC integrator. The complete one-loop EW corrections can be separated in the code
into weak, QED FSR, QED ISR and interference (IFI) contributions. The dominant part of the higher-order EW corrections is QED FSR
which can be modelled by PHOTOS program. The effects of QED FSR calculated by PHOTOS and ReneSANCe agree very well. The K-factors for
the remaining parts of NLO EW can be computed with ReneSANCe for various differential cross sections, $A_{FB}$ and charge asymmetries:
$
K^{\text{EW}} = \frac{d\sigma_{\text{LO QCD}}^{\text{NLO EW}}}{d\sigma_{\text{LO QCD}}^{\text{LO EW}}}.
$
In factorized approach this K-factors applies to NNLO QCD cross-sections:
$
d\sigma_{\text{NNLO QCD}}^{\text{NLO EW}} = d\sigma_{\text{NNLO QCD}}^{\text{LO EW}} \cdot K^{\text{EW}}.
$

In the case of NC DY the separation of the FSR contribution is straightforward and appears at the level of Feynman diagrams. But the corresponding separation in the case of CC DY is not trivial and it is even not gauge invariant. For the sake of agreement with PHOTOS, a special prescription for this separation was introduced in SANC modules.

A formal separation of the pure weak (PW) and QED contributions $\delta^{PW}$ and $\delta^{QED}$ to the total $W^+ \to u+\bar{d}$ decay width $\Gamma_W^{PW+QED} = \Gamma_W^{LO}(\delta^{PW}+\delta^{QED})$ is depend on 't Hooft scale parameter $\mu_{PW}$ and $\delta^{QED} = \frac{\alpha}{pi}\left[Q_W^2(\frac{11}{6}-\frac{\pi^2}{3}) + (Q_u^2+Q_d^2)(\frac{11}{8}-\frac{3}{4}\log\frac{M_W^2}{\mu_{PW}^2})\right]$. In order to separate the FSR QED we can choose $\mu_{PW} = M_W\exp(-\frac{11}{12})$. This is in agreement with the corresponding treatment in PHOTOS.

\section{Overview of recent xFitter results}

\subsection{W and Z Boson Production}
\begin{figure}[tbh]
\includegraphics[width=0.40\textwidth]{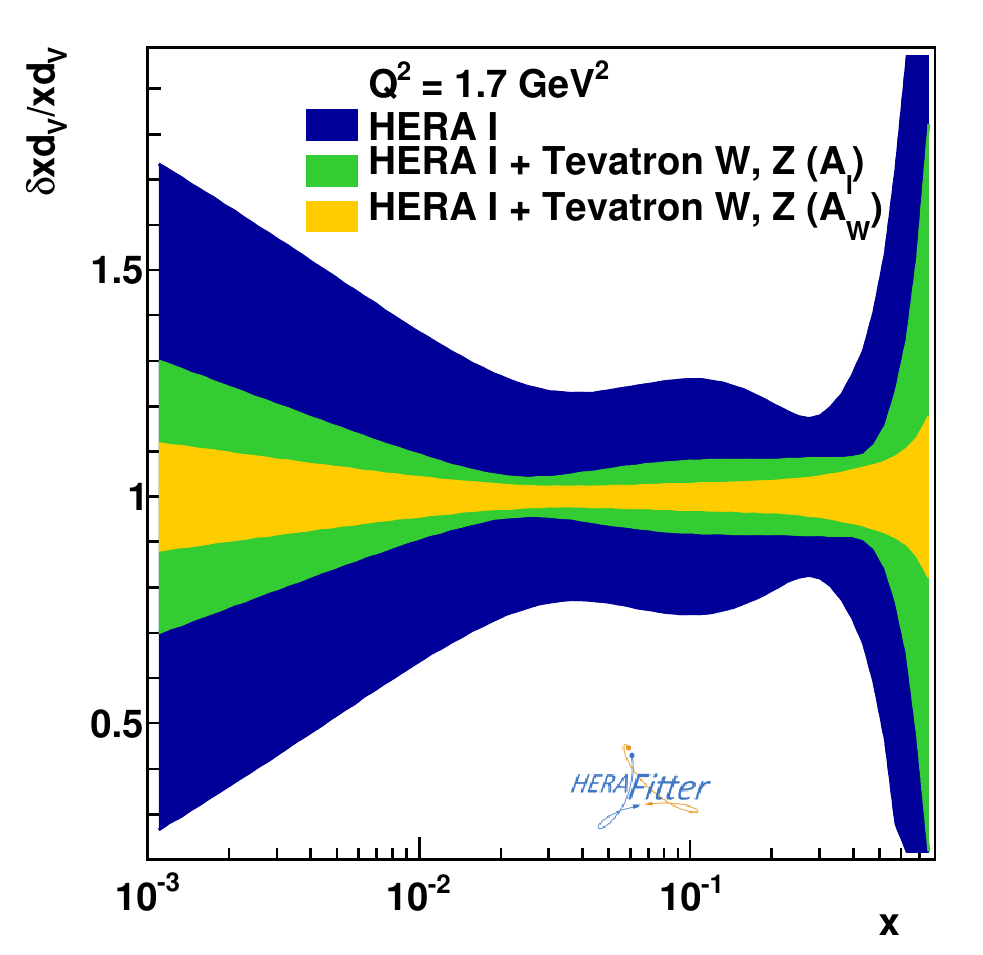}
\caption{ 
We show the d-valence PDF relative  uncertainty 
at $Q^2 = 1.7~{\rm GeV}^2$ as a function of the Bjorken-$x$
determined with a fit to the HERA~I data (blue), 
HERA~I and Tevatron W-boson asymmetry and Z-boson data (yellow), 
and  HERA I and Tevatron W-boson lepton asymmetry and Z-boson data (green).
Figure from Ref.~\cite{HERAFitterdevelopersTeam:2015cre}.
}
\label{fig:wzprod}
\end{figure}

In Reference~\cite{HERAFitterdevelopersTeam:2015cre}, the xFitter collaboration
(formerly HERAFitter) analyzed  measurements of the W-boson charge asymmetry and of the Z-boson production cross sections, performed at the Tevatron collider in Run II by the D0 and CDF collaborations.
Figure~\ref{fig:wzprod}
shows the comparison of the relative uncertainty of the
PDFs using the HERA~I data alone, and together with the Tevatron $W/Z$ data. 
This new data significantly reduces the PDF uncertainty, and
this is most noticeable for the case of the $d$-valence, as shown.

To study the possible model dependence due to the
W-boson rapidity reconstruction, an alternative fit was
performed in which the W-boson charge asymmetries
measured by CDF and D0 were excluded, and the latest
D0 measurement of the electron asymmetry was included.
In Fig.~\ref{fig:wzprod},  the fit to the lepton
asymmetry data (green band) yields very compatible results
to the W-boson asymmetry data (yellow band), 
but the uncertainties on the $d_v$ PDF are up to twice as large.

These findings highlight the importance of the Tevatron W- and Z-boson production
data to constrain the PDFs, 
and illustrate the utility of xFitter to easily demonstrate the impact of individual data sets. 
All the supporting material to allow fits
of the Tevatron data, including the updated correlation
model and the grid files for fast theory calculations, are
publicly available on the xFitter web page.

\subsection{The Photon PDF}
\begin{figure}[tbh]
\includegraphics[width=0.40\textwidth]{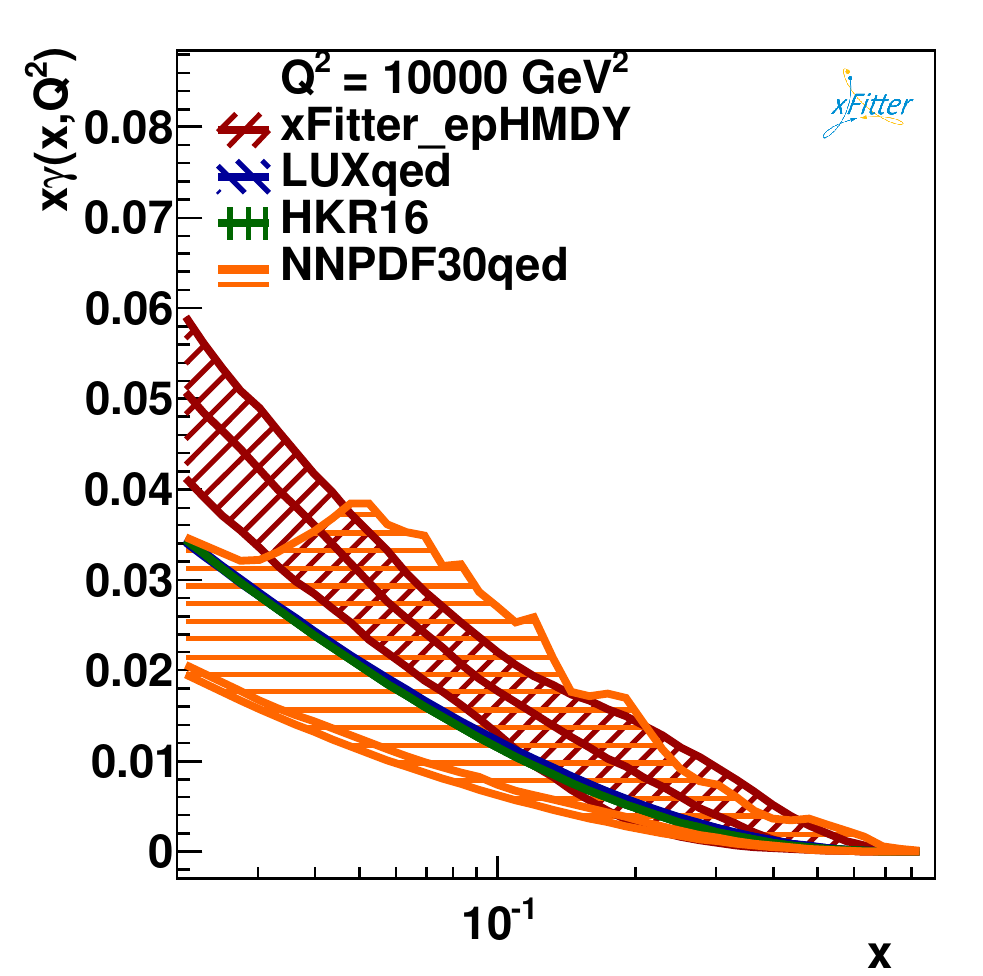}
\caption{Comparison between the photon PDF $x \gamma(x,Q^2)$
at $Q^2{=}10^4\,{\rm GeV}^2$  from the present NNLO analysis
(xFitter\_epHMDY) with the corresponding results from 
NNPDF3.0qed~\cite{NNPDF:2014otw}, 
LUXqed~\cite{Manohar:2016nzj} and 
HKR16~\cite{Harland-Lang:2016apc}. 
The PDF uncertainties are shown at the 68\%~CL obtained from the MC method, while model and parametrisation uncertainties are addressed separately.
For HKR16 only the central value is shown, 
while for LUXqed the associated PDF uncertainty band  is included.
Figure from Ref.~\cite{xFitterDevelopersTeam:2017fxf}.
}
\label{fig:photon}
\end{figure}
\begin{figure*}[!tbh]
\mbox{
\includegraphics[width=0.33\textwidth]{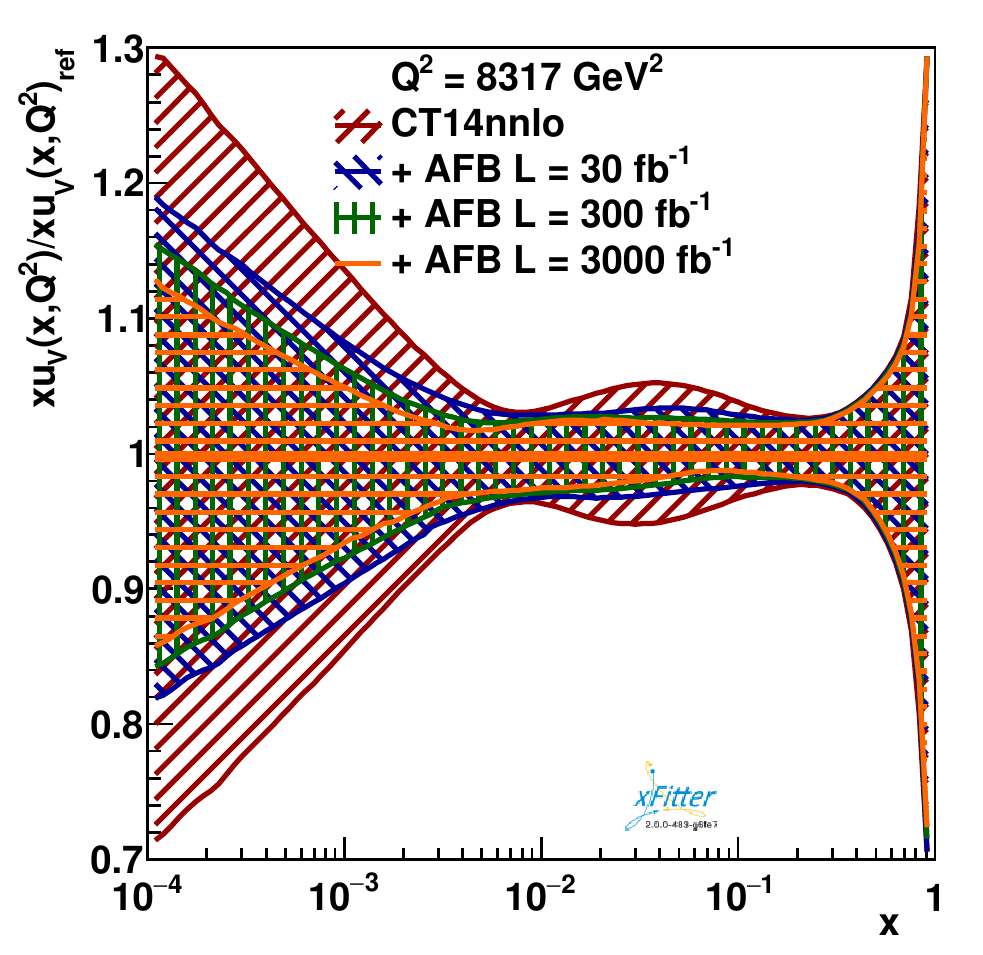}
\includegraphics[width=0.33\textwidth]{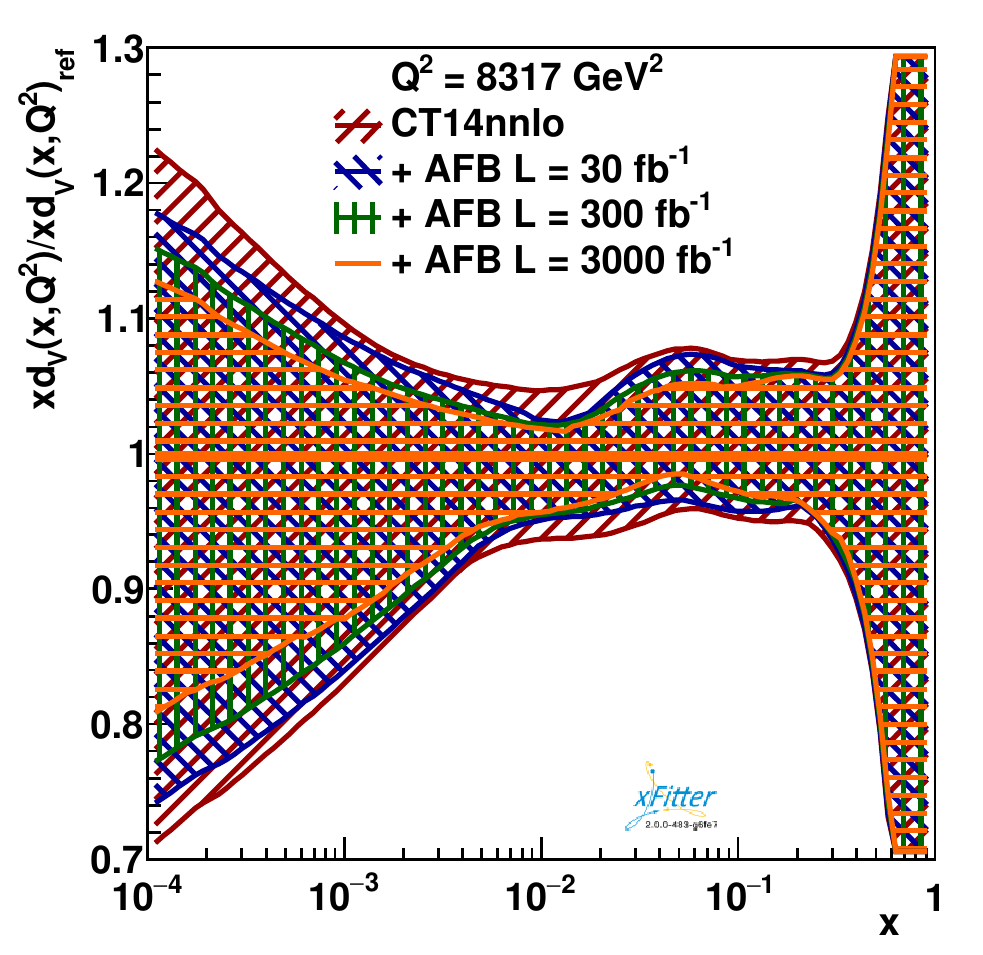}
\includegraphics[width=0.33\textwidth]{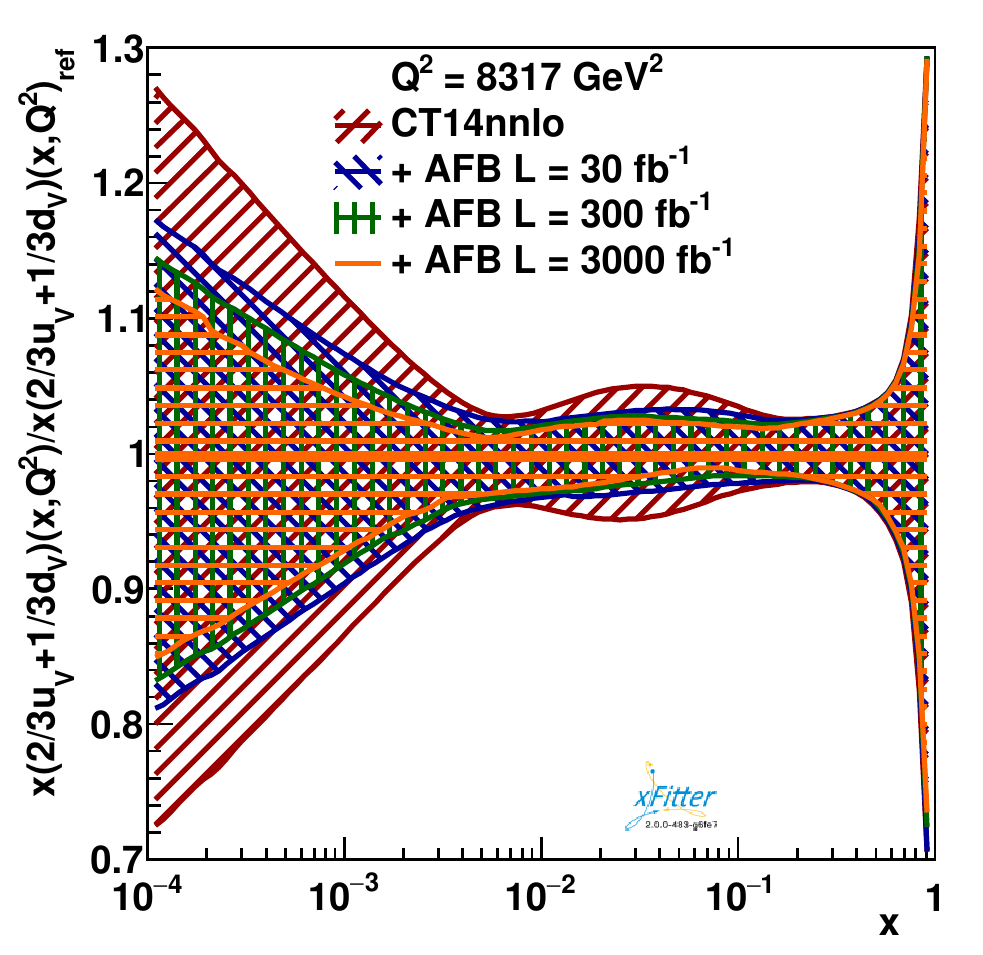}
}
\caption{A sample profiling comparison illustrating the impact of the $A_{FB}$ data on the valence quark PDFs and their weighted sum for the original (red) and profiled (blue, green, orange) CT14nnlo~\cite{Dulat:2015mca} set.
Figure from Ref.~\cite{Accomando:2019vqt}.
}
\label{fig:afb}
\end{figure*}

Achieving the highest precision for theoretical predictions  requires that the calculations  include perturbative QCD corrections up to (N)NNLO, and electroweak (EW) corrections up to NLO. %
Working at this level of precision demands the QED effects are fully included, 
and this requires the introduction of the   photon parton distribution of the proton,  $x \gamma(x,Q^2)$. 

To demonstrate these capabilities, Ref.~\cite{xFitterDevelopersTeam:2017fxf}
analyzed 
recent ATLAS measurements of high-mass Drell-Yan dilepton production at $\sqrt{s}=8$~TeV  
to determine the photon PDF,
and to compare it with some of the existing determinations from the literature. 

To  include the photon PDF, 
xFitter links to   the APFEL program~\cite{Bertone:2013vaa} to incorporate  NLO QED effects. 
The inclusion of NLO QED evolution effects are cross checked using the independent QEDEVOL~\cite{Sadykov:2014aua}  code based on the QCDNUM~\cite{Botje:2010ay} evolution program.
Additionally, the  aMCfast interface~\cite{Bertone:2014zva} is used to include   the photon-initiated contributions in the EW calculations within MadGraph5\_aMC@NLO~\cite{Alwall:2014hca}.

The resulting photon PDF determination  represents an important validation of our understanding on the nature and implications of the photon PDF.
Fig.~\ref{fig:photon} shows 
the results as compared with other recent QED fits.
For $x\leq 0.1$, the four determinations of the photon PDF are consistent within  uncertainties.
For smaller values of $x$, the photon PDF from LUXqed and HKR16 is somewhat smaller than \mbox{xFitter\_epHMDY}, but
still in agreement at the 2-$\sigma$ level. This agreement is further improved if the PDF uncertainties in \mbox{xFitter\_epHMDY}
arising from variations of the input parametrisation are added to experimental uncertainties.
Moreover, the results of this work and NNPDF3.0QED agree at the 68\% CL for $x\leq 0.03$, and the agreement extends
to smaller values of $x$ once the parametrisation uncertainties in xFitter\_epHMDY are accounted for. The LUXqed and
the HKR16 calculations of $x\gamma(x, Q^2)$ are very close to each other across the entire range of $x$.

The results of this study have been made possible by a
number of technical developments that should be of direct application for future PDF fits accounting for QED corrections. 
These technical improvements will certainly be helpful for future
studies of the photon PDF of the proton.
\subsection{The Forward-Backward Asymmetry in Neutral Current Drell-Yan Production}

The DY-induced lepton charge asymmetry in charged current (CC) processes has been an effective way to constrain PDFs.
Ref.~\cite{Accomando:2019vqt} examines the use of neutral current (NC) lepton charge asymmetry measurements, which are traditionally used in the context of precision determinations of the weak mixing angle $\theta_W$,
to obtain improved PDF constraints.

Specifically, a forward-backward asymmetry $A^{*}_{FB}$ is  computed at NLO
using the \mbox{MadGraph5\_aMC@NLO} program~\cite{Alwall:2014hca} interfaced to APPLgrid~\cite{Carli:2010rw} through aMCfast~\cite{Bertone:2014zva}.
The $A^{*}_{FB}$ is sensitive to valence quark PDFs through the combination of chiral couplings $(2/3 u_V + 1/3 d_V)$.
The valence quark PDFs in turn also influence the sea PDF constraints via the sum rule relations.

For this study,  three sets of pseudodata relevant for LHC Run~2, 3,~and HL-LHC were considered. 
Fig~\ref{fig:afb} displays the impact of the data on the valence quark PDFs and their combination weighted by their electric charges, for 30, 300 and 3000$\,{\rm fb}^{-1}$.
Comparing those profiled error bands, we note a visible improvement in the
distribution of the valence quark PDFs, especially in the region of small $x$.
Some improvement in the high $x$ region of quark PDFs can be obtained employing sufficiently large data samples and applying suitable rapidity cuts on the $A^{*}_{FB}$ observable.

This analysis illustrates how high-statistics measurements, both cross section and asymmetry distributions, from the LHC Runs 2, 3 and the HL-LHC stage can be exploited to place constraints on the PDFs. 
The PDF profiling calculations in the xFitter framework can flexibly be used to investigate the impact of pseudodata on PDF determinations.

\subsection{Impact of low-x resummation on QCD analysis of HERA data}

\begin{figure}[tb]
\includegraphics[width=0.49\textwidth]{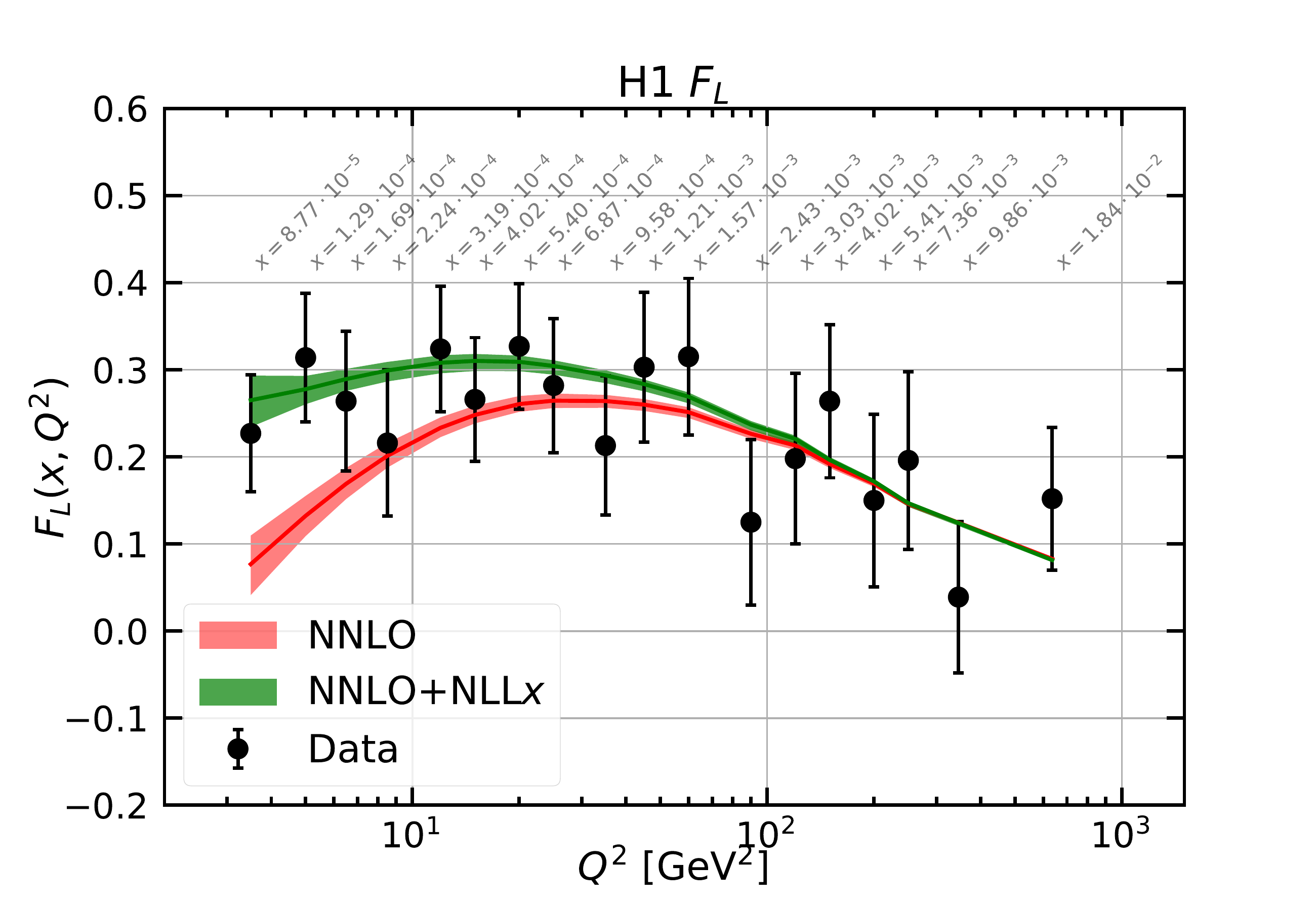}
\caption{The H1 extraction of $F_L$ compared to the predictions with
and without the  $\ln(1/x)$ resummation (NLLx) contributions.
Figure from Ref.~\cite{xFitterDevelopersTeam:2018hym}.
}
\label{fig:FLH1}
\end{figure}

In Ref.~\cite{xFitterDevelopersTeam:2018hym}, xFitter explored the 
impact of resumming the Next-to-Leading-Log $\ln(1/x)$  (NLLx) contributions in the small~$x$ region
using a  QCD analysis of HERA data.
This work was prompted, in part, by the observation that the 
$\chi^2$  of the NNLO fits was not improved as compared with the NLO 
fit for low values of $Q^2$.  

This analysis used the  final combined $e^\pm p$  cross-section measurements of
H1 and ZEUS~\cite{H1:2015ubc}  which covered the kinematic range of $Q^2$ from
0.045~GeV${}^2$ to 50000~GeV${}^2$ and of Bjorken $x$ from 0.65
down to $6{\times}10^{-7}$. 
Fits to this final combined HERA DIS cross-section data within the conventional DGLAP framework of QCD have shown some tension at low~$x$ and low~$Q^2$. 
Our goal was to determine whether  incorporating the $\ln(1/x)$-resummation terms into the HERAPDF fits 
might help resolve these tensions.

While the details are presented in Ref.~\cite{xFitterDevelopersTeam:2018hym}, 
in brief it was observed that the total $\chi^2$ for the fits improved from 1468/1207 for the NNLO fit,
to 1394/1207 for the NNLO+NNLLx fit.
The nominal improvement of the total $\chi^2$ reflects the fact that the small~$x$ data is only a portion of the 
full data set;
however, 
the impact in the small~$x$ region is significant 
as indicated in Fig.~\ref{fig:FLH1} where we see the $\ln(1/x)$ contributions yield an
improved fit for the longitudinal structure function
in the small~$x$ and low~$Q^2$ region.
The precise   kinematic region where this resummation is important is delineated 
in Fig.~11 of Ref.~\cite{xFitterDevelopersTeam:2018hym}.

Additionally, the resummation yields a larger gluon PDF in the small~$x$ and low~$Q^2$ region,
and this avoids potential problems due to a negative gluon PDF in forward physics. 
These features make PDFs with ln(1/x) resummation much
more suitable for use in MC generators, such as Sherpa~\cite{Gleisberg:2008ta}.

In conclusion, $\ln(1/x)$ resummation provides substantial improvement in the description of the precise \hbox{HERA1+2}
combined data. 
The NLLx contributions yield an improved description of the data, especially the longitudinal structure function $F_L$
in the small~$x$ region, as compared to both the (un-resummed) NLO and NNLO analysis, and 
it also avoids potential problems of the negative gluon PDF at low~$x$ and $Q^2$.
\subsection{Heavy quark matching scales: Unifying the FFNS and VFNS}

\begin{figure}[!htb]
\includegraphics[width=0.40\textwidth]{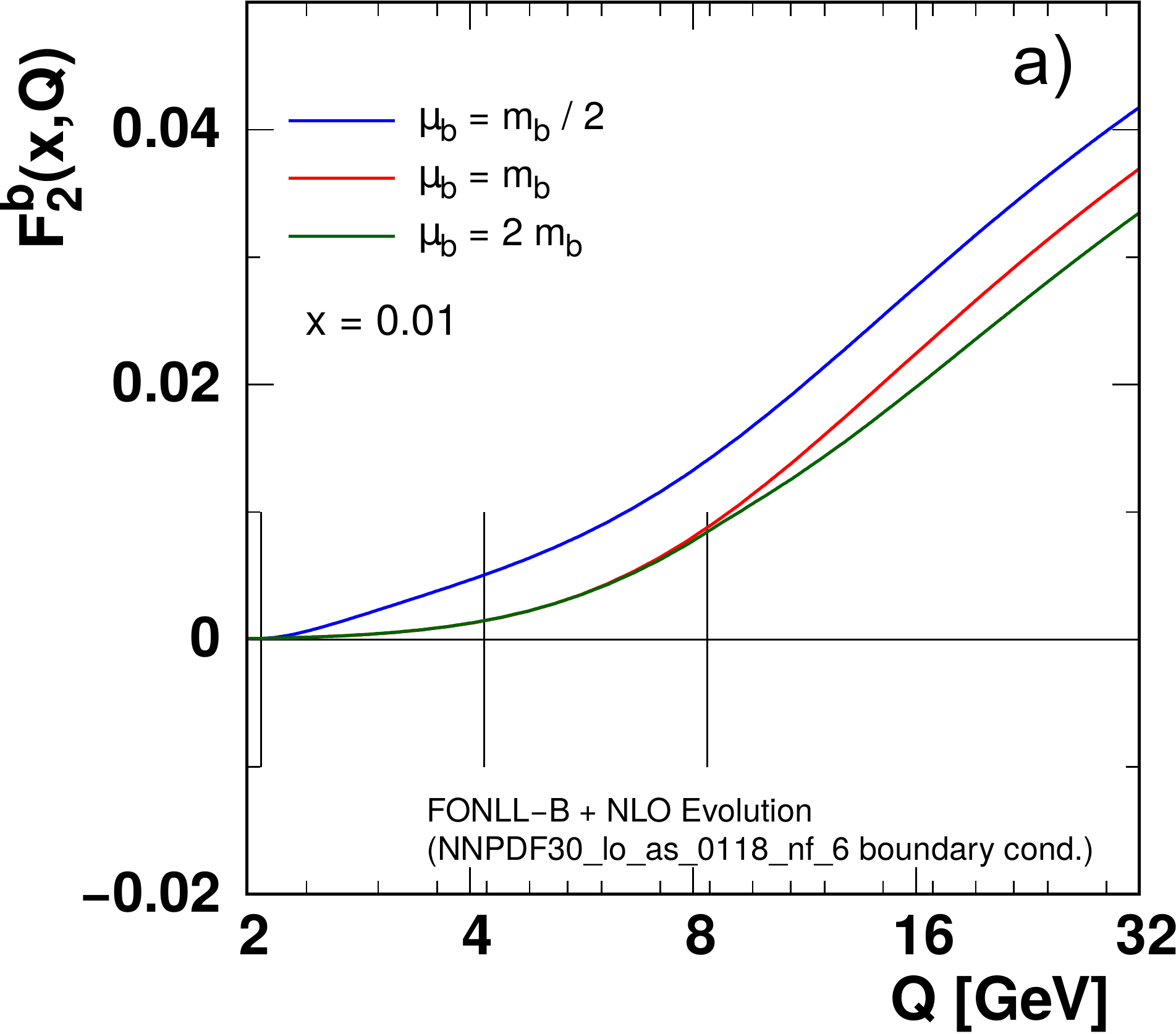} 
\hfil
\includegraphics[width=0.40\textwidth]{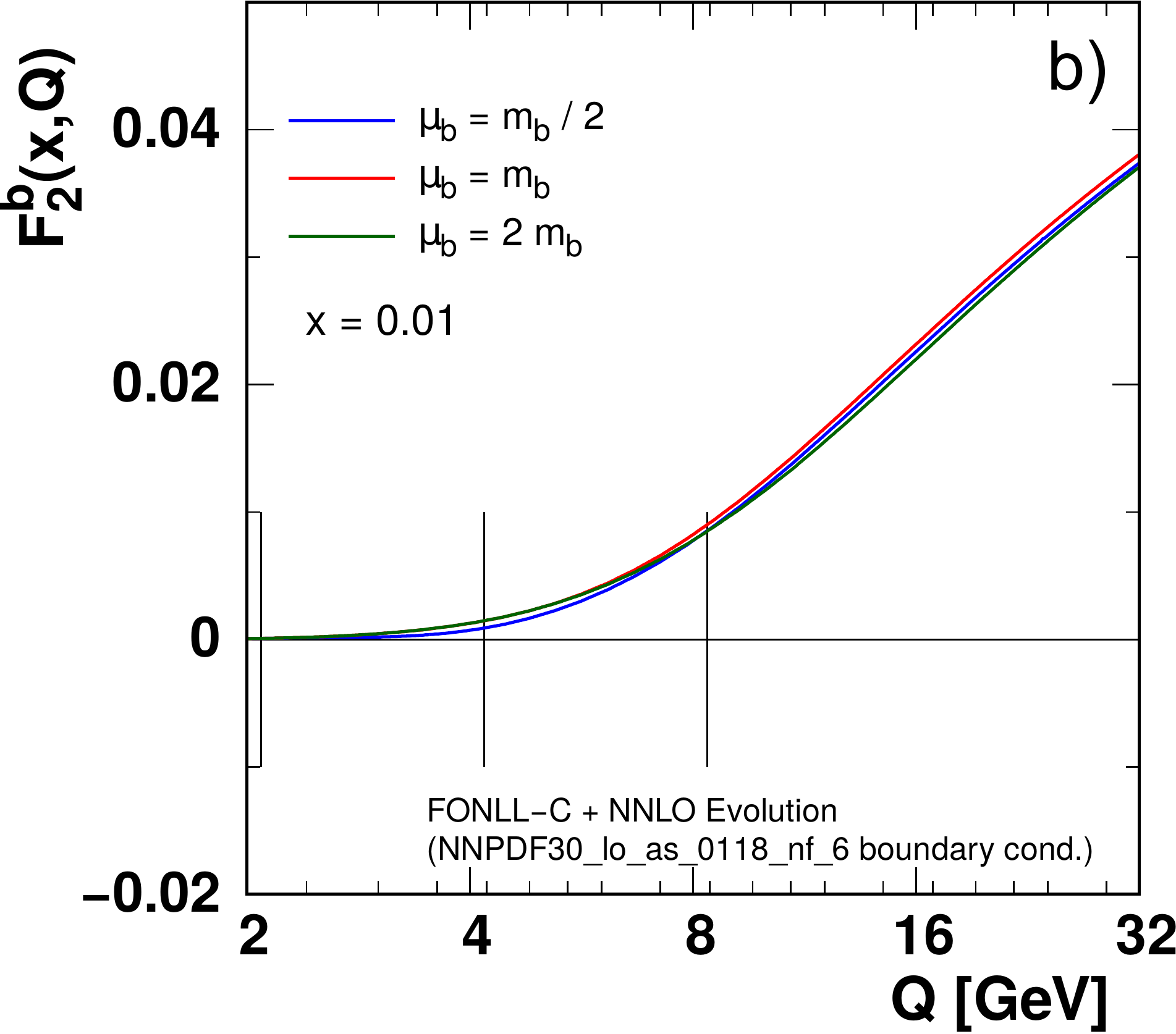}
\caption{The b-quark structure function
$F_2^b(x,\mu)$  for different choices of the matching scales $\mu_m {=} \{m_b /2, m_b , 2m_b \}$ 
(indicated by the vertical lines) computed at NLO (Fig.-a) and NNLO (Fig.-b).
Figure from Ref.~\cite{xFitterDevelopersTeam:2017fzy}.
}
\label{fig:matching}
\end{figure}

High precision phenomenological analysis of DIS data requires 
a proper treatment of the heavy quarks mass scales. 
The Variable Flavor Number Scheme (VFNS) 
incorporates  the heavy quark mass scale across the
full kinematic range by varying the number of active
flavors ($N_F$) in the DGLAP QCD evolution.
In the VFNS,  the matching scale $\mu_m$  determines  the transition 
from $N_F$ to $N_F+1$ active flavors.
Historically, $\mu_m$   was taken to be
the heavy quark mass $m_H$; however, 
this choice is not required, and there are advantages to using a flexible
matching scale. 
This feature is implemented in xFitter via the APFEL code, 
and we demonstrate the benefits   in Ref.~\cite{xFitterDevelopersTeam:2017fzy}.

An essential element of this analysis is to implement the correction matching of the PDFs
across the flavor threshold. 
The matching conditions have been computed to NLO and NNLO within the xFitter code,
and the impact of this choice on the  physical structure function $F_2^b(x,Q)$
is illustrated in  Fig.~\ref{fig:matching}.  
We emphasize some key features illustrated in the matching conditions. 

At NLO (Fig.~\ref{fig:matching}-a), we observe that there are sizeable differences for the three choices 
of matching scales, $\mu_b=\{m_b/2, m_b, 2m_b\}$. 
As the NLO  matching conditions incorporate the ${\cal O}(\alpha_S)$ 
DGLAP evolution contributions, the differences of these curves simply reflects the NNLO 
${\cal O}(\alpha_S^2)$ correction. 
Thus, when we compute the NNLO result (Fig.~\ref{fig:matching}-b), 
we find the differences due to the matching scale are now 
significantly reduced as they are N${}^3$LO ${\cal O}(\alpha_S^3)$.

Ref.~\cite{xFitterDevelopersTeam:2017fzy}  illustrates the use of
the variable matching scale for the case of heavy quark (charm and bottom) production at HERA.
The flexibility of choosing the $\mu_m$  matching scale provides a number of advantages. 
By adjusting the matching scale, we can effectively transition between a VFNS and a FFNS in a seamless manner. 
For example, $\mu_m{=}m_b$ would correspond to the traditional VFNS, and $\mu_m{\to} \infty$
would correspond to the traditional FFNS. Furthermore we can choose any scale in between.
Thus, the  variable heavy flavor matching scale $\mu_m$
generalizes   the transition between the
FFNS and the VFNS.

On a more practical level, this flexibility allows one to shift the matching scale 
so that the 
discontinuities  associated with the $N_F$ to $N_F+1$ transition
do not lie in the middle of a specific data set.
In summary,  the flexibility of choosing the heavy flavor matching scale $\mu_m$
generalizes the transition between a
FFNS and a VFNS, and provides a theoretical
``laboratory'' which can quantitatively test proposed
implementations. 
The ability to vary the
heavy flavor matching scales  not only provides
new insights into the intricacies of QCD, but also has
practical advantages for PDF fits
\subsection{Constraining the Strange PDF}

\begin{figure}[tbh]
\includegraphics[width=0.40\textwidth]{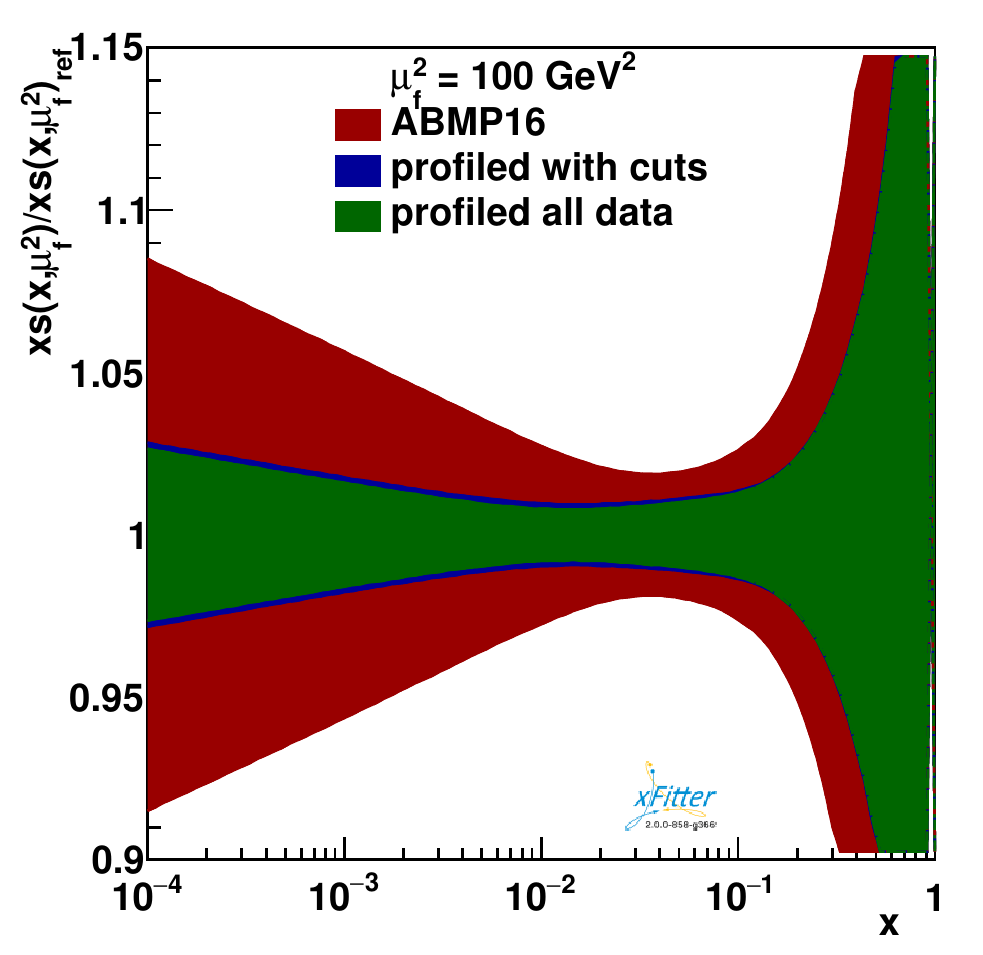}
\caption{
The relative strange  PDF uncertainty at 
$\mu_f^2{=}100$~GeV${}^2$ of the original and
profiled ABMP16 PDF set.
Figure from Ref.~\cite{xFitterDevelopersTeam:2019ygc}.
}
\label{fig:strange}
\end{figure}

The study of heavy quark production plays a critical role enabling us to fully 
characterize the properties of the SM. 
The charm quark is
especially useful in this respect as the combination of 
fixed-target and collider data allow us to investigate the 
full kinematic spectrum from the  threshold 
region ($m_c\sim Q$) to the asymptotic high energy limit ($m_c\ll Q$).
Additionally, the charged-current charm production provides 
direct access to the strange quark distribution. 
The strange PDF is of particular interest because it still has large 
uncertainties despite being extensively investigated in a number
of experiments such as inclusive $W/Z$ production and $W+c$ associated production.
Looking to the future, it is clearly important to reduce
the uncertainty of the PDFs in general, and the 
strange-quark in particular, as we strive
to make increasingly precise tests of the SM.

In Ref.~\cite{xFitterDevelopersTeam:2019ygc}, 
we used the xFitter framework to study charm production in charged-current deep-inelastic scattering (DIS) using  LHeC pseudodata to estimate the potential improvement from a future DIS facility. As xFitter implements both fixed-flavor- and variable-flavor-number schemes, we also compared the impact of these different theoretical choices to highlight several interesting aspects of multi-scale calculations.

For the LHeC parameters, we assumed a
7~TeV proton beam on a 60~GeV electron
beam yielding $\sqrt{s}\sim 1.3$~TeV
with a nominal design
luminosity of $10^{33}{\rm cm}^{-2}\, {\rm s}^{-1}$.
Compared to HERA, 
this extends the covered kinematic range by an order
of magnitude in both $x_{Bj}$ and $Q^2$. 
The predictions are
provided for unpolarized beams in the kinematic range
$10^2 < Q^2 < 10^5 \, {\rm GeV}^2$, 
$10^{-4} < x_{Bj} < 0.25$,
and 
$0.0024 < y < 0.76$.

Using the high-statistics pseudodata with the xFitter analysis, 
we find strong constraints on the strange-quark PDF,
especially in the previously unexplored small-$x_{Bj}$ region.
Figure~\ref{fig:strange} illustrates the relative improvement of the 
strange PDF uncertainty using the ABMP16 PDFs~\cite{Alekhin:2018pai}.
The general reduction of the PDF uncertainties is evident across the full $x$ range. 
We also performed calculations using the NNPDF3.1 PDFs~\cite{NNPDF:2017mvq} and obtained similar conclusions. 

As xFitter can compute in both the Fixed Flavor Number Scheme (FFNS)
Variable Flavor Number Scheme (VFNS),
we investigated the differences due the the scheme choice. 
We computed the reduced cross section using both the FFNS and VFNS, 
and then cut out those pseudodata where the scheme uncertainty was larger 
than the PDF uncertainty. In Fig.~\ref{fig:strange} this result  (in blue) is 
labeled as ``profiled with cuts."
If we include all the data (labeled, ``profiled all data'') we obtain 
the uncertainty band labeled in green. 
Comparing these two results, we find that the pseudodata can impose very strong 
constraints on the PDFs and this is  independent
of the particular heavy-flavor scheme.

\subsection{The Pion PDF}

\begin{figure*}[tbh]
\includegraphics[width=0.85\textwidth]{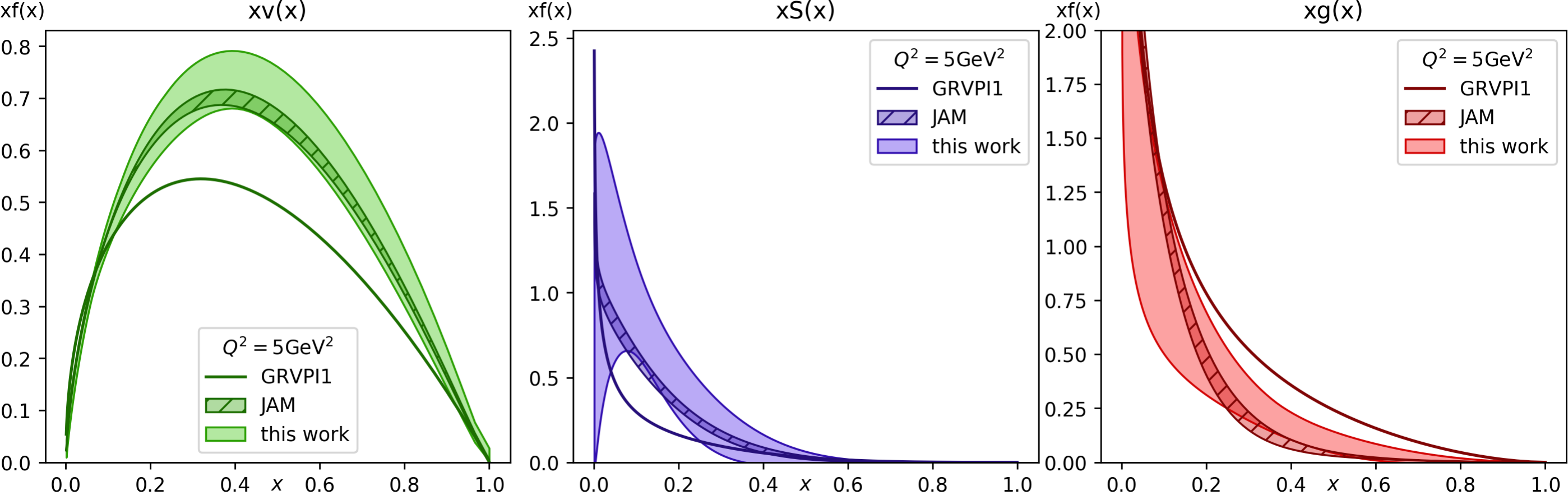}
\caption{Comparison between the pion PDFs obtained in this work, a determination by the JAM collaboration~\cite{Barry:2018ort},
and the GRVPI1 pion PDF set~\cite{Gluck:1991ey}.
Figure from Ref.~\cite{Novikov:2020snp}.
}
\label{fig:pionPDF}
\end{figure*}

To further demonstrate the versatility of xFitter, 
Ref.~\cite{Novikov:2020snp}  presented the first open-source analysis 
of parton distribution functions (PDFs) of charged pions.
The pion is the  simplest $q \bar{q}$ state in the quark-parton model
of hadrons. However, despite this apparent simplicity, 
the pion structure is currently poorly understood,
especially compared to the proton. 

Experimentally, the pion PDF is known mostly from
QCD analyses of Drell-Yan (DY) and prompt photon
production data.
This analysis used a 
combination of Drell-Yan (E615 and NA10) and prompt photon
(WA70) data to  provide constraints on both the quarks and
gluons across the kinematic range.

The calculations are implemented at next-to-leading order 
(NLO) using APPLgrids generated by MCFM generator which  allows
for efficient numerical computations; additionally, 
modifications were made to APPLgrid which facilitate both
meson and hadron PDFs in the initial state.

Fig.~\ref{fig:pionPDF} presents the fitted pion PDFs 
and compares with results from JAM~\cite{Barry:2018ort} and GRVPI1~\cite{Gluck:1991ey}.
In general, the results obtained from xFitter compare favorably to JAM within uncertainties,
but do show differences with the older GRV analysis. 
For the uncertainty bands,  the $\mu_R$ and $\mu_F$ renormalization scales were varied, 
as well as the PDF parameter values. 
For the  valence distribution, the dominant uncertainty is the scale variation,
and this is rather well constrained. 
For the sea and gluon distributions, 
the calculation only included   the direct photons in the model; hence, 
the missing fragmentation contribution to the prompt
photon production process is the dominant uncertainty source. 
This is certainly an issue to be improved in a future analysis. 

To further illustrate the flexibility of the xFitter framework,
both a \mbox{3-parameter} and \mbox{4-parameter} form for the initial valence distribution was used.
This type of parameter flexibility is crucial as it allows, for example, 
the investigation of the PDF slope in the large $x$ limit. 

While this study demonstrated the versatility of xFitter to perform meson PDF analysis, 
there are numerous directions this work can be extended. 
For example, new $J/\Psi$ data  could play an important
role in constraining the gluon~\cite{Chang:2020rdy},
and data   from
future experiments such as COMPASS++/AMBER~\cite{Adams:2018pwt},
may allow for   more flexible parameterizations and improved 
constraints.

\subsection{The Pion Fragmentation Function}
\begin{figure}[tbh]
\mbox{
\includegraphics[width=0.24\textwidth]{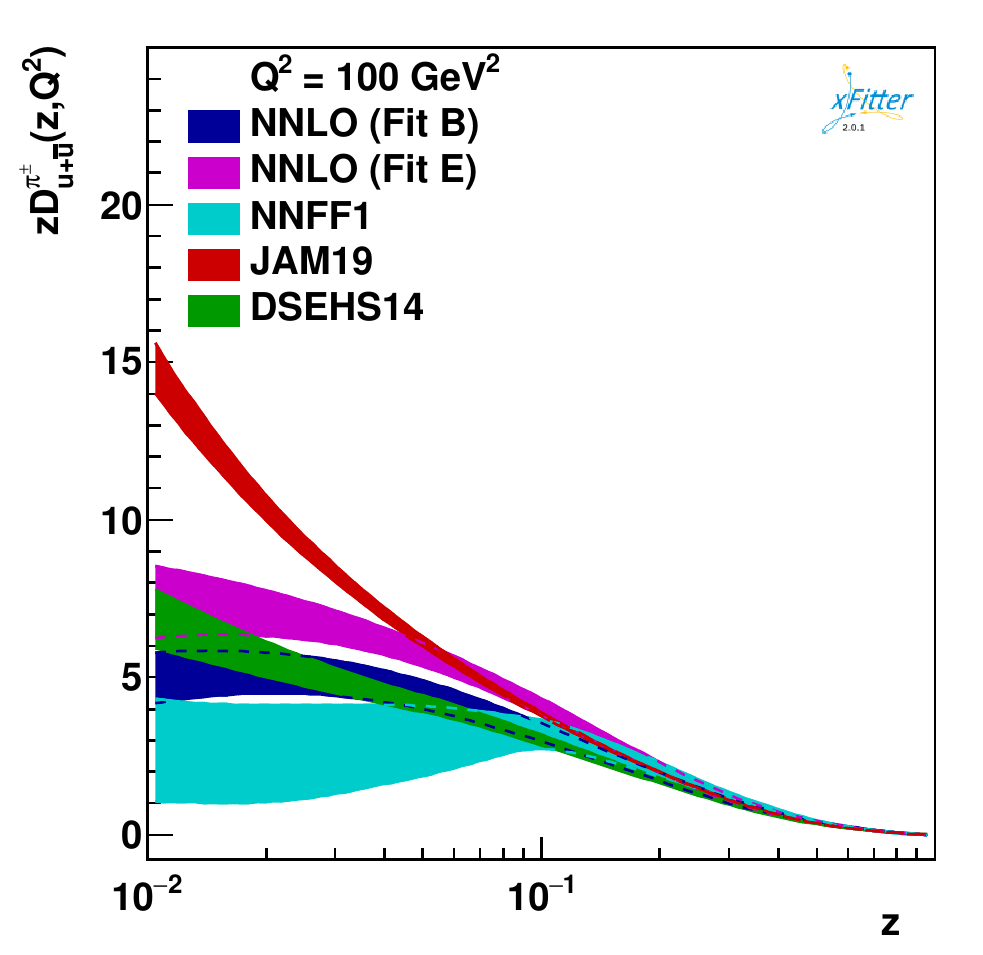}
\includegraphics[width=0.24\textwidth]{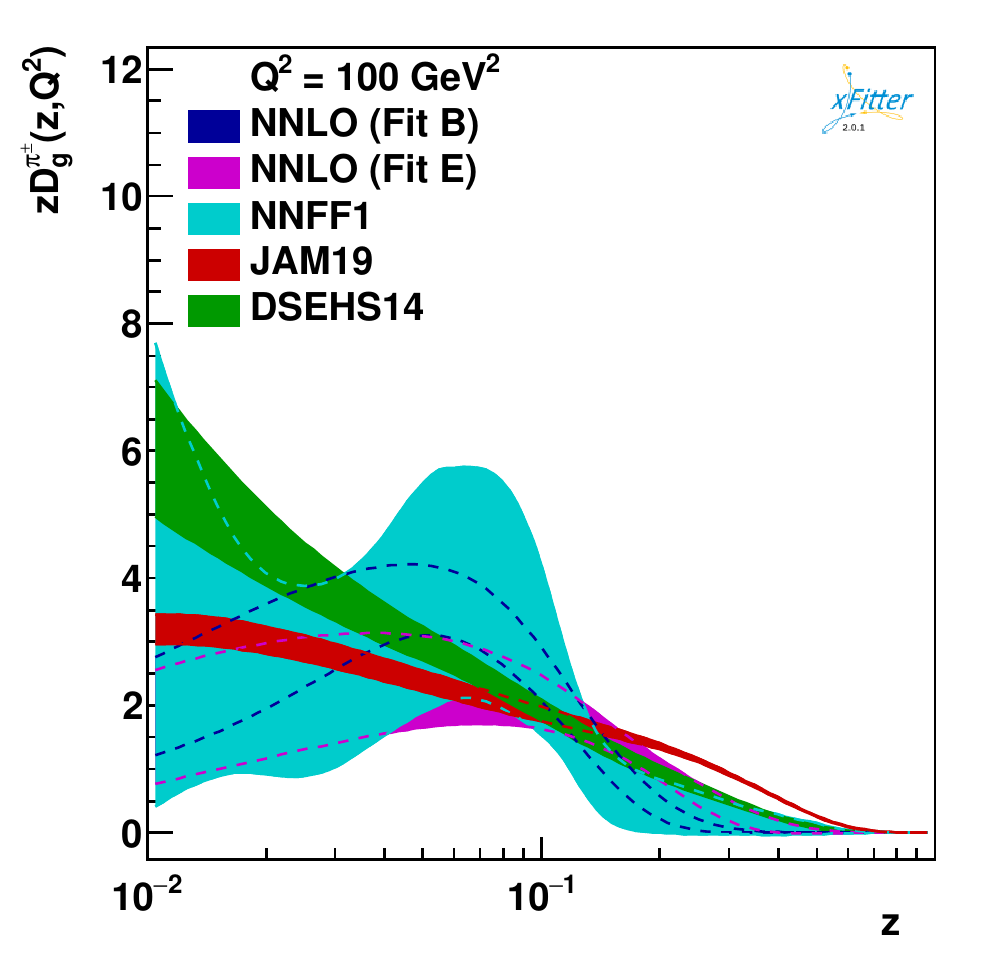}
}
\caption{
Comparison of the preferred Fit~E~[IPMx] 
as well as Fit~B for charged pion FFs ($\pi^+ {+} \pi^-$) at NNLO with 
NNFF1 [4] at NNLO, JAM19 [13] at NLO, DSEHS [5] at NLO
for   $Q^2 {=} 100\, {\rm GeV}^2$. 
Note, discretion is necessary when interpreting the very low z region; see text.
Figure from Ref.~\cite{Abdolmaleki:2021yjf}.
}
\label{fig:pionFF}
\end{figure}

In addition to pion PDFs, xFitter can also compute pion fragmentation functions. 
Reference~\cite{Abdolmaleki:2021yjf} presents the first open-source analysis of fragmentation functions  of charged pions (entitled IPM-xFitter) computed at next-to-leading order (NLO) and next-to-next-to-leading order (NNLO) accuracy in perturbative QCD using the xFitter framework.
This study incorporated a comprehensive  set of pion production data from single-inclusive annihilation (SIA) processes, as well as the most recent measurements of inclusive cross-sections of single pion by the BELLE collaboration. 
 
A primary  goal of this analysis was to investigate the influence of the 
BELLE13~\cite{Belle:2013lfg}, BELLE20~\cite{Belle:2020pvy},  and BaBar~\cite{BaBar:2013yrg} data sets on the resulting fragmentation functions.
A total of five fits were generated using different data combinations and different kinematic cuts. 
1)~Fit~A focused on the impact of the BELLE13 data set without BELLE20;
2)~Fit~B used the BELLE20 data set without BELLE13;
3)~Fit~C used the BELLE20 data set without BaBar;
4)~Fit~D focused the impact of the BELLE20 data set without either BELLE13 or BaBar, 
and imposed a $z{>}0.2$ cut on the BELLE20;
5)~Fit~E excluded BELLE13 but included  BELLE20 with a $z{>}0.2$ cut, and 
 BaBar with a $z{>}0.1$ cut.

Comparisons of selected fits are displayed in Fig.~\ref{fig:pionFF} along with 
results from the literature. The different fits are in reasonable agreement for larger $z$ values, 
but begin to diverge for smaller $z$.
Note, discretion is necessary when interpreting the very low z region as the extrapolation of
the fragmentation function grids extends beyond the region fitted in the individual analyses. For example, the JAM19 focus was on SIDIS in the
region \mbox{$z \gtrsim 0.2$}, and NNFF1 used a lower kinematic cut of \mbox{$z_{min} {=}0.02$} for $Q=M_Z$ and 0.075 for $Q {<} M_Z$. 
While Fit~E is our preferred fit, we also display Fit~B to highlight the impact of the low z cuts.

This study generally found  good quality for the fits across most of the $z$ range, 
but  the description of the data in the low-z region remains an unresolved puzzle.
The resulting NLO and NNLO pion FFs provide valuable insights for applications in present and future high-energy analysis of pion final state processes.

Contemporaneous with the above study, 
a related investigation was reported in Ref.~\cite{Khalek:2021gxf} using a
complementary neural-network approach with different
data sets; this opens new avenues for future study 
and may help resolve some of the issues in the small $z$ region.

\section{Applications of \texorpdfstring{\lowercase{x}F\lowercase{itter}}{xFitter}}

\subsection{Usage of \texorpdfstring{\lowercase{x}F\lowercase{itter}}{xFitter} by LHC Collaborations}

xFitter is the standard tool for QCD analyses by the ATLAS and CMS Collaborations at the LHC,
resulting in a number of different studies.

xFitter has been used to quantify the sensitivity to PDFs of differential W and Z bosons~\cite{ATLAS:2012sjl,CMS:2013pzl,ATLAS:2016nqi,CMS:2016qqr}
and of W+charm quark \cite{CMS:2018dxg,CMS:2021oxn} cross-section measurements, 
showing how they provide important constraints on the strangeness content of the proton. 
Other studies have illustrated the gluon PDF sensitivity of $t\bar{t}$ cross-sections measurements~\cite{CMS:2017iqf,CMS:2017zpm,ATL-PHYS-PUB-2018-017}.
An ATLAS study~\cite{ATLAS:2016oxs} highlighted cross-section ratios, 
such as for $t\bar{t}/Z$ measured at different energies,
as a powerful quantity  to probe the gluon/sea PDF ratio free of large experimental uncertainties.

The xFitter capabilities to perform coherent analysis of PDFs plus additional parameters of interest, 
has been exploited to measure SM parameters which correlate with PDFs.
The  $t\bar{t}$ cross-sections have been used for extractions of the top quark mass~\cite{ATLAS:2017dhr,CMS:2018fks,CMS:2019esx}, 
which for the CMS analyses has been done simultaneously 
with a determination of the strong coupling and of the PDFs.
Simultaneous extractions of the strong coupling with the PDFs have been performed by CMS also from inclusive jets~\cite{CMS:2016lna,CMS:2021yzl} and dijet~\cite{CMS:2017jfq} cross-sections,
recently extended to determinations of the contribution of SMEFT operators 
introducing new four-fermion interactions simultaneously with the PDFs~\cite{CMS:2021yzl}.

In Ref.~\cite{ATLAS:2016nqi}, high precision measurements of the inclusive differential $W^{\pm}$ and $Z/\gamma^{*}$ boson cross sections at 7~TeV were added to the HERA data, resulting in the PDF set \textbf{ATLASepWZ16}, which improved on the HERAPDF2.0 set in various respects. Firstly, the strange content of the sea was determined rather than assumed to be a fixed fraction of the light sea. Indeed, compared to previous determinations, the strange sea was found to be enhanced at low $x$. Secondly, the accuracy of the valence quark distributions for \mbox{$x <$ 0.1} was improved.\\ 
In Ref.~\cite{ATLAS:2021qnl}, the ATLAS Collaboration performed a PDF fit including for the first time measurements for the production of W and Z boson in association with a jet, resulting in the \textbf{ATLASepWZVjets20} PDF set. The $V$ + jets data are sensitive to partons at higher $x$ than can be accessed by inclusive $W,Z/\gamma^{*}$ data and, in particular, they constrain the $\bar{d}$ and $\bar{s}$ quarks at higher $x$.\\
And more recently  an ATLAS PDF fit~\cite{ATLAS:2021vod} was performed including a variety of measurements of different production processes (Drell-Yan, V+jets, direct photon production, $t\bar{t}$ and inclusive jets data) at different center-of-mass energies. The resulting set of PDFs is called \textbf{ATLASpdf21}.

\subsection{Nuclear PDFs with xFitter}
\begin{figure}[tbh]
\vspace*{-0.5cm}
\includegraphics[width=0.40\textwidth]{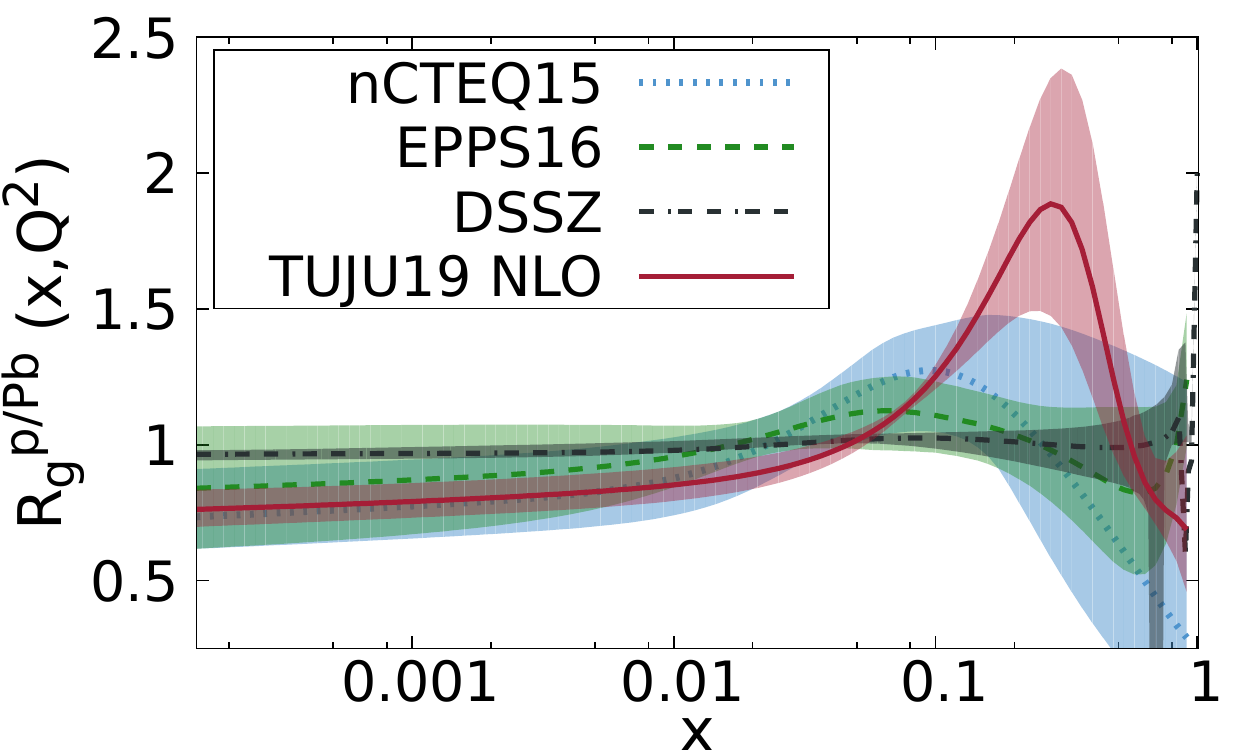}
\includegraphics[width=0.40\textwidth]{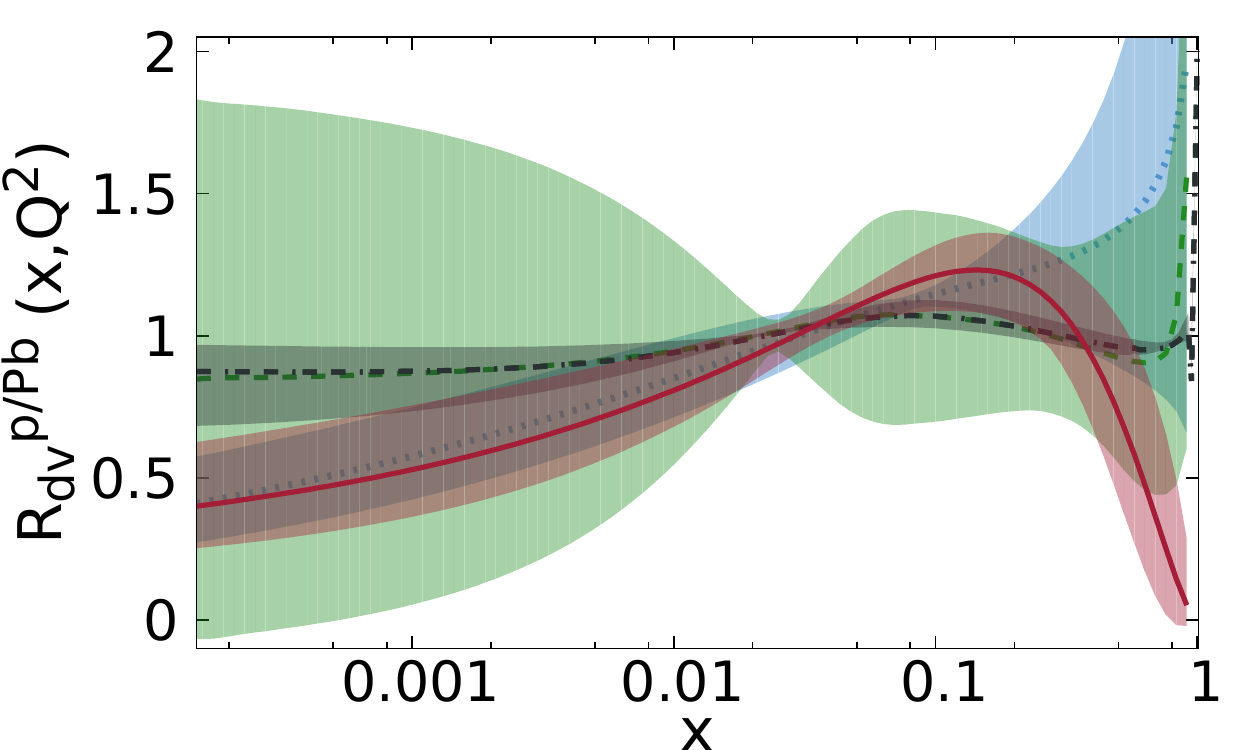}
\caption{Selected 
nuclear PDF ratios for lead as compared to proton  at NLO 
for  TUJU19~\cite{Walt:2019slu}, 
nCTEQ15~\cite{Kovarik:2015cma},
EPPS16~\cite{Eskola:2016oht}, 
and DSSZ~\cite{deFlorian:2011fp} at  $Q^2 = 100~{\rm GeV}^2$. 
Figures from Ref.~\cite{Walt:2019slu}.
}
\label{fig:npdf}
\end{figure}

Reference~\cite{Walt:2019slu}  extended xFitter 
to include the capability of performing fits to nuclear PDFs, 
and generated a new set of nPDFs (TUJU19) at both NLO and NNLO. 
While this extension is capable of performing a simultaneous fit to 
both the proton and nuclear degrees of freedom, 
to ensure stability they first generated a proton baseline 
derived with a very similar setup as for the HERA2.0 PDFs~\cite{H1:2015ubc},
and then used this as the starting point for the nPDF investigation. 
Both charged-lepton and neutrino DIS data sets wer included, and  
and   isoscalar corrections and correlated uncertainties were also incorporated. 

The necessary modifications to the xFitter code were extensive. 
The PDF parameterization was extended to include the nuclear A dimension;
this increased the number of fitting parameters, and these changes
were propagated throughout the code (e.g., in the steering file and the MINUIT interface).
The nuclear data sets depend on the nuclear A and Z, so this information was provided 
within the data files and the input/output routines were modified accordingly.  
Finally, the DGLAP evolution code was modified to evolve  nuclear PDFs covering
different combinations of A and Z individually. 
These modifications included a new numerical integration routine
to accommodate the  flexibility of an \mbox{A-dependent} normalization. 

This work highlights the versatility and flexibility of the xFitter program, 
and this study serves as a foundation for a wide variety other investigations
involving nuclear interactions. 
The details of this work are presented in Ref.~\cite{Walt:2019slu},
and the code is available on GitLab.\footnote{
The xFitter code for the TUJU19 nPDFs is available at:\hfil
\href{https://gitlab.com/fitters/xfitter/-/tree/NuclearPDFs_2.0.1}{\mbox{https://gitlab.com/fitters/xfitter/-/tree/NuclearPDFs\_2.0.1}}}

\subsection{Higgs Physics with \texorpdfstring{\lowercase{x}F\lowercase{itter}}{xFitter}}

Precision studies in the Higgs sector of the Standard Model (SM) are central to 
current~\cite{LHCHiggsCrossSectionWorkingGroup:2016ypw} 
and forthcoming~\cite{Cepeda:2019klc} physics 
programs at the LHC. The dominant mechanism for the 
production of Higgs bosons in $pp$ collisions at the LHC is given by gluon fusion. With 
the very high accuracy reached in perturbative QCD calculations of 
Higgs production cross sections, currently of next-to-next-to-next-to-leading 
order (N$^3$LO)~\cite{Chen:2021isd} in the QCD coupling $\alpha_s$,  
the theoretical systematic uncertainties on the predictions for Higgs boson  
production are strongly influenced by the gluon PDF, as well as the 
sea-quark PDFs  coupled to gluons through   initial-state QCD evolution.  
The PDF  contribution is estimated~\cite{Cepeda:2019klc} to be 
about 30$ \% $ of the total uncertainty, including $\alpha_s$ and scale variations.

The primary source of knowledge of the gluon PDF is provided at present 
by HERA deep inelastic scattering (DIS) experimental measurements.  
While future DIS experiments~\cite{LHeC:2020van,Accardi:2012qut,Proceedings:2020eah} are proposed 
to extend the range and accuracy of our current knowledge of the gluon PDF,  
 substantial progress can also come from measurements at the LHC itself, particularly 
in the forthcoming high-luminosity phase HL-LHC~\cite{Azzi:2019yne}. 
Gluon PDF studies have been considered so far 
from the analysis of light-quark jets~\cite{AbdulKhalek:2020jut}, 
open~\cite{Zenaiev:2019ktw,Cacciari:2015fta} 
and bound-state~\cite{Flett:2019pux} charm and bottom quark production, 
top quark production~\cite{Czakon:2016olj}.  These studies rely  on 
parton-level processes with colored particles in the lowest-order final state, which are 
influenced by large radiative corrections.       
 Ref.~\cite{Amoroso:2020fjw} proposes an alternative approach, based on considering   
color-singlet  production at the LHC,  
and (analogously to the case of DIS at HERA)   achieving sensitivity to the gluon PDF 
through ${\cal O} ( \alpha_s)$ contributions, guided by criteria of 
perturbative stability and experimental precision. 

Drell-Yan (DY) lepton pair production via electroweak vector boson exchange is one of the 
most precisely measured processes at the LHC. The DY cross section summed over the 
vector-boson polarizations  is sensitive to the gluon PDF for finite vector-boson transverse momenta 
$p_T$. However, in the $p_T$ region where the cross section is  largest, it is affected by 
large perturbative corrections to all orders in $\alpha_s$ (see e.g.~\cite{Angeles-Martinez:2015sea}, and 
references therein).    Ref.~\cite{Amoroso:2020fjw} therefore turns to the contributions of the 
individual polarizations of the vector boson.  It exploits the sensitivity of the DY angular 
coefficient $A_0$ associated with the longitudinal vector-boson polarization to the gluon PDF  
in order to constrain the Higgs boson cross section from gluon fusion. 
The coefficient $A_0$ is  perturbatively stable,  as illustrated by the smallness of its 
 next-to-leading-order (NLO) and  
next-to-next-to-leading-order (NNLO)~\cite{Gauld:2017tww} radiative corrections for finite $p_T$, and 
precisely measured at the LHC~\cite{CMS:2015cyj,ATLAS:2016rnf}. 
With xFitter, we can investigate in detail the impact of precision measurements of the 
longitudinally-polarized vector-boson coefficient $A_0$ on the theoretical predictions 
for the Higgs boson production cross section. 
These studies can further be extended, using xFitter, to other DY angular coefficients, since 
additional sensitivity may be gained  from longitudinal-transverse polarization interferences, 
as in the parity-conserving $A_1$ and parity-violating $A_3$ coefficients. 

\begin{figure}[t!]
\begin{center}
\includegraphics[width=0.43\textwidth]{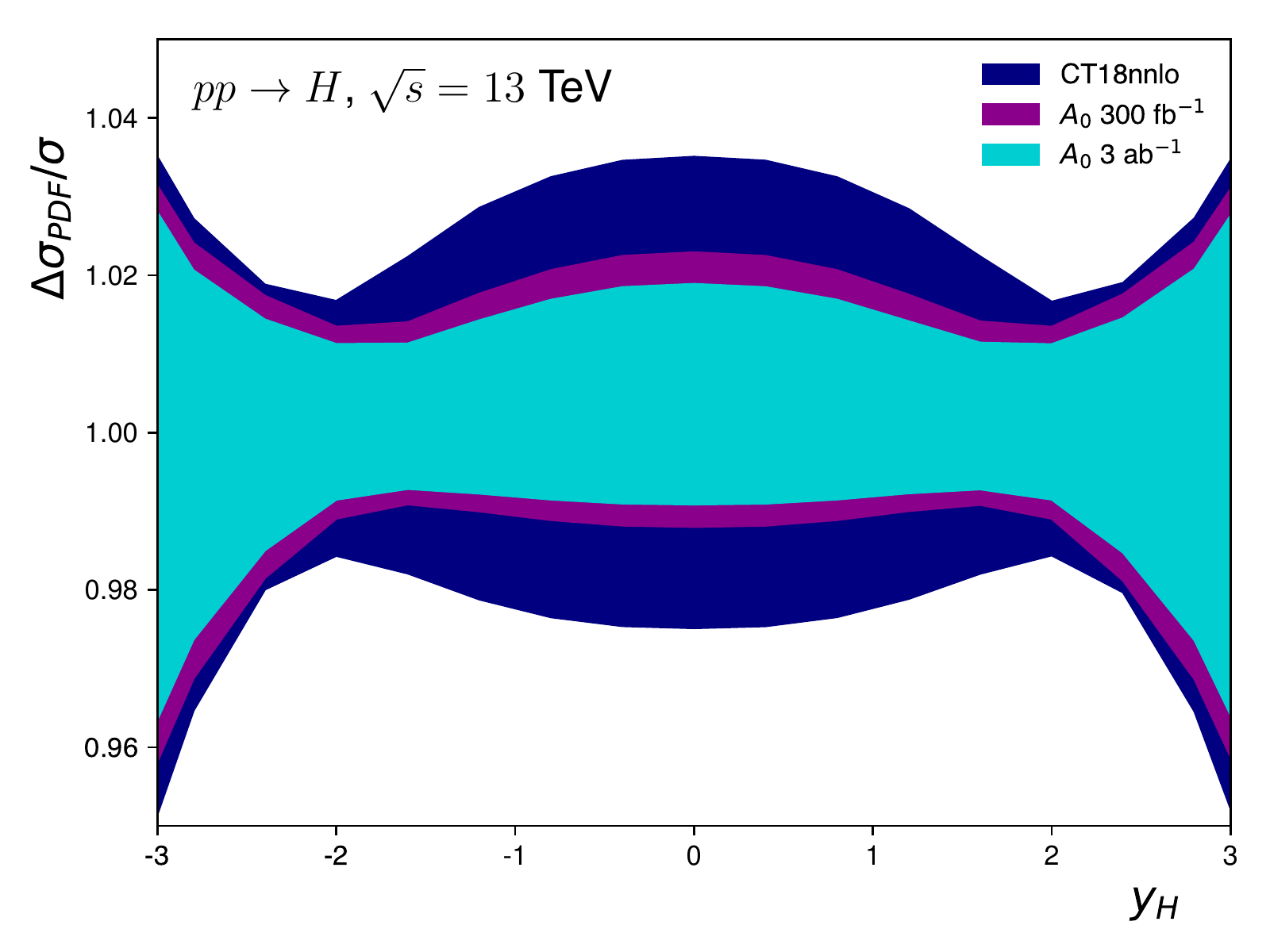}
\end{center}
\caption{Ratio of PDF uncertainties for the gluon-gluon fusion SM Higgs boson cross-section 
in $pp$ collisions at $\sqrt{s}=$13~TeV as a function of the Higgs rapidity~\cite{Amoroso:2020fjw}. The blue band 
shows the uncertainties of the CT18nnlo PDF set~\cite{Hou:2019efy}, reduced to 68\% CL coverage. The red 
and green bands show the uncertainties of
the CT18nnlo including constraints from the $A_0$ measurement
and assuming 300~fb$^{-1}$ and 3~ab$^{-1}$, respectively.}
\label{fig:HiggsRapidity}
\end{figure}

An example illustrating the implications of the longitudinally polarized $A_0$ coefficient  
on the Higgs boson production is shown in 
Fig.~\ref{fig:HiggsRapidity}~\cite{Amoroso:2020fjw} versus the Higgs boson rapidity $y_H$. 
The SM Higgs boson production is computed  in the gluon fusion  mode for 
$\sqrt{s}=13$ TeV $ pp$ collisions, using the MCFM code~\cite{Campbell:2019dru}  
at NLO in QCD perturbation theory. We evaluate PDF uncertainties on the Higgs cross section 
including constraints obtained with xFitter from  $A_0$ profiling near the $Z$-boson mass. 
We see that in the region $ - 2 
{\raisebox{-.6ex}{\rlap{$\,\sim\,$}} \raisebox{.4ex}{$\,<\,$}}
 y_H 
 {\raisebox{-.6ex}{\rlap{$\,\sim\,$}} \raisebox{.4ex}{$\,<\,$}}
   2 $ the uncertainty is reduced by 
about 30 - 40~$ \% $ in the Run III scenario, and a further reduction to about 50~$ \% $ takes 
place in the HL-LHC scenario.

\begin{figure}[t!]
\begin{center}
\includegraphics[width=0.43\textwidth]{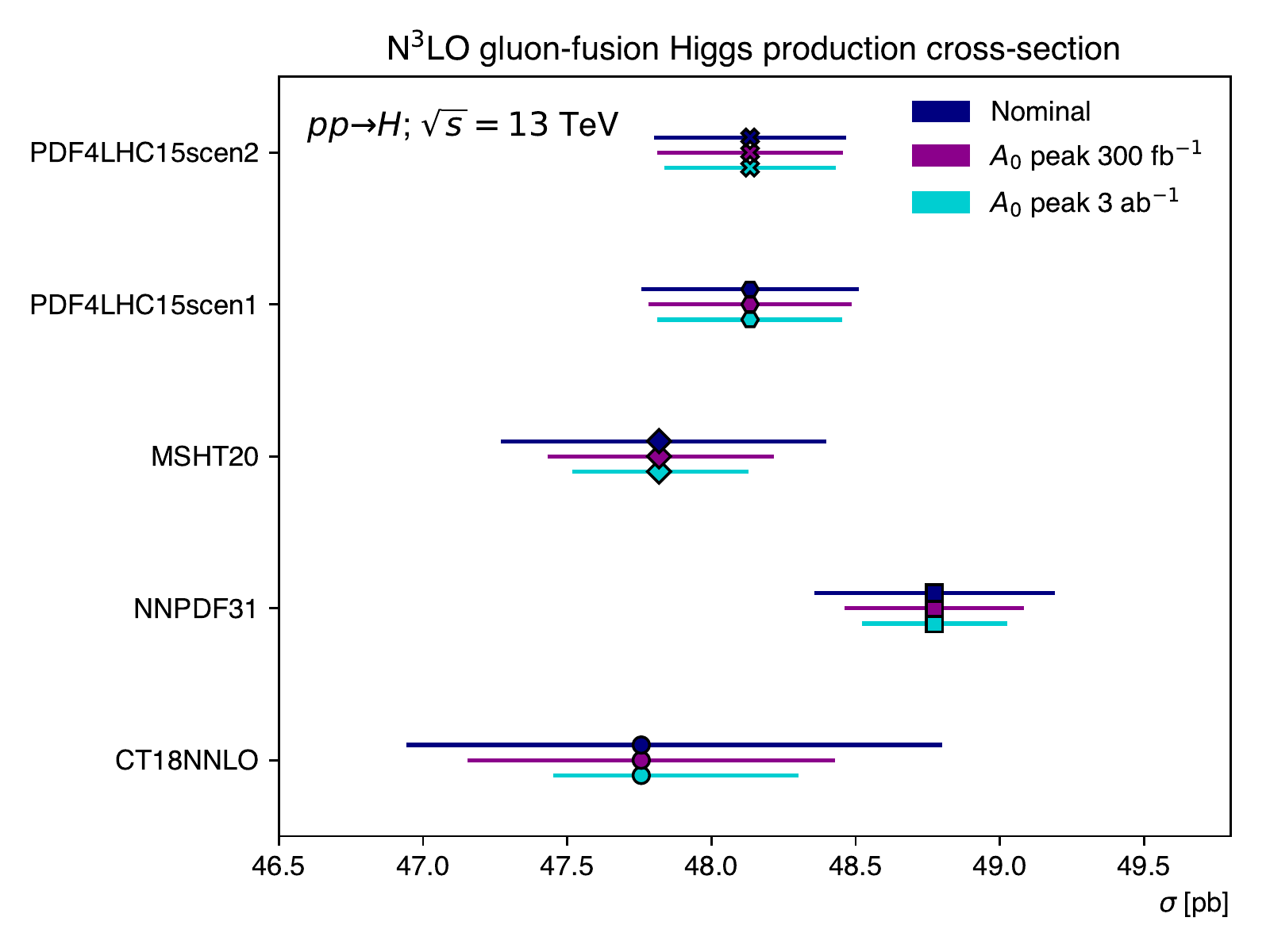}
\end{center}
\caption{The gluon-gluon fusion Higgs boson production cross-section~\cite{Amoroso:2020fjw} 
at N$^3$LO for different PDFs, showing the 
uncertainty from PDFs and their expected reduction including constraints from the $A_0$ measurement 
assuming 300~fb$^{-1}$ and 3~ab$^{-1}$, respectively.}
\label{fig:HiggsSigma}
\end{figure}

In   Fig.~\ref{fig:HiggsSigma}~\cite{Amoroso:2020fjw}  
we perform a higher-order N$^3$LO calculation for the  Higgs boson total cross section 
using the code  {\tt{ggHiggs}}~\cite{Bonvini:2014jma,Bonvini:2018ixe}. We report the 
result for the cross section and its uncertainty in the cases of the 
current CT18nnlo~\cite{Hou:2019efy},  NNPDF3.1nnlo~\cite{NNPDF:2017mvq} and MSHT20nnlo~\cite{Bailey:2020ooq} global sets as well as projected sets, based on complete LHC data 
sample~\cite{AbdulKhalek:2018rok}. 
The  PDF4LHC15scen1/2 sets, which are PDF projections including HL-LHC pseudodata, also show a smaller, but not negligible, reduction in uncertainties. 
Notwithstanding the numerical differences, the behavior is qualitatively similar for the different sets, and  provides further support at N$^3$LO to the picture given in Fig.~\ref{fig:HiggsRapidity} for the NLO Higgs boson rapidity cross section.

The  approach illustrated above can be   extended to mass regions away from the $Z$ peak, where it 
has the potential to provide complementary physics information.  High-mass DY angular distributions 
allow the region of larger $x$ momentum fractions to be accessed and will be relevant for associated 
Higgs boson production with a gauge/Higgs boson or heavy-flavour quarks.  Measurements of $A_0$ 
at low masses  may be used to probe $p_T$ dependent gluon PDF effects,  
influencing the Higgs boson $p_T$ spectrum for low transverse momenta 
and ratios of Higgs to DY $p_T$ spectra~\cite{Cipriano:2013ooa}, as well as  the 
small-$x$ regime~\cite{Bonvini:2018ixe,Hautmann:2002tu} of Higgs boson production relevant 
to the highest energy frontier~\cite{Contino:2016spe}.

\subsection{BSM Physics with \texorpdfstring{\lowercase{x}F\lowercase{itter}}{xFitter}}

The capabilities of xFitter can be exploited for studies of 
new physics beyond the Standard Model (BSM). 

Consider for instance BSM experimental searches at the LHC in 
dielectron/dimuon channels and charged lepton plus missing transverse 
energy channels. These are classic methods to search for new 
$Z^\prime / W^\prime$ gauge bosons: see for 
example the analyses by ATLAS in Refs.~\cite{ATLAS:2019erb,ATLAS:2019lsy} and 
by CMS in Refs.~\cite{CMS:2018ipm,CMS:2021ctt}.  Such  
searches will be further pursued at the 
High-Luminosity LHC (HL-LHC)~\cite{CidVidal:2018eel}. 

In the case of BSM scenarios with narrow vector resonances one 
can rely on traditional ``bump search" analyses based on 
the Breit-Wigner (BW) lineshape. In the case of BSM scenarios featuring vector 
resonances with large width, on the other hand, the 
default bump searches are likely 
 not sufficient: instead of  an easily observable narrow BW lineshape, the 
 resonance appears as a broad shoulder spreading over the SM background. 
 Alternative experimental approaches can be applied to the case of 
 wide resonances, in which one has to exploit the tails of the measured mass 
 distributions. Examples of such approaches include for instance ``counting strategy" 
 analyses. These approaches are much more dependent than bump searches 
 on the modeling of the dominant production process~\cite{Accomando:2019ahs}, for 
 both signal and background. 

Then, in the case of wide-resonance searches the role of PDFs  becomes 
prominent as one of the main sources of theoretical systematic uncertainties, 
because it directly affects our ability to test BSM scenarios and 
the experimental sensitivity to new physics. This thus influences the potential 
of experimental searches for discovering or setting exclusion bounds on 
heavy BSM states. 

In the multi-TeV region of $W^\prime / Z^\prime$ masses defined 
by the 
 current LHC exclusion 
 limits~\cite{ATLAS:2019erb,ATLAS:2019lsy,CMS:2018ipm,CMS:2021ctt}, 
   quark distributions in the valence sector dominate the 
 PDF systematics. 
With an xFitter profiling analysis, 
 Ref.~\cite{Fiaschi:2021okg} finds that  the valence quark systematics 
 can be 
 improved by combining high-precision measurements of 
 Charged Current (CC) and Neutral Current (NC)  
 Drell-Yan asymmetries in the mass region near the SM weak boson poles. 
 This exploits 
 the sensitivity of the NC forward-backward asymmetry $A_{\rm{FB}}$ 
 to the charge-weighted linear 
 combination $ (2/3) u_V + (1/3) d_V$ of up-quark and down-quark 
 valence distributions~\cite{Accomando:2019vqt} and the sensitivity of 
 the CC lepton-charge asymmetry $A_W$ to the difference $ u_V - d_V$.
The constraints from the $A_{\rm{FB}}$ and $A_W$ combination at the SM 
weak boson mass scale, 
examined in the two projected luminosity scenarios of 
300 fb$^{-1}$ (for the LHC Run 3) and of 3000 fb$^{-1}$ 
(for the HL-LHC~\cite{CidVidal:2018eel}), turn out to improve the relative 
PDF uncertainties by up to around 20\% \cite{Fiaschi:2021okg} in the region 
of the invariant and transverse mass spectra between 2 TeV and 6 TeV, in 
which evidence for $W^\prime$ and/or $ Z^\prime$ states with large widths 
could first be observed. 

We can then analyze quantitatively how the ``improved PDFs", obtained from the 
reduction of the valence PDF uncertainty 
due to  $A_{\rm{FB}}$ and $A_W$ precision measurements at the SM 
weak boson mass scale, result  into an enhancement of the  experimental sensitivity 
to BSM searches at the TeV scale. In the next subsection we illustrate this 
with  a specific example. 

In this kind of analysis,  an important feature,  stemming from multi-resonant profiles  
of  mass spectra and  influencing  the 
experimental search strategies,   is the following.  
Strong interference effects between the BSM resonances 
themselves and between the BSM and SM states can give
rise to  a statistically significant 
depletion of events  below the BW  peak in the 
invariant  mass distribution, 
leading to the appearance of a pronounced 
{\em dip}~\cite{Accomando:2019ahs,Fiaschi:2021okg}. 
It is possible to define the significance of the depletion of events 
in a manner similar to that for the excess of events of the peak, 
which can be used to extract model dependent exclusion and 
discovery limits in the model's parameter space.  One can then 
present such limits resulting from the analysis of the spectra for 
either the peak or the dip. The xFitter framework enables  one to   
investigate whether the peak or the dip analyses especially benefit 
from  the ``improved  PDFs".   

We next turn to the specific example of 
BSM $W^\prime / Z^\prime$ states in composite Higgs models, 
considering broad-resonance searches in the leptonic channels. 
For sensitivity studies comparing leptonic with heavy-quark channels in 
composite Higgs models see e.g.~\cite{Liu:2019bua}.

\subsubsection{Enhancing the sensitivity of BSM searches with improved PDFs}

New gauge sectors in strongly-coupled models of electroweak symmetry 
breaking~\cite{Panico:2015jxa} based on composite Nambu-Goldstone 
Higgs~\cite{Kaplan:1983fs,Kaplan:1983sm} feature 
multiple $W^\prime $ and $ Z^\prime$ broad resonances, and  interference 
effects of the heavy bosons with each other and with SM gauge bosons. 
An example is the 4-Dimensional Composite Higgs Model (4DCHM) 
realization~\cite{DeCurtis:2011yx} of the minimal composite Higgs model of 
Ref.~\cite{Agashe:2004rs}. The parameter space of the model can be 
characterized in terms of two parameters, the compositeness scale $f$ and 
the coupling $g_\rho$ of the new resonances, with the resonances mass scale 
being of order $M \sim f g_\rho$~\cite{Panico:2015jxa,Giudice:2007fh}. 
Ref.~\cite{Fiaschi:2021sin}  selects two 4DCHM benchmarks, denoted by 
A and B,  each characterized by 
specific values of $f$, $g_\rho$ and the resonance masses. 
We next present an example of the results for 
exclusion and discovery limits in the parameter space of these 
two 4DCHM benchmarks which can be obtained by 
using the ``improved  PDFs"~\cite{Fiaschi:2021okg}.

\begin{figure}[htb]
\begin{center}
\includegraphics[width=0.35\textwidth]{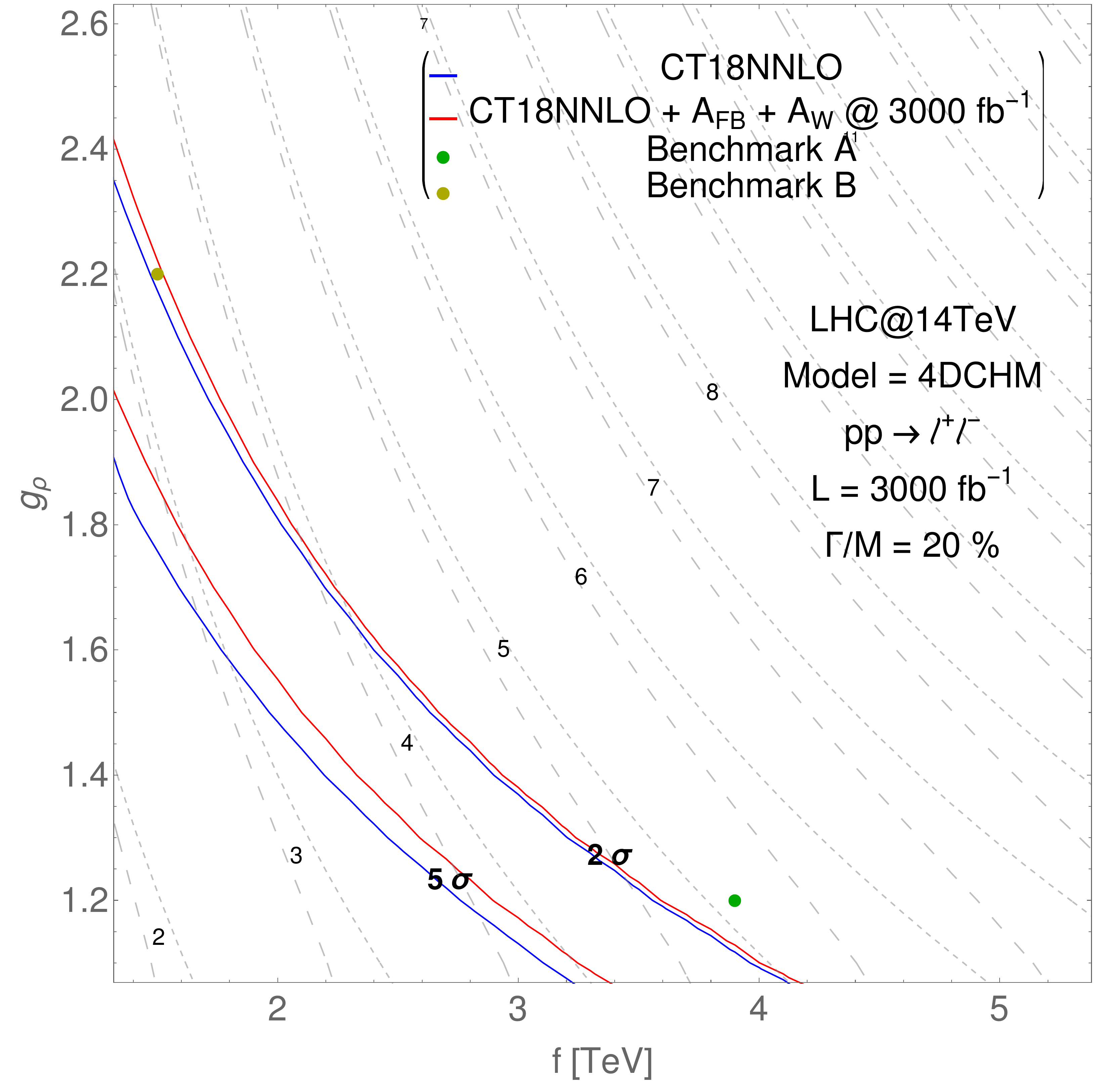}
\includegraphics[width=0.35\textwidth]{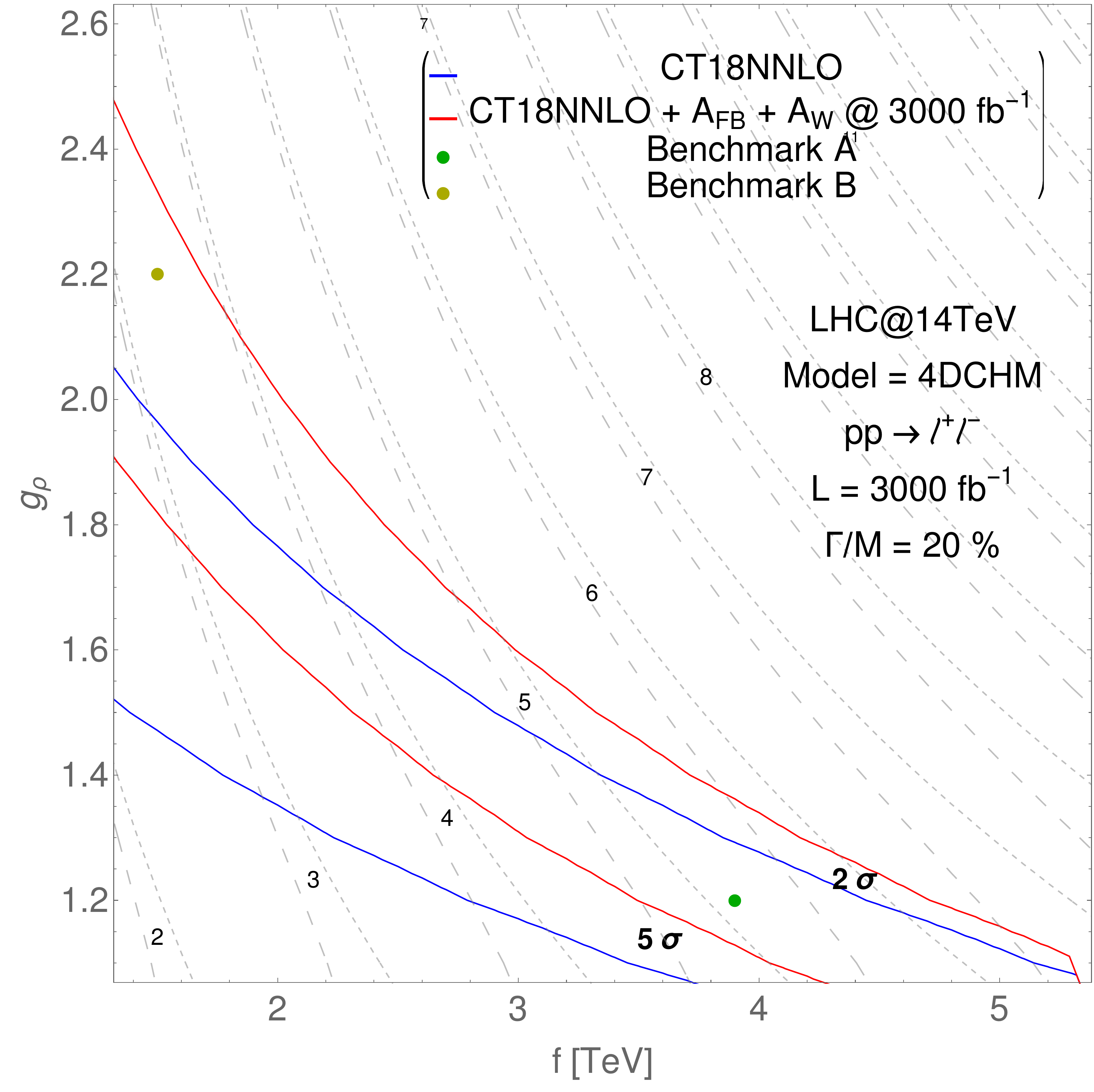}
\end{center}
\caption{Exclusion and discovery limits at 3000 fb$^{-1}$ for the peak (top) and for the dip (bottom) for $Z^{\prime}$ resonances 
with $\Gamma / M$ = 20\% \cite{Fiaschi:2021sin}.
The short (long) dashed contours give the BSM boson mass $M_{Z_2}$ ($M_{Z_3} \simeq M_{W_2}$) in TeV.}
\label{fig:Contour_20_3000}
\end{figure}

In Fig.~\ref{fig:Contour_20_3000}~\cite{Fiaschi:2021sin} we show the limits from NC Drell-Yan 
on the model parameter space, described in terms of $f$ and $g_\rho$, 
for the HL-LHC  stage with center-of-mass 
energy of 14 TeV and an integrated luminosity of 3000 fb$^{-1}$. 
In the background we give  contour plots for the masses of the new gauge bosons. 
The blue and red curves are obtained respectively 
with the baseline CT18NNLO PDF set~\cite{Hou:2019efy} and with 
the profiled PDFs~\cite{Fiaschi:2021okg} using the $A_{\rm{FB}}$ and $A_W$ 
combination with 3000 fb$^{-1}$ of integrated luminosity.   
The top panel in Fig.~\ref{fig:Contour_20_3000} is for the peak analysis, 
the bottom panel is for the dip analysis.
In this setup, the peak of benchmark A would 
still be below the experimental sensitivity, while the peak of benchmark B, if 
the improved PDFs are employed, would be right below the 2$\sigma$ exclusion.
When exploiting the depletion of events in the dip below the peak, the sensitivity on the model increases remarkably.
Furthermore, as visible from the bottom plot of Fig.~\ref{fig:Contour_20_3000}, the improvement 
on the PDF also has a very large impact particularly in the region of small $f$ and large $g_\rho$, where 
the sensitivity on the dip can overtake the LHC reach in the peak region once the profiled PDFs are employed. 
Taking into account the reduction of PDF uncertainty, benchmark A would be now at the edge of the 5$\sigma$ discovery, 
while the sensitivity on benchmark B would almost reach 3$\sigma$.  

Tab.~\ref{tab:AB_dip_neutral} \cite{Fiaschi:2021sin}  lists,  in the case of the two benchmarks A and B, the integration intervals in 
invariant mass for the dip region as well as the resulting cross sections for the SM background and the 
complete 4DCHM, together with the associated PDF uncertainty using the baseline 
CT18NNLO PDF set~\cite{Hou:2019efy} and the profiled version~\cite{Fiaschi:2021okg} using 
the $A_{\rm{FB}}$ and $A_W$ combination with 3000 fb$^{-1}$ of integrated luminosity. It reports also the 
obtained significances for an integrated luminosity of 3000 fb$^{-1}$ employing the two PDF errors.   The improvement in significance 
due to the profiled PDFs is  sizeable, as  significances grow by an amount from 40\% to 90\% (i.e., by a factor of 
about 10 in comparison) for the benchmarks A and B, respectively.   
A more modest improvement in significance is found \cite{Fiaschi:2021sin} in the case of the peak.

\begin{table}[htb]
\begin{center}
\begin{tabular}{|c|c|c|c|}
\hline
\multicolumn{4}{|c|}{Benchmark A}\\
\hline
inf [TeV] & sup [TeV] & $\sigma_{\rm SM}$ [fb] & $\sigma_{\rm SM+BSM}$ [fb] \\
2.06 & 4.99 & 1.69 $\cdot$ 10$^{-1}$ & 1.42 $\cdot$ 10$^{-1}$ \\
\hline
$\Delta_{\rm PDF}$ base [fb] & $\Delta_{\rm PDF}$ profiled [fb] & $\alpha$ (base) & $\alpha$ (profiled)\\
9.5 $\cdot$ 10$^{-3}$ & 4.6 $\cdot$ 10$^{-3}$ & 3.34 & 4.82 \\
\hline
\end{tabular}
\\[1cm]
\begin{tabular}{|c|c|c|c|}
\hline
\multicolumn{4}{|c|}{Benchmark B}\\
\hline
inf [TeV] & sup [TeV] & $\sigma_{\rm SM}$ [fb] & $\sigma_{\rm SM+BSM}$ [fb] \\
1.36 & 3.36 & 1.53 & 1.45 \\
\hline
$\Delta_{\rm PDF}$ base [fb] & $\Delta_{\rm PDF}$ profiled [fb] & $\alpha$ (base) & $\alpha$ (profiled)\\
6.8 $\cdot$ 10$^{-2}$ & 3.1 $\cdot$ 10$^{-2}$ & 1.53 & 2.91 \\
\hline
\end{tabular}
\end{center}
\vspace*{-0.3cm}
\caption{Integration limits for the dip region, integrated cross section for the SM background 
and the complete model and its PDF uncertainty with the baseline CT18NNLO 
PDF set and the profiled PDF set using $A_{\rm FB} + A_W$ pseudodata as well as the 
significances $\alpha$ employing the two PDF errors for an integrated luminosity of 3000 fb$^{-1}$ for 
the benchmarks A and B \cite{Fiaschi:2021sin}.}
\label{tab:AB_dip_neutral}
\end{table}

The above studies can be extended to CC Drell-Yan.  In fact, it is noted 
 in~ \cite{Fiaschi:2021sin}  that, 
for cases in which both neutral and charged  new 
states appear and are correlated by theory,  
 the  reduction of systematic PDF error  can profitably be applied to combined $W^\prime$ and $Z^\prime$
  searches. For example,  in the 4DCHM 
  a direct $W^\prime$ exclusion (or indeed discovery) achieved in the CC channel can be used indirectly to 
  probe the existence of a $Z^\prime$ better than this can be done with direct searches in the NC channel.
The analysis of the dip often provides more stringent limits than the signal coming from the peak, with the 
reduction of systematic PDF errors playing a crucial part in this conclusion.

\subsubsection{Applications with SMEFT}

In the discussion of the previous subsection, we study the impact of 
``improved" PDFs on BSM physics at the multi-TeV scale, while the 
improvement itself in  the PDFs comes from  processes at the 
SM weak boson mass scale.  The  asymmetry pseudodata employed in 
the profiling~\cite{Fiaschi:2021okg}   are centered around the SM 
vector boson peaks and take into account only SM contributions. 
Since we consider TeV scale BSM resonances, the new physics 
contribution to the forward-backward asymmetry 
$A_{FB}$ at the weak scale, which is due to the 
interference between SM and BSM process, is 
small, and a similar conclusion holds for 
the charged-current differential cross section distribution. 

On the other hand, a different scenario arises if one includes 
data or pseudodata in the PDF determination which do contain 
contamination from BSM contributions. This is the case, for example, 
of data or pseudodata at high mass scales in the multi-TeV region. 
In this case, assuming  purely SM predictions would lead to biased 
estimations on new physics constraints: see for instance the 
recent studies~\cite{Greljo:2021kvv,CMS:2021yzl}. 

This bias can be avoided if 
 the PDFs are extracted simultaneously with the BSM parameters. 
Such non-biased approaches are feasible with xFitter via 
interfaces to BSM computations, for instance by using SM effective field theory 
(SMEFT) expansions. For example, the 
CIJET~\cite{Gao:2012qpa,Gao:2013kp} interface allows 
extending jet cross sections with dimension-6 SMEFT operators for 
purely left-handed, vector-like or axial vector-like color-singlet exchanges. 
An xFitter analysis in this direction is underway.

\section{Conclusion}

The xFitter program is a versatile, flexible, modular, and
comprehensive tool that can facilitate analyses of the experimental
data and theoretical calculations.

It is a valuable framework for bench-marking and understanding differences
between PDF fits, and it can provide impact studies for possible
future facilities including HL-LHC, EIC, LHeC, DUNE, and UHE Cosmic Ray experiments.  
We encourage use of xFitter, and welcome new contributions from the
community to ensure xFitter continues to incorporate the latest
theoretical advances and precision experimental data.

\bibliographystyle{utphys}
\bibliography{refs,extra}

\providecommand{\href}[2]{#2}\begingroup\raggedright\begin{thebibliography}{100}

\bibitem{Alekhin:2014irh}
S.~Alekhin {\em et~al.}, ``{HERAFitter},''
  \href{http://dx.doi.org/10.1140/epjc/s10052-015-3480-z}{{\em Eur. Phys. J. C}
  {\bfseries 75} no.~7, (2015) 304},
  \href{http://arxiv.org/abs/1410.4412}{{\ttfamily arXiv:1410.4412 [hep-ph]}}.

\bibitem{HERAFitterdevelopersTeam:2015cre}
{\bfseries HERAFitter developers' Team} Collaboration, S.~Camarda {\em et~al.},
  ``{QCD analysis of $W$- and $Z$-boson production at Tevatron},''
  \href{http://dx.doi.org/10.1140/epjc/s10052-015-3655-7}{{\em Eur. Phys. J. C}
  {\bfseries 75} no.~9, (2015) 458},
  \href{http://arxiv.org/abs/1503.05221}{{\ttfamily arXiv:1503.05221
  [hep-ph]}}.

\bibitem{Accomando:2019vqt}
E.~Accomando {\em et~al.}, ``{PDF Profiling Using the Forward-Backward
  Asymmetry in Neutral Current Drell-Yan Production},''
  \href{http://dx.doi.org/10.1007/JHEP10(2019)176}{{\em JHEP} {\bfseries 10}
  (2019) 176}, \href{http://arxiv.org/abs/1907.07727}{{\ttfamily
  arXiv:1907.07727 [hep-ph]}}.

\bibitem{Amoroso:2020fjw}
S.~Amoroso, J.~Fiaschi, F.~Giuli, A.~Glazov, F.~Hautmann, and O.~Zenaiev,
  ``{Longitudinal Z-boson polarization and the Higgs boson production cross
  section at the Large Hadron Collider},''
  \href{http://dx.doi.org/10.1016/j.physletb.2021.136613}{{\em Phys. Lett. B}
  {\bfseries 821} (2021) 136613},
  \href{http://arxiv.org/abs/2012.10298}{{\ttfamily arXiv:2012.10298
  [hep-ph]}}.

\bibitem{Bertone:2017tig}
{\bfseries xFitter Developers' Team} Collaboration, V.~Bertone {\em et~al.},
  ``{xFitter 2.0.0: An Open Source QCD Fit Framework},''
  \href{http://dx.doi.org/10.22323/1.297.0203}{{\em PoS} {\bfseries DIS2017}
  (2018) 203}, \href{http://arxiv.org/abs/1709.01151}{{\ttfamily
  arXiv:1709.01151 [hep-ph]}}.

\bibitem{Botje:2010ay}
M.~Botje, ``{QCDNUM: Fast QCD Evolution and Convolution},''
  \href{http://dx.doi.org/10.1016/j.cpc.2010.10.020}{{\em Comput. Phys.
  Commun.} {\bfseries 182} (2011) 490--532},
  \href{http://arxiv.org/abs/1005.1481}{{\ttfamily arXiv:1005.1481 [hep-ph]}}.

\bibitem{Bertone:2013vaa}
V.~Bertone, S.~Carrazza, and J.~Rojo, ``{APFEL: A PDF Evolution Library with
  QED corrections},'' \href{http://dx.doi.org/10.1016/j.cpc.2014.03.007}{{\em
  Comput. Phys. Commun.} {\bfseries 185} (2014) 1647--1668},
  \href{http://arxiv.org/abs/1310.1394}{{\ttfamily arXiv:1310.1394 [hep-ph]}}.

\bibitem{Bertone:2017gds}
V.~Bertone, ``{APFEL++: A new PDF evolution library in C++},''
  \href{http://dx.doi.org/10.22323/1.297.0201}{{\em PoS} {\bfseries DIS2017}
  (2018) 201}, \href{http://arxiv.org/abs/1708.00911}{{\ttfamily
  arXiv:1708.00911 [hep-ph]}}.

\bibitem{Buckley:2014ana}
A.~Buckley, J.~Ferrando, S.~Lloyd, K.~Nordstr\"om, B.~Page, M.~R\"ufenacht,
  M.~Sch\"onherr, and G.~Watt, ``{LHAPDF6: parton density access in the LHC
  precision era},''
  \href{http://dx.doi.org/10.1140/epjc/s10052-015-3318-8}{{\em Eur. Phys. J. C}
  {\bfseries 75} (2015) 132}, \href{http://arxiv.org/abs/1412.7420}{{\ttfamily
  arXiv:1412.7420 [hep-ph]}}.

\bibitem{Carli:2010rw}
T.~Carli, D.~Clements, A.~Cooper-Sarkar, C.~Gwenlan, G.~P. Salam, F.~Siegert,
  P.~Starovoitov, and M.~Sutton, ``{A posteriori inclusion of parton density
  functions in NLO QCD final-state calculations at hadron colliders: The
  APPLGRID Project},''
  \href{http://dx.doi.org/10.1140/epjc/s10052-010-1255-0}{{\em Eur. Phys. J. C}
  {\bfseries 66} (2010) 503--524},
  \href{http://arxiv.org/abs/0911.2985}{{\ttfamily arXiv:0911.2985 [hep-ph]}}.

\bibitem{Bertone:2016lga}
V.~Bertone, S.~Carrazza, and N.~P. Hartland, ``{APFELgrid: a high performance
  tool for parton density determinations},''
  \href{http://dx.doi.org/10.1016/j.cpc.2016.10.006}{{\em Comput. Phys.
  Commun.} {\bfseries 212} (2017) 205--209},
  \href{http://arxiv.org/abs/1605.02070}{{\ttfamily arXiv:1605.02070
  [hep-ph]}}.

\bibitem{Stober:2015nlg}
D.~Britzger, G.~S. Klaus~Rabbertz, F.~Stober, and M.~Wobisch, ``{Recent
  Developments of the fastNLO Toolkit},''
  \href{http://dx.doi.org/10.22323/1.247.0055}{{\em PoS} {\bfseries DIS2015}
  (2015) 055}.

\bibitem{Aliev:2010zk}
M.~Aliev, H.~Lacker, U.~Langenfeld, S.~Moch, P.~Uwer, and M.~Wiedermann,
  ``{HATHOR: HAdronic Top and Heavy quarks crOss section calculatoR},''
  \href{http://dx.doi.org/10.1016/j.cpc.2010.12.040}{{\em Comput. Phys.
  Commun.} {\bfseries 182} (2011) 1034--1046},
  \href{http://arxiv.org/abs/1007.1327}{{\ttfamily arXiv:1007.1327 [hep-ph]}}.

\bibitem{LHeC:2020van}
{\bfseries LHeC, FCC-he Study Group} Collaboration, P.~Agostini {\em et~al.},
  ``{The Large Hadron-Electron Collider at the HL-LHC},''
  \href{http://dx.doi.org/10.1088/1361-6471/abf3ba}{{\em J. Phys. G} {\bfseries
  48} no.~11, (2021) 110501}, \href{http://arxiv.org/abs/2007.14491}{{\ttfamily
  arXiv:2007.14491 [hep-ex]}}.

\bibitem{Accardi:2012qut}
A.~Accardi {\em et~al.}, ``{Electron Ion Collider: The Next QCD Frontier}:
  {Understanding the glue that binds us all},''
  \href{http://dx.doi.org/10.1140/epja/i2016-16268-9}{{\em Eur. Phys. J. A}
  {\bfseries 52} no.~9, (2016) 268},
  \href{http://arxiv.org/abs/1212.1701}{{\ttfamily arXiv:1212.1701 [nucl-ex]}}.

\bibitem{Proceedings:2020eah}
\href{http://dx.doi.org/10.1142/11684}{{\em {Proceedings, Probing Nucleons and
  Nuclei in High Energy Collisions: Dedicated to the Physics of the Electron
  Ion Collider}: {Seattle (WA), United States, October 1 - November 16,
  2018}}}.
\newblock WSP, 2, 2020.
\newblock \href{http://arxiv.org/abs/2002.12333}{{\ttfamily arXiv:2002.12333
  [hep-ph]}}.

\bibitem{Walt:2019slu}
M.~Walt, I.~Helenius, and W.~Vogelsang, ``{Open-source QCD analysis of nuclear
  parton distribution functions at NLO and NNLO},''
  \href{http://dx.doi.org/10.1103/PhysRevD.100.096015}{{\em Phys. Rev. D}
  {\bfseries 100} no.~9, (2019) 096015},
  \href{http://arxiv.org/abs/1908.03355}{{\ttfamily arXiv:1908.03355
  [hep-ph]}}.

\bibitem{Novikov:2020snp}
I.~Novikov {\em et~al.}, ``{Parton Distribution Functions of the Charged Pion
  Within The xFitter Framework},''
  \href{http://dx.doi.org/10.1103/PhysRevD.102.014040}{{\em Phys. Rev. D}
  {\bfseries 102} no.~1, (2020) 014040},
  \href{http://arxiv.org/abs/2002.02902}{{\ttfamily arXiv:2002.02902
  [hep-ph]}}.

\bibitem{Abdolmaleki:2019acd}
{\bfseries xFitter Developers' Team} Collaboration, H.~Abdolmaleki {\em
  et~al.}, ``{Probing the strange content of the proton with charm production
  in charged current at LHeC},''
  \href{http://dx.doi.org/10.1140/epjc/s10052-019-7362-7}{{\em Eur. Phys. J. C}
  {\bfseries 79} no.~10, (2019) 864},
  \href{http://arxiv.org/abs/1907.01014}{{\ttfamily arXiv:1907.01014
  [hep-ph]}}.

\bibitem{Abdolmaleki:2019ubu}
H.~Abdolmaleki {\em et~al.}, ``{Forward-Backward Drell-Yan Asymmetry and PDF
  Determination},'' in {\em {54th Rencontres de Moriond on QCD and High Energy
  Interactions}}, pp.~211--214.
\newblock ARISF, 7, 2019.
\newblock \href{http://arxiv.org/abs/1907.08301}{{\ttfamily arXiv:1907.08301
  [hep-ph]}}.

\bibitem{Accomando:2018nig}
E.~Accomando, J.~Fiaschi, F.~Hautmann, and S.~Moretti, ``{Neutral current
  forward\textendash{}backward asymmetry: from $\theta _W$ to PDF
  determinations},''
  \href{http://dx.doi.org/10.1140/epjc/s10052-018-6120-6}{{\em Eur. Phys. J. C}
  {\bfseries 78} no.~8, (2018) 663},
  \href{http://arxiv.org/abs/1805.09239}{{\ttfamily arXiv:1805.09239
  [hep-ph]}}. [Erratum: Eur.Phys.J.C 79, 453 (2019)].

\bibitem{Accomando:2017scx}
E.~Accomando, J.~Fiaschi, F.~Hautmann, and S.~Moretti, ``{Constraining Parton
  Distribution Functions from Neutral Current Drell-Yan Measurements},''
  \href{http://dx.doi.org/10.1103/PhysRevD.98.013003}{{\em Phys. Rev. D}
  {\bfseries 98} no.~1, (2018) 013003},
  \href{http://arxiv.org/abs/1712.06318}{{\ttfamily arXiv:1712.06318
  [hep-ph]}}. [Erratum: Phys.Rev.D 99, 079902 (2019)].

\bibitem{Accomando:2016tah}
E.~Accomando, J.~Fiaschi, F.~Hautmann, S.~Moretti, and C.~H.
  Shepherd-Themistocleous, ``{Photon-initiated production of a dilepton final
  state at the LHC: Cross section versus forward-backward asymmetry studies},''
  \href{http://dx.doi.org/10.1103/PhysRevD.95.035014}{{\em Phys. Rev. D}
  {\bfseries 95} no.~3, (2017) 035014},
  \href{http://arxiv.org/abs/1606.06646}{{\ttfamily arXiv:1606.06646
  [hep-ph]}}.

\bibitem{Accomando:2016ehi}
E.~Accomando, J.~Fiaschi, F.~Hautmann, S.~Moretti, and C.~H.
  Shepherd-Themistocleous, ``{The effect of real and virtual photons in the
  di-lepton channel at the LHC},''
  \href{http://dx.doi.org/10.1016/j.physletb.2017.04.025}{{\em Phys. Lett. B}
  {\bfseries 770} (2017) 1--7},
  \href{http://arxiv.org/abs/1612.08168}{{\ttfamily arXiv:1612.08168
  [hep-ph]}}.

\bibitem{xFitterDevelopersTeam:2017fxf}
{\bfseries xFitter Developers' Team} Collaboration, F.~Giuli {\em et~al.},
  ``{The photon PDF from high-mass Drell\textendash{}Yan data at the LHC},''
  \href{http://dx.doi.org/10.1140/epjc/s10052-017-4931-5}{{\em Eur. Phys. J. C}
  {\bfseries 77} no.~6, (2017) 400},
  \href{http://arxiv.org/abs/1701.08553}{{\ttfamily arXiv:1701.08553
  [hep-ph]}}.

\bibitem{Angeles-Martinez:2015sea}
R.~Angeles-Martinez {\em et~al.}, ``{Transverse Momentum Dependent (TMD) parton
  distribution functions: status and prospects},''
  \href{http://dx.doi.org/10.5506/APhysPolB.46.2501}{{\em Acta Phys. Polon. B}
  {\bfseries 46} no.~12, (2015) 2501--2534},
  \href{http://arxiv.org/abs/1507.05267}{{\ttfamily arXiv:1507.05267
  [hep-ph]}}.

\bibitem{Hautmann:2013tba}
F.~Hautmann and H.~Jung, ``{Transverse momentum dependent gluon density from
  DIS precision data},''
  \href{http://dx.doi.org/10.1016/j.nuclphysb.2014.03.014}{{\em Nucl. Phys. B}
  {\bfseries 883} (2014) 1--19},
  \href{http://arxiv.org/abs/1312.7875}{{\ttfamily arXiv:1312.7875 [hep-ph]}}.

\bibitem{Dooling:2014kia}
S.~Dooling, F.~Hautmann, and H.~Jung, ``{Hadroproduction of electroweak gauge
  boson plus jets and TMD parton density functions},''
  \href{http://dx.doi.org/10.1016/j.physletb.2014.07.035}{{\em Phys. Lett. B}
  {\bfseries 736} (2014) 293--298},
  \href{http://arxiv.org/abs/1406.2994}{{\ttfamily arXiv:1406.2994 [hep-ph]}}.

\bibitem{Hautmann:2014uua}
F.~Hautmann, H.~Jung, and S.~T. Monfared, ``{The CCFM uPDF evolution uPDFevolv
  Version 1.0.00},''
  \href{http://dx.doi.org/10.1140/epjc/s10052-014-3082-1}{{\em Eur. Phys. J. C}
  {\bfseries 74} (2014) 3082}, \href{http://arxiv.org/abs/1407.5935}{{\ttfamily
  arXiv:1407.5935 [hep-ph]}}.

\bibitem{Martinez:2018jxt}
A.~Bermudez~Martinez, P.~Connor, H.~Jung, A.~Lelek, R.~\v{Z}leb\v{c}\'\i{}k,
  F.~Hautmann, and V.~Radescu, ``{Collinear and TMD parton densities from fits
  to precision DIS measurements in the parton branching method},''
  \href{http://dx.doi.org/10.1103/PhysRevD.99.074008}{{\em Phys. Rev. D}
  {\bfseries 99} no.~7, (2019) 074008},
  \href{http://arxiv.org/abs/1804.11152}{{\ttfamily arXiv:1804.11152
  [hep-ph]}}.

\bibitem{Hautmann:2017fcj}
F.~Hautmann, H.~Jung, A.~Lelek, V.~Radescu, and R.~Zlebcik, ``{Collinear and
  TMD Quark and Gluon Densities from Parton Branching Solution of QCD Evolution
  Equations},'' \href{http://dx.doi.org/10.1007/JHEP01(2018)070}{{\em JHEP}
  {\bfseries 01} (2018) 070}, \href{http://arxiv.org/abs/1708.03279}{{\ttfamily
  arXiv:1708.03279 [hep-ph]}}.

\bibitem{Hautmann:2017xtx}
F.~Hautmann, H.~Jung, A.~Lelek, V.~Radescu, and R.~Zlebcik, ``{Soft-gluon
  resolution scale in QCD evolution equations},''
  \href{http://dx.doi.org/10.1016/j.physletb.2017.07.005}{{\em Phys. Lett. B}
  {\bfseries 772} (2017) 446--451},
  \href{http://arxiv.org/abs/1704.01757}{{\ttfamily arXiv:1704.01757
  [hep-ph]}}.

\bibitem{Hautmann:2014kza}
F.~Hautmann, H.~Jung, M.~Kr\"amer, P.~J. Mulders, E.~R. Nocera, T.~C. Rogers,
  and A.~Signori, ``{TMDlib and TMDplotter: library and plotting tools for
  transverse-momentum-dependent parton distributions},''
  \href{http://dx.doi.org/10.1140/epjc/s10052-014-3220-9}{{\em Eur. Phys. J. C}
  {\bfseries 74} (2014) 3220}, \href{http://arxiv.org/abs/1408.3015}{{\ttfamily
  arXiv:1408.3015 [hep-ph]}}.

\bibitem{Abdulov:2021ivr}
N.~A. Abdulov {\em et~al.}, ``{TMDlib2 and TMDplotter: a platform for 3D hadron
  structure studies},''
  \href{http://dx.doi.org/10.1140/epjc/s10052-021-09508-8}{{\em Eur. Phys. J.
  C} {\bfseries 81} no.~8, (2021) 752},
  \href{http://arxiv.org/abs/2103.09741}{{\ttfamily arXiv:2103.09741
  [hep-ph]}}.

\bibitem{Catani:1994sq}
S.~Catani and F.~Hautmann, ``{High-energy factorization and small x deep
  inelastic scattering beyond leading order},''
  \href{http://dx.doi.org/10.1016/0550-3213(94)90636-X}{{\em Nucl. Phys. B}
  {\bfseries 427} (1994) 475--524},
  \href{http://arxiv.org/abs/hep-ph/9405388}{{\ttfamily arXiv:hep-ph/9405388}}.

\bibitem{Kuraev:1976ge}
E.~A. Kuraev, L.~N. Lipatov, and V.~S. Fadin, ``{Multi - Reggeon Processes in
  the Yang-Mills Theory},'' {\em Sov. Phys. JETP} {\bfseries 44} (1976)
  443--450.

\bibitem{Kuraev:1977fs}
E.~A. Kuraev, L.~N. Lipatov, and V.~S. Fadin, ``{The Pomeranchuk Singularity in
  Nonabelian Gauge Theories},'' {\em Sov. Phys. JETP} {\bfseries 45} (1977)
  199--204.

\bibitem{Balitsky:1978ic}
I.~I. Balitsky and L.~N. Lipatov, ``{The Pomeranchuk Singularity in Quantum
  Chromodynamics},'' {\em Sov. J. Nucl. Phys.} {\bfseries 28} (1978) 822--829.

\bibitem{Jaroszewicz:1982gr}
T.~Jaroszewicz, ``{Gluonic Regge Singularities and Anomalous Dimensions in
  QCD},'' \href{http://dx.doi.org/10.1016/0370-2693(82)90345-8}{{\em Phys.
  Lett. B} {\bfseries 116} (1982) 291--294}.

\bibitem{Catani:1993rn}
S.~Catani and F.~Hautmann, ``{Quark anomalous dimensions at small x},''
  \href{http://dx.doi.org/10.1016/0370-2693(93)90174-G}{{\em Phys. Lett. B}
  {\bfseries 315} (1993) 157--163}.

\bibitem{Fadin:1998py}
V.~S. Fadin and L.~N. Lipatov, ``{BFKL pomeron in the next-to-leading
  approximation},'' \href{http://dx.doi.org/10.1016/S0370-2693(98)00473-0}{{\em
  Phys. Lett. B} {\bfseries 429} (1998) 127--134},
  \href{http://arxiv.org/abs/hep-ph/9802290}{{\ttfamily arXiv:hep-ph/9802290}}.

\bibitem{Ciafaloni:1998gs}
M.~Ciafaloni and G.~Camici, ``{Energy scale(s) and next-to-leading BFKL
  equation},'' \href{http://dx.doi.org/10.1016/S0370-2693(98)00551-6}{{\em
  Phys. Lett. B} {\bfseries 430} (1998) 349--354},
  \href{http://arxiv.org/abs/hep-ph/9803389}{{\ttfamily arXiv:hep-ph/9803389}}.

\bibitem{Bonvini:2016wki}
M.~Bonvini, S.~Marzani, and T.~Peraro, ``{Small-$x$ resummation from HELL},''
  \href{http://dx.doi.org/10.1140/epjc/s10052-016-4445-6}{{\em Eur. Phys. J. C}
  {\bfseries 76} no.~11, (2016) 597},
  \href{http://arxiv.org/abs/1607.02153}{{\ttfamily arXiv:1607.02153
  [hep-ph]}}.

\bibitem{Bonvini:2017ogt}
M.~Bonvini, S.~Marzani, and C.~Muselli, ``{Towards parton distribution
  functions with small-$x$ resummation: HELL 2.0},''
  \href{http://dx.doi.org/10.1007/JHEP12(2017)117}{{\em JHEP} {\bfseries 12}
  (2017) 117}, \href{http://arxiv.org/abs/1708.07510}{{\ttfamily
  arXiv:1708.07510 [hep-ph]}}.

\bibitem{xFitterDevelopersTeam:2018hym}
{\bfseries xFitter Developers' Team} Collaboration, H.~Abdolmaleki {\em
  et~al.}, ``{Impact of low-$x$ resummation on QCD analysis of HERA data},''
  \href{http://dx.doi.org/10.1140/epjc/s10052-018-6090-8}{{\em Eur. Phys. J. C}
  {\bfseries 78} no.~8, (2018) 621},
  \href{http://arxiv.org/abs/1802.00064}{{\ttfamily arXiv:1802.00064
  [hep-ph]}}.

\bibitem{Bonvini:2019wxf}
M.~Bonvini and F.~Giuli, ``{A new simple PDF parametrization: improved
  description of the HERA data},''
  \href{http://dx.doi.org/10.1140/epjp/i2019-12872-x}{{\em Eur. Phys. J. Plus}
  {\bfseries 134} no.~10, (2019) 531},
  \href{http://arxiv.org/abs/1902.11125}{{\ttfamily arXiv:1902.11125
  [hep-ph]}}.

\bibitem{Bertone:2016ywq}
{\bfseries xFitter Developers' Team} Collaboration, V.~Bertone {\em et~al.},
  ``{A determination of $m_c(m_c)$ from HERA data using a matched heavy-flavor
  scheme},'' \href{http://dx.doi.org/10.1007/JHEP08(2016)050}{{\em JHEP}
  {\bfseries 08} (2016) 050}, \href{http://arxiv.org/abs/1605.01946}{{\ttfamily
  arXiv:1605.01946 [hep-ph]}}.

\bibitem{GolecBiernat:1998js}
K.~J. Golec-Biernat and M.~Wusthoff, ``{Saturation effects in deep inelastic
  scattering at low Q**2 and its implications on diffraction},''
  \href{http://dx.doi.org/10.1103/PhysRevD.59.014017}{{\em Phys. Rev. D}
  {\bfseries 59} (1998) 014017},
  \href{http://arxiv.org/abs/hep-ph/9807513}{{\ttfamily arXiv:hep-ph/9807513}}.

\bibitem{Bartels:2002cj}
J.~Bartels, K.~J. Golec-Biernat, and H.~Kowalski, ``{A modification of the
  saturation model: DGLAP evolution},''
  \href{http://dx.doi.org/10.1103/PhysRevD.66.014001}{{\em Phys. Rev. D}
  {\bfseries 66} (2002) 014001},
  \href{http://arxiv.org/abs/hep-ph/0203258}{{\ttfamily arXiv:hep-ph/0203258}}.

\bibitem{Iancu:2003ge}
E.~Iancu, K.~Itakura, and S.~Munier, ``{Saturation and BFKL dynamics in the
  HERA data at small x},''
  \href{http://dx.doi.org/10.1016/j.physletb.2004.02.040}{{\em Phys. Lett. B}
  {\bfseries 590} (2004) 199--208},
  \href{http://arxiv.org/abs/hep-ph/0310338}{{\ttfamily arXiv:hep-ph/0310338}}.

\bibitem{Luszczak:2013rxa}
A.~Luszczak and H.~Kowalski, ``{Dipole model analysis of high precision HERA
  data},'' \href{http://dx.doi.org/10.1103/PhysRevD.89.074051}{{\em Phys. Rev.
  D} {\bfseries 89} no.~7, (2014) 074051},
  \href{http://arxiv.org/abs/1312.4060}{{\ttfamily arXiv:1312.4060 [hep-ph]}}.

\bibitem{Luszczak:2016bxd}
A.~Luszczak and H.~Kowalski, ``{Dipole model analysis of highest precision HERA
  data, including very low $Q^{2}$'s},''
  \href{http://dx.doi.org/10.1103/PhysRevD.95.014030}{{\em Phys. Rev. D}
  {\bfseries 95} no.~1, (2017) 014030},
  \href{http://arxiv.org/abs/1611.10100}{{\ttfamily arXiv:1611.10100
  [hep-ph]}}.

\bibitem{Luszczak:2017dwf}
A.~\L{}uszczak and W.~Sch\"afer, ``{Incoherent diffractive photoproduction of
  $J/\psi$ and $\Upsilon$ on heavy nuclei in the color dipole approach},''
  \href{http://dx.doi.org/10.1103/PhysRevC.97.024903}{{\em Phys. Rev. C}
  {\bfseries 97} no.~2, (2018) 024903},
  \href{http://arxiv.org/abs/1712.04502}{{\ttfamily arXiv:1712.04502
  [hep-ph]}}.

\bibitem{Luszczak:2019vdc}
A.~\L{}uszczak and W.~Sch\"afer, ``{Coherent photoproduction of $J/\psi$ in
  nucleus-nucleus collisions in the color dipole approach},''
  \href{http://dx.doi.org/10.1103/PhysRevC.99.044905}{{\em Phys. Rev. C}
  {\bfseries 99} no.~4, (2019) 044905},
  \href{http://arxiv.org/abs/1901.07989}{{\ttfamily arXiv:1901.07989
  [hep-ph]}}.

\bibitem{Cooper-Sarkar:2018ufj}
A.~M. Cooper-Sarkar and K.~Wichmann, ``{QCD analysis of the ATLAS and CMS
  $W^{\pm}$ and $Z$ cross-section measurements and implications for the strange
  sea density},'' \href{http://dx.doi.org/10.1103/PhysRevD.98.014027}{{\em
  Phys. Rev. D} {\bfseries 98} no.~1, (2018) 014027},
  \href{http://arxiv.org/abs/1803.00968}{{\ttfamily arXiv:1803.00968
  [hep-ex]}}.

\bibitem{Aaboud:2016btc}
{\bfseries ATLAS} Collaboration, M.~Aaboud {\em et~al.}, ``{Precision
  measurement and interpretation of inclusive $W^+$ , $W^-$ and $Z/\gamma ^*$
  production cross sections with the ATLAS detector},''
  \href{http://dx.doi.org/10.1140/epjc/s10052-017-4911-9}{{\em Eur. Phys. J. C}
  {\bfseries 77} no.~6, (2017) 367},
  \href{http://arxiv.org/abs/1612.03016}{{\ttfamily arXiv:1612.03016
  [hep-ex]}}.

\bibitem{Aad:2014xca}
{\bfseries ATLAS} Collaboration, G.~Aad {\em et~al.}, ``{Measurement of the
  production of a $W$ boson in association with a charm quark in $pp$
  collisions at $\sqrt{s} =$ 7 TeV with the ATLAS detector},''
  \href{http://dx.doi.org/10.1007/JHEP05(2014)068}{{\em JHEP} {\bfseries 05}
  (2014) 068}, \href{http://arxiv.org/abs/1402.6263}{{\ttfamily arXiv:1402.6263
  [hep-ex]}}.

\bibitem{Chatrchyan:2013uja}
{\bfseries CMS} Collaboration, S.~Chatrchyan {\em et~al.}, ``{Measurement of
  Associated W + Charm Production in pp Collisions at $\sqrt{s}$ = 7 TeV},''
  \href{http://dx.doi.org/10.1007/JHEP02(2014)013}{{\em JHEP} {\bfseries 02}
  (2014) 013}, \href{http://arxiv.org/abs/1310.1138}{{\ttfamily arXiv:1310.1138
  [hep-ex]}}.

\bibitem{AbdulKhalek:2021gbh}
R.~Abdul~Khalek {\em et~al.}, ``{Science Requirements and Detector Concepts for
  the Electron-Ion Collider: EIC Yellow Report},''
  \href{http://arxiv.org/abs/2103.05419}{{\ttfamily arXiv:2103.05419
  [physics.ins-det]}}.

\bibitem{Arguelles:2019xgp}
C.~A. Arg\"uelles {\em et~al.}, ``{New opportunities at the next-generation
  neutrino experiments I: BSM neutrino physics and dark matter},''
  \href{http://dx.doi.org/10.1088/1361-6633/ab9d12}{{\em Rept. Prog. Phys.}
  {\bfseries 83} no.~12, (2020) 124201},
  \href{http://arxiv.org/abs/1907.08311}{{\ttfamily arXiv:1907.08311
  [hep-ph]}}.

\bibitem{Alvarez-Ruso:2017oui}
{\bfseries NuSTEC} Collaboration, L.~Alvarez-Ruso {\em et~al.}, ``{NuSTEC White
  Paper: Status and challenges of neutrino\textendash{}nucleus scattering},''
  \href{http://dx.doi.org/10.1016/j.ppnp.2018.01.006}{{\em Prog. Part. Nucl.
  Phys.} {\bfseries 100} (2018) 1--68},
  \href{http://arxiv.org/abs/1706.03621}{{\ttfamily arXiv:1706.03621
  [hep-ph]}}.

\bibitem{Aartsen:2016xlq}
{\bfseries IceCube} Collaboration, M.~G. Aartsen {\em et~al.}, ``{Observation
  and Characterization of a Cosmic Muon Neutrino Flux from the Northern
  Hemisphere using six years of IceCube data},''
  \href{http://dx.doi.org/10.3847/0004-637X/833/1/3}{{\em Astrophys. J.}
  {\bfseries 833} no.~1, (2016) 3},
  \href{http://arxiv.org/abs/1607.08006}{{\ttfamily arXiv:1607.08006
  [astro-ph.HE]}}.

\bibitem{Aartsen:2013jdh}
{\bfseries IceCube} Collaboration, M.~G. Aartsen {\em et~al.}, ``{Evidence for
  High-Energy Extraterrestrial Neutrinos at the IceCube Detector},''
  \href{http://dx.doi.org/10.1126/science.1242856}{{\em Science} {\bfseries
  342} (2013) 1242856}, \href{http://arxiv.org/abs/1311.5238}{{\ttfamily
  arXiv:1311.5238 [astro-ph.HE]}}.

\bibitem{Bhattacharya:2016jce}
A.~Bhattacharya, R.~Enberg, Y.~S. Jeong, C.~S. Kim, M.~H. Reno, I.~Sarcevic,
  and A.~Stasto, ``{Prompt atmospheric neutrino fluxes: perturbative QCD models
  and nuclear effects},'' \href{http://dx.doi.org/10.1007/JHEP11(2016)167}{{\em
  JHEP} {\bfseries 11} (2016) 167},
  \href{http://arxiv.org/abs/1607.00193}{{\ttfamily arXiv:1607.00193
  [hep-ph]}}.

\bibitem{Zenaiev:2019ktw}
{\bfseries PROSA} Collaboration, O.~Zenaiev, M.~V. Garzelli, K.~Lipka, S.~O.
  Moch, A.~Cooper-Sarkar, F.~Olness, A.~Geiser, and G.~Sigl, ``{Improved
  constraints on parton distributions using LHCb, ALICE and HERA heavy-flavour
  measurements and implications for the predictions for prompt
  atmospheric-neutrino fluxes},''
  \href{http://dx.doi.org/10.1007/JHEP04(2020)118}{{\em JHEP} {\bfseries 04}
  (2020) 118}, \href{http://arxiv.org/abs/1911.13164}{{\ttfamily
  arXiv:1911.13164 [hep-ph]}}.

\bibitem{ceres-solver}
S.~Agarwal, K.~Mierle, and Others, ``Ceres solver.''
  \url{http://ceres-solver.org}.

\bibitem{H1:2018mkk}
{\bfseries H1} Collaboration, V.~Andreev {\em et~al.}, ``{Determination of
  electroweak parameters in polarised deep-inelastic scattering at HERA},''
  \href{http://dx.doi.org/10.1140/epjc/s10052-018-6236-8}{{\em Eur. Phys. J. C}
  {\bfseries 78} no.~9, (2018) 777},
  \href{http://arxiv.org/abs/1806.01176}{{\ttfamily arXiv:1806.01176
  [hep-ex]}}.

\bibitem{Britzger:2020kgg}
D.~Britzger, M.~Klein, and H.~Spiesberger, ``{Electroweak physics in inclusive
  deep inelastic scattering at the LHeC},''
  \href{http://dx.doi.org/10.1140/epjc/s10052-020-8367-y}{{\em Eur. Phys. J. C}
  {\bfseries 80} no.~9, (2020) 831},
  \href{http://arxiv.org/abs/2007.11799}{{\ttfamily arXiv:2007.11799
  [hep-ph]}}.

\bibitem{Hautmann:2019biw}
F.~Hautmann, L.~Keersmaekers, A.~Lelek, and A.~M. Van~Kampen, ``{Dynamical
  resolution scale in transverse momentum distributions at the LHC},''
  \href{http://dx.doi.org/10.1016/j.nuclphysb.2019.114795}{{\em Nucl. Phys. B}
  {\bfseries 949} (2019) 114795},
  \href{http://arxiv.org/abs/1908.08524}{{\ttfamily arXiv:1908.08524
  [hep-ph]}}.

\bibitem{Camarda:2019zyx}
S.~Camarda {\em et~al.}, ``{DYTurbo: Fast predictions for Drell-Yan
  processes},'' \href{http://dx.doi.org/10.1140/epjc/s10052-020-7757-5}{{\em
  Eur. Phys. J. C} {\bfseries 80} no.~3, (2020) 251},
  \href{http://arxiv.org/abs/1910.07049}{{\ttfamily arXiv:1910.07049
  [hep-ph]}}. [Erratum: Eur.Phys.J.C 80, 440 (2020)].

\bibitem{Bertone:2022sso}
V.~Bertone, G.~Bozzi, and F.~Hautmann, ``{Perturbative hysteresis and emergent
  resummation scales},''
  \href{http://dx.doi.org/10.1103/PhysRevD.105.096003}{{\em Phys. Rev. D}
  {\bfseries 105} no.~9, (2022) 096003},
  \href{http://arxiv.org/abs/2202.03380}{{\ttfamily arXiv:2202.03380
  [hep-ph]}}.

\bibitem{Bertone:2022ope}
V.~Bertone, G.~Bozzi, and F.~Hautmann, ``{Resummation Scales and the Assessment
  of Theoretical Uncertainties in Parton Distribution Functions},'' in {\em
  {29th International Workshop on Deep-Inelastic Scattering and Related
  Subjects}}.
\newblock 5, 2022.
\newblock \href{http://arxiv.org/abs/2205.15900}{{\ttfamily arXiv:2205.15900
  [hep-ph]}}.

\bibitem{NNPDF:2014otw}
{\bfseries NNPDF} Collaboration, R.~D. Ball {\em et~al.}, ``{Parton
  distributions for the LHC Run II},''
  \href{http://dx.doi.org/10.1007/JHEP04(2015)040}{{\em JHEP} {\bfseries 04}
  (2015) 040}, \href{http://arxiv.org/abs/1410.8849}{{\ttfamily arXiv:1410.8849
  [hep-ph]}}.

\bibitem{Manohar:2016nzj}
A.~Manohar, P.~Nason, G.~P. Salam, and G.~Zanderighi, ``{How bright is the
  proton? A precise determination of the photon parton distribution
  function},'' \href{http://dx.doi.org/10.1103/PhysRevLett.117.242002}{{\em
  Phys. Rev. Lett.} {\bfseries 117} no.~24, (2016) 242002},
  \href{http://arxiv.org/abs/1607.04266}{{\ttfamily arXiv:1607.04266
  [hep-ph]}}.

\bibitem{Harland-Lang:2016apc}
L.~A. Harland-Lang, V.~A. Khoze, and M.~G. Ryskin, ``{The photon PDF in events
  with rapidity gaps},''
  \href{http://dx.doi.org/10.1140/epjc/s10052-016-4100-2}{{\em Eur. Phys. J. C}
  {\bfseries 76} no.~5, (2016) 255},
  \href{http://arxiv.org/abs/1601.03772}{{\ttfamily arXiv:1601.03772
  [hep-ph]}}.

\bibitem{Dulat:2015mca}
S.~Dulat, T.-J. Hou, J.~Gao, M.~Guzzi, J.~Huston, P.~Nadolsky, J.~Pumplin,
  C.~Schmidt, D.~Stump, and C.~P. Yuan, ``{New parton distribution functions
  from a global analysis of quantum chromodynamics},''
  \href{http://dx.doi.org/10.1103/PhysRevD.93.033006}{{\em Phys. Rev. D}
  {\bfseries 93} no.~3, (2016) 033006},
  \href{http://arxiv.org/abs/1506.07443}{{\ttfamily arXiv:1506.07443
  [hep-ph]}}.

\bibitem{Sadykov:2014aua}
R.~Sadykov, ``{Impact of QED radiative corrections on Parton Distribution
  Functions},'' \href{http://arxiv.org/abs/1401.1133}{{\ttfamily
  arXiv:1401.1133 [hep-ph]}}.

\bibitem{Bertone:2014zva}
V.~Bertone, R.~Frederix, S.~Frixione, J.~Rojo, and M.~Sutton, ``{aMCfast:
  automation of fast NLO computations for PDF fits},''
  \href{http://dx.doi.org/10.1007/JHEP08(2014)166}{{\em JHEP} {\bfseries 08}
  (2014) 166}, \href{http://arxiv.org/abs/1406.7693}{{\ttfamily arXiv:1406.7693
  [hep-ph]}}.

\bibitem{Alwall:2014hca}
J.~Alwall, R.~Frederix, S.~Frixione, V.~Hirschi, F.~Maltoni, O.~Mattelaer,
  H.~S. Shao, T.~Stelzer, P.~Torrielli, and M.~Zaro, ``{The automated
  computation of tree-level and next-to-leading order differential cross
  sections, and their matching to parton shower simulations},''
  \href{http://dx.doi.org/10.1007/JHEP07(2014)079}{{\em JHEP} {\bfseries 07}
  (2014) 079}, \href{http://arxiv.org/abs/1405.0301}{{\ttfamily arXiv:1405.0301
  [hep-ph]}}.

\bibitem{H1:2015ubc}
{\bfseries H1, ZEUS} Collaboration, H.~Abramowicz {\em et~al.}, ``{Combination
  of measurements of inclusive deep inelastic ${e^{\pm }p}$ scattering cross
  sections and QCD analysis of HERA data},''
  \href{http://dx.doi.org/10.1140/epjc/s10052-015-3710-4}{{\em Eur. Phys. J. C}
  {\bfseries 75} no.~12, (2015) 580},
  \href{http://arxiv.org/abs/1506.06042}{{\ttfamily arXiv:1506.06042
  [hep-ex]}}.

\bibitem{Gleisberg:2008ta}
T.~Gleisberg, S.~Hoeche, F.~Krauss, M.~Schonherr, S.~Schumann, F.~Siegert, and
  J.~Winter, ``{Event generation with SHERPA 1.1},''
  \href{http://dx.doi.org/10.1088/1126-6708/2009/02/007}{{\em JHEP} {\bfseries
  02} (2009) 007}, \href{http://arxiv.org/abs/0811.4622}{{\ttfamily
  arXiv:0811.4622 [hep-ph]}}.

\bibitem{xFitterDevelopersTeam:2017fzy}
{\bfseries xFitter Developers Team} Collaboration, V.~Bertone {\em et~al.},
  ``{Impact of the heavy quark matching scales in PDF fits},''
  \href{http://dx.doi.org/10.1140/epjc/s10052-017-5407-3}{{\em Eur. Phys. J. C}
  {\bfseries 77} no.~12, (2017) 837},
  \href{http://arxiv.org/abs/1707.05343}{{\ttfamily arXiv:1707.05343
  [hep-ph]}}.

\bibitem{xFitterDevelopersTeam:2019ygc}
{\bfseries xFitter Developers' Team} Collaboration, H.~Abdolmaleki {\em
  et~al.}, ``{Probing the strange content of the proton with charm production
  in charged current at LHeC},''
  \href{http://dx.doi.org/10.1140/epjc/s10052-019-7362-7}{{\em Eur. Phys. J. C}
  {\bfseries 79} no.~10, (2019) 864},
  \href{http://arxiv.org/abs/1907.01014}{{\ttfamily arXiv:1907.01014
  [hep-ph]}}.

\bibitem{Alekhin:2018pai}
S.~Alekhin, J.~Bl\"umlein, and S.~Moch, ``{NLO PDFs from the ABMP16 fit},''
  \href{http://dx.doi.org/10.1140/epjc/s10052-018-5947-1}{{\em Eur. Phys. J. C}
  {\bfseries 78} no.~6, (2018) 477},
  \href{http://arxiv.org/abs/1803.07537}{{\ttfamily arXiv:1803.07537
  [hep-ph]}}.

\bibitem{NNPDF:2017mvq}
{\bfseries NNPDF} Collaboration, R.~D. Ball {\em et~al.}, ``{Parton
  distributions from high-precision collider data},''
  \href{http://dx.doi.org/10.1140/epjc/s10052-017-5199-5}{{\em Eur. Phys. J. C}
  {\bfseries 77} no.~10, (2017) 663},
  \href{http://arxiv.org/abs/1706.00428}{{\ttfamily arXiv:1706.00428
  [hep-ph]}}.

\bibitem{Barry:2018ort}
P.~C. Barry, N.~Sato, W.~Melnitchouk, and C.-R. Ji, ``{First Monte Carlo Global
  QCD Analysis of Pion Parton Distributions},''
  \href{http://dx.doi.org/10.1103/PhysRevLett.121.152001}{{\em Phys. Rev.
  Lett.} {\bfseries 121} no.~15, (2018) 152001},
  \href{http://arxiv.org/abs/1804.01965}{{\ttfamily arXiv:1804.01965
  [hep-ph]}}.

\bibitem{Gluck:1991ey}
M.~Gluck, E.~Reya, and A.~Vogt, ``{Pionic parton distributions},''
  \href{http://dx.doi.org/10.1007/BF01559743}{{\em Z. Phys. C} {\bfseries 53}
  (1992) 651--656}.

\bibitem{Chang:2020rdy}
W.-C. Chang, J.-C. Peng, S.~Platchkov, and T.~Sawada, ``{Constraining gluon
  density of pions at large $x$ by pion-induced $J/\psi$ production},''
  \href{http://dx.doi.org/10.1103/PhysRevD.102.054024}{{\em Phys. Rev. D}
  {\bfseries 102} no.~5, (2020) 054024},
  \href{http://arxiv.org/abs/2006.06947}{{\ttfamily arXiv:2006.06947
  [hep-ph]}}.

\bibitem{Adams:2018pwt}
B.~Adams {\em et~al.}, ``{Letter of Intent: A New QCD facility at the M2 beam
  line of the CERN SPS (COMPASS++/AMBER)},''
  \href{http://arxiv.org/abs/1808.00848}{{\ttfamily arXiv:1808.00848
  [hep-ex]}}.

\bibitem{Abdolmaleki:2021yjf}
{\bfseries xfitter Developers\textquoteright{} Team} Collaboration,
  H.~Abdolmaleki, M.~Soleymaninia, H.~Khanpour, S.~Amoroso, F.~Giuli,
  A.~Glazov, A.~Luszczak, F.~Olness, and O.~Zenaiev, ``{QCD analysis of pion
  fragmentation functions in the xFitter framework},''
  \href{http://dx.doi.org/10.1103/PhysRevD.104.056019}{{\em Phys. Rev. D}
  {\bfseries 104} no.~5, (2021) 056019},
  \href{http://arxiv.org/abs/2105.11306}{{\ttfamily arXiv:2105.11306
  [hep-ph]}}.

\bibitem{Belle:2013lfg}
{\bfseries Belle} Collaboration, M.~Leitgab {\em et~al.}, ``{Precision
  Measurement of Charged Pion and Kaon Differential Cross Sections in e+e-
  Annihilation at s=10.52 GeV},''
  \href{http://dx.doi.org/10.1103/PhysRevLett.111.062002}{{\em Phys. Rev.
  Lett.} {\bfseries 111} (2013) 062002},
  \href{http://arxiv.org/abs/1301.6183}{{\ttfamily arXiv:1301.6183 [hep-ex]}}.

\bibitem{Belle:2020pvy}
{\bfseries Belle} Collaboration, R.~Seidl {\em et~al.}, ``{Update of inclusive
  cross sections of single and pairs of identified light charged hadrons},''
  \href{http://dx.doi.org/10.1103/PhysRevD.101.092004}{{\em Phys. Rev. D}
  {\bfseries 101} no.~9, (2020) 092004},
  \href{http://arxiv.org/abs/2001.10194}{{\ttfamily arXiv:2001.10194
  [hep-ex]}}.

\bibitem{BaBar:2013yrg}
{\bfseries BaBar} Collaboration, J.~P. Lees {\em et~al.}, ``{Production of
  charged pions, kaons, and protons in $e^+e^-$ annihilations into hadrons at
  $\sqrt{s}$=10.54 GeV},''
  \href{http://dx.doi.org/10.1103/PhysRevD.88.032011}{{\em Phys. Rev. D}
  {\bfseries 88} (2013) 032011},
  \href{http://arxiv.org/abs/1306.2895}{{\ttfamily arXiv:1306.2895 [hep-ex]}}.

\bibitem{Khalek:2021gxf}
R.~A. Khalek, V.~Bertone, and E.~R. Nocera, ``{Determination of unpolarized
  pion fragmentation functions using semi-inclusive deep-inelastic-scattering
  data},'' \href{http://dx.doi.org/10.1103/PhysRevD.104.034007}{{\em Phys. Rev.
  D} {\bfseries 104} no.~3, (2021) 034007},
  \href{http://arxiv.org/abs/2105.08725}{{\ttfamily arXiv:2105.08725
  [hep-ph]}}.

\bibitem{ATLAS:2012sjl}
{\bfseries ATLAS} Collaboration, G.~Aad {\em et~al.}, ``{Determination of the
  strange quark density of the proton from ATLAS measurements of the $W \to
  \ell \nu$ and $Z \to \ell\ell$ cross sections},''
  \href{http://dx.doi.org/10.1103/PhysRevLett.109.012001}{{\em Phys. Rev.
  Lett.} {\bfseries 109} (2012) 012001},
  \href{http://arxiv.org/abs/1203.4051}{{\ttfamily arXiv:1203.4051 [hep-ex]}}.

\bibitem{CMS:2013pzl}
{\bfseries CMS} Collaboration, S.~Chatrchyan {\em et~al.}, ``{Measurement of
  the Muon Charge Asymmetry in Inclusive $pp \to W+X$ Production at $\sqrt s =$
  7 TeV and an Improved Determination of Light Parton Distribution
  Functions},'' \href{http://dx.doi.org/10.1103/PhysRevD.90.032004}{{\em Phys.
  Rev. D} {\bfseries 90} no.~3, (2014) 032004},
  \href{http://arxiv.org/abs/1312.6283}{{\ttfamily arXiv:1312.6283 [hep-ex]}}.

\bibitem{ATLAS:2016nqi}
{\bfseries ATLAS} Collaboration, M.~Aaboud {\em et~al.}, ``{Precision
  measurement and interpretation of inclusive $W^+$ , $W^-$ and $Z/\gamma ^*$
  production cross sections with the ATLAS detector},''
  \href{http://dx.doi.org/10.1140/epjc/s10052-017-4911-9}{{\em Eur. Phys. J. C}
  {\bfseries 77} no.~6, (2017) 367},
  \href{http://arxiv.org/abs/1612.03016}{{\ttfamily arXiv:1612.03016
  [hep-ex]}}.

\bibitem{CMS:2016qqr}
{\bfseries CMS} Collaboration, V.~Khachatryan {\em et~al.}, ``{Measurement of
  the differential cross section and charge asymmetry for inclusive $\mathrm
  {p}\mathrm {p}\rightarrow \mathrm {W}^{\pm }+X$ production at ${\sqrt{s}} =
  8$ TeV},'' \href{http://dx.doi.org/10.1140/epjc/s10052-016-4293-4}{{\em Eur.
  Phys. J. C} {\bfseries 76} no.~8, (2016) 469},
  \href{http://arxiv.org/abs/1603.01803}{{\ttfamily arXiv:1603.01803
  [hep-ex]}}.

\bibitem{CMS:2018dxg}
{\bfseries CMS} Collaboration, A.~M. Sirunyan {\em et~al.}, ``{Measurement of
  associated production of a W boson and a charm quark in proton-proton
  collisions at $\sqrt{s} =$ 13 TeV},''
  \href{http://dx.doi.org/10.1140/epjc/s10052-019-6752-1}{{\em Eur. Phys. J. C}
  {\bfseries 79} no.~3, (2019) 269},
  \href{http://arxiv.org/abs/1811.10021}{{\ttfamily arXiv:1811.10021
  [hep-ex]}}.

\bibitem{CMS:2021oxn}
{\bfseries CMS} Collaboration, A.~Tumasyan {\em et~al.}, ``{Measurements of the
  associated production of a W boson and a charm quark in proton-proton
  collisions at $\sqrt{s}$ = 8 TeV},''
  \href{http://arxiv.org/abs/2112.00895}{{\ttfamily arXiv:2112.00895
  [hep-ex]}}.

\bibitem{CMS:2017iqf}
{\bfseries CMS} Collaboration, A.~M. Sirunyan {\em et~al.}, ``{Measurement of
  double-differential cross sections for top quark pair production in pp
  collisions at $\sqrt{s} = 8$ $\,\text {TeV}$ and impact on parton
  distribution functions},''
  \href{http://dx.doi.org/10.1140/epjc/s10052-017-4984-5}{{\em Eur. Phys. J. C}
  {\bfseries 77} no.~7, (2017) 459},
  \href{http://arxiv.org/abs/1703.01630}{{\ttfamily arXiv:1703.01630
  [hep-ex]}}.

\bibitem{CMS:2017zpm}
{\bfseries CMS} Collaboration, A.~M. Sirunyan {\em et~al.}, ``{Measurement of
  the inclusive $ \mathrm{t}\overline{\mathrm{t}} $ cross section in pp
  collisions at $ \sqrt{s}=5.02 $ TeV using final states with at least one
  charged lepton},'' \href{http://dx.doi.org/10.1007/JHEP03(2018)115}{{\em
  JHEP} {\bfseries 03} (2018) 115},
  \href{http://arxiv.org/abs/1711.03143}{{\ttfamily arXiv:1711.03143
  [hep-ex]}}.

\bibitem{ATL-PHYS-PUB-2018-017}
{\bfseries ATLAS} Collaboration, ``{Determination of the parton distribution
  functions of the proton from ATLAS measurements of differential $W$ and
  $Z/\gamma^*$ and $t\bar{t}$ cross sections},''.

\bibitem{ATLAS:2016oxs}
{\bfseries ATLAS} Collaboration, M.~Aaboud {\em et~al.}, ``{Measurements of
  top-quark pair to $Z$-boson cross-section ratios at $\sqrt s = 13, 8, 7$ TeV
  with the ATLAS detector},''
  \href{http://dx.doi.org/10.1007/JHEP02(2017)117}{{\em JHEP} {\bfseries 02}
  (2017) 117}, \href{http://arxiv.org/abs/1612.03636}{{\ttfamily
  arXiv:1612.03636 [hep-ex]}}.

\bibitem{ATLAS:2017dhr}
{\bfseries ATLAS} Collaboration, M.~Aaboud {\em et~al.}, ``{Measurement of
  lepton differential distributions and the top quark mass in $t\bar{t}$
  production in $pp$ collisions at $\sqrt{s}=8$ TeV with the ATLAS detector},''
  \href{http://dx.doi.org/10.1140/epjc/s10052-017-5349-9}{{\em Eur. Phys. J. C}
  {\bfseries 77} no.~11, (2017) 804},
  \href{http://arxiv.org/abs/1709.09407}{{\ttfamily arXiv:1709.09407
  [hep-ex]}}.

\bibitem{CMS:2018fks}
{\bfseries CMS} Collaboration, A.~M. Sirunyan {\em et~al.}, ``{Measurement of
  the $\mathrm{t}\overline{\mathrm{t}}$ production cross section, the top quark
  mass, and the strong coupling constant using dilepton events in pp collisions
  at $\sqrt{s} =$ 13 TeV},''
  \href{http://dx.doi.org/10.1140/epjc/s10052-019-6863-8}{{\em Eur. Phys. J. C}
  {\bfseries 79} no.~5, (2019) 368},
  \href{http://arxiv.org/abs/1812.10505}{{\ttfamily arXiv:1812.10505
  [hep-ex]}}.

\bibitem{CMS:2019esx}
{\bfseries CMS} Collaboration, A.~M. Sirunyan {\em et~al.}, ``{Measurement of
  $\mathrm{t\bar t}$ normalised multi-differential cross sections in pp
  collisions at $\sqrt s=13$ TeV, and simultaneous determination of the strong
  coupling strength, top quark pole mass, and parton distribution functions},''
  \href{http://dx.doi.org/10.1140/epjc/s10052-020-7917-7}{{\em Eur. Phys. J. C}
  {\bfseries 80} no.~7, (2020) 658},
  \href{http://arxiv.org/abs/1904.05237}{{\ttfamily arXiv:1904.05237
  [hep-ex]}}.

\bibitem{CMS:2016lna}
{\bfseries CMS} Collaboration, V.~Khachatryan {\em et~al.}, ``{Measurement and
  QCD analysis of double-differential inclusive jet cross sections in pp
  collisions at $ \sqrt{s}=8 $ TeV and cross section ratios to 2.76 and 7
  TeV},'' \href{http://dx.doi.org/10.1007/JHEP03(2017)156}{{\em JHEP}
  {\bfseries 03} (2017) 156}, \href{http://arxiv.org/abs/1609.05331}{{\ttfamily
  arXiv:1609.05331 [hep-ex]}}.

\bibitem{CMS:2021yzl}
{\bfseries CMS} Collaboration, A.~Tumasyan {\em et~al.}, ``{Measurement and QCD
  analysis of double-differential inclusive jet cross sections in proton-proton
  collisions at $ \sqrt{s} $ = 13 TeV},''
  \href{http://dx.doi.org/10.1007/JHEP02(2022)142}{{\em JHEP} {\bfseries 02}
  (2022) 142}, \href{http://arxiv.org/abs/2111.10431}{{\ttfamily
  arXiv:2111.10431 [hep-ex]}}.

\bibitem{CMS:2017jfq}
{\bfseries CMS} Collaboration, A.~M. Sirunyan {\em et~al.}, ``{Measurement of
  the triple-differential dijet cross section in proton-proton collisions at
  $\sqrt{s}=8\,\text {TeV} $ and constraints on parton distribution
  functions},'' \href{http://dx.doi.org/10.1140/epjc/s10052-017-5286-7}{{\em
  Eur. Phys. J. C} {\bfseries 77} no.~11, (2017) 746},
  \href{http://arxiv.org/abs/1705.02628}{{\ttfamily arXiv:1705.02628
  [hep-ex]}}.

\bibitem{ATLAS:2021qnl}
{\bfseries ATLAS} Collaboration, G.~Aad {\em et~al.}, ``{Determination of the
  parton distribution functions of the proton from ATLAS measurements of
  differential W$^{\pm}$ and Z boson production in association with jets},''
  \href{http://dx.doi.org/10.1007/JHEP07(2021)223}{{\em JHEP} {\bfseries 07}
  (2021) 223}, \href{http://arxiv.org/abs/2101.05095}{{\ttfamily
  arXiv:2101.05095 [hep-ex]}}.

\bibitem{ATLAS:2021vod}
{\bfseries ATLAS} Collaboration, G.~Aad {\em et~al.}, ``{Determination of the
  parton distribution functions of the proton using diverse ATLAS data from
  $pp$ collisions at $\sqrt{s} = 7$, 8 and 13~TeV},''
  \href{http://dx.doi.org/10.1140/epjc/s10052-022-10217-z}{{\em Eur. Phys. J.
  C} {\bfseries 82} no.~5, (2022) 438},
  \href{http://arxiv.org/abs/2112.11266}{{\ttfamily arXiv:2112.11266
  [hep-ex]}}.

\bibitem{Kovarik:2015cma}
K.~Kovarik {\em et~al.}, ``{nCTEQ15 - Global analysis of nuclear parton
  distributions with uncertainties in the CTEQ framework},''
  \href{http://dx.doi.org/10.1103/PhysRevD.93.085037}{{\em Phys. Rev. D}
  {\bfseries 93} no.~8, (2016) 085037},
  \href{http://arxiv.org/abs/1509.00792}{{\ttfamily arXiv:1509.00792
  [hep-ph]}}.

\bibitem{Eskola:2016oht}
K.~J. Eskola, P.~Paakkinen, H.~Paukkunen, and C.~A. Salgado, ``{EPPS16: Nuclear
  parton distributions with LHC data},''
  \href{http://dx.doi.org/10.1140/epjc/s10052-017-4725-9}{{\em Eur. Phys. J. C}
  {\bfseries 77} no.~3, (2017) 163},
  \href{http://arxiv.org/abs/1612.05741}{{\ttfamily arXiv:1612.05741
  [hep-ph]}}.

\bibitem{deFlorian:2011fp}
D.~de~Florian, R.~Sassot, P.~Zurita, and M.~Stratmann, ``{Global Analysis of
  Nuclear Parton Distributions},''
  \href{http://dx.doi.org/10.1103/PhysRevD.85.074028}{{\em Phys. Rev. D}
  {\bfseries 85} (2012) 074028},
  \href{http://arxiv.org/abs/1112.6324}{{\ttfamily arXiv:1112.6324 [hep-ph]}}.

\bibitem{LHCHiggsCrossSectionWorkingGroup:2016ypw}
{\bfseries LHC Higgs Cross Section Working Group} Collaboration, D.~de~Florian
  {\em et~al.}, ``{Handbook of LHC Higgs Cross Sections: 4. Deciphering the
  Nature of the Higgs Sector},''
  \href{http://arxiv.org/abs/1610.07922}{{\ttfamily arXiv:1610.07922
  [hep-ph]}}.

\bibitem{Cepeda:2019klc}
M.~Cepeda {\em et~al.}, ``{Report from Working Group 2}: {Higgs Physics at the
  HL-LHC and HE-LHC},''
  \href{http://dx.doi.org/10.23731/CYRM-2019-007.221}{{\em CERN Yellow Rep.
  Monogr.} {\bfseries 7} (2019) 221--584},
  \href{http://arxiv.org/abs/1902.00134}{{\ttfamily arXiv:1902.00134
  [hep-ph]}}.

\bibitem{Chen:2021isd}
X.~Chen, T.~Gehrmann, E.~W.~N. Glover, A.~Huss, B.~Mistlberger, and A.~Pelloni,
  ``{Fully Differential Higgs Boson Production to Third Order in QCD},''
  \href{http://dx.doi.org/10.1103/PhysRevLett.127.072002}{{\em Phys. Rev.
  Lett.} {\bfseries 127} no.~7, (2021) 072002},
  \href{http://arxiv.org/abs/2102.07607}{{\ttfamily arXiv:2102.07607
  [hep-ph]}}.

\bibitem{Azzi:2019yne}
P.~Azzi {\em et~al.}, ``{Report from Working Group 1}: {Standard Model Physics
  at the HL-LHC and HE-LHC},''
  \href{http://dx.doi.org/10.23731/CYRM-2019-007.1}{{\em CERN Yellow Rep.
  Monogr.} {\bfseries 7} (2019) 1--220},
  \href{http://arxiv.org/abs/1902.04070}{{\ttfamily arXiv:1902.04070
  [hep-ph]}}.

\bibitem{AbdulKhalek:2020jut}
R.~Abdul~Khalek {\em et~al.}, ``{Phenomenology of NNLO jet production at the
  LHC and its impact on parton distributions},''
  \href{http://dx.doi.org/10.1140/epjc/s10052-020-8328-5}{{\em Eur. Phys. J. C}
  {\bfseries 80} no.~8, (2020) 797},
  \href{http://arxiv.org/abs/2005.11327}{{\ttfamily arXiv:2005.11327
  [hep-ph]}}.

\bibitem{Cacciari:2015fta}
M.~Cacciari, M.~L. Mangano, and P.~Nason, ``{Gluon PDF constraints from the
  ratio of forward heavy-quark production at the LHC at $\sqrt{S}=7$ and 13
  TeV},'' \href{http://dx.doi.org/10.1140/epjc/s10052-015-3814-x}{{\em Eur.
  Phys. J. C} {\bfseries 75} no.~12, (2015) 610},
  \href{http://arxiv.org/abs/1507.06197}{{\ttfamily arXiv:1507.06197
  [hep-ph]}}.

\bibitem{Flett:2019pux}
C.~A. Flett, S.~P. Jones, A.~D. Martin, M.~G. Ryskin, and T.~Teubner, ``{How to
  include exclusive $J/\psi$ production data in global PDF analyses},''
  \href{http://dx.doi.org/10.1103/PhysRevD.101.094011}{{\em Phys. Rev. D}
  {\bfseries 101} no.~9, (2020) 094011},
  \href{http://arxiv.org/abs/1908.08398}{{\ttfamily arXiv:1908.08398
  [hep-ph]}}.

\bibitem{Czakon:2016olj}
M.~Czakon, N.~P. Hartland, A.~Mitov, E.~R. Nocera, and J.~Rojo, ``{Pinning down
  the large-x gluon with NNLO top-quark pair differential distributions},''
  \href{http://dx.doi.org/10.1007/JHEP04(2017)044}{{\em JHEP} {\bfseries 04}
  (2017) 044}, \href{http://arxiv.org/abs/1611.08609}{{\ttfamily
  arXiv:1611.08609 [hep-ph]}}.

\bibitem{Gauld:2017tww}
R.~Gauld, A.~Gehrmann-De~Ridder, T.~Gehrmann, E.~W.~N. Glover, and A.~Huss,
  ``{Precise predictions for the angular coefficients in Z-boson production at
  the LHC},'' \href{http://dx.doi.org/10.1007/JHEP11(2017)003}{{\em JHEP}
  {\bfseries 11} (2017) 003}, \href{http://arxiv.org/abs/1708.00008}{{\ttfamily
  arXiv:1708.00008 [hep-ph]}}.

\bibitem{CMS:2015cyj}
{\bfseries CMS} Collaboration, V.~Khachatryan {\em et~al.}, ``{Angular
  coefficients of Z bosons produced in pp collisions at $\sqrt{s}$ = 8 TeV and
  decaying to $\mu^+ \mu^-$ as a function of transverse momentum and
  rapidity},'' \href{http://dx.doi.org/10.1016/j.physletb.2015.08.061}{{\em
  Phys. Lett. B} {\bfseries 750} (2015) 154--175},
  \href{http://arxiv.org/abs/1504.03512}{{\ttfamily arXiv:1504.03512
  [hep-ex]}}.

\bibitem{ATLAS:2016rnf}
{\bfseries ATLAS} Collaboration, G.~Aad {\em et~al.}, ``{Measurement of the
  angular coefficients in $Z$-boson events using electron and muon pairs from
  data taken at $\sqrt{s}=8$ TeV with the ATLAS detector},''
  \href{http://dx.doi.org/10.1007/JHEP08(2016)159}{{\em JHEP} {\bfseries 08}
  (2016) 159}, \href{http://arxiv.org/abs/1606.00689}{{\ttfamily
  arXiv:1606.00689 [hep-ex]}}.

\bibitem{Campbell:2019dru}
J.~Campbell and T.~Neumann, ``{Precision Phenomenology with MCFM},''
  \href{http://dx.doi.org/10.1007/JHEP12(2019)034}{{\em JHEP} {\bfseries 12}
  (2019) 034}, \href{http://arxiv.org/abs/1909.09117}{{\ttfamily
  arXiv:1909.09117 [hep-ph]}}.

\bibitem{Hou:2019efy}
T.-J. Hou {\em et~al.}, ``{New CTEQ global analysis of quantum chromodynamics
  with high-precision data from the LHC},''
  \href{http://dx.doi.org/10.1103/PhysRevD.103.014013}{{\em Phys. Rev. D}
  {\bfseries 103} no.~1, (2021) 014013},
  \href{http://arxiv.org/abs/1912.10053}{{\ttfamily arXiv:1912.10053
  [hep-ph]}}.

\bibitem{Bonvini:2014jma}
M.~Bonvini, R.~D. Ball, S.~Forte, S.~Marzani, and G.~Ridolfi, ``{Updated Higgs
  cross section at approximate N$^3$LO},''
  \href{http://dx.doi.org/10.1088/0954-3899/41/9/095002}{{\em J. Phys. G}
  {\bfseries 41} (2014) 095002},
  \href{http://arxiv.org/abs/1404.3204}{{\ttfamily arXiv:1404.3204 [hep-ph]}}.

\bibitem{Bonvini:2018ixe}
M.~Bonvini and S.~Marzani, ``{Double resummation for Higgs production},''
  \href{http://dx.doi.org/10.1103/PhysRevLett.120.202003}{{\em Phys. Rev.
  Lett.} {\bfseries 120} no.~20, (2018) 202003},
  \href{http://arxiv.org/abs/1802.07758}{{\ttfamily arXiv:1802.07758
  [hep-ph]}}.

\bibitem{Bailey:2020ooq}
S.~Bailey, T.~Cridge, L.~A. Harland-Lang, A.~D. Martin, and R.~S. Thorne,
  ``{Parton distributions from LHC, HERA, Tevatron and fixed target data:
  MSHT20 PDFs},'' \href{http://dx.doi.org/10.1140/epjc/s10052-021-09057-0}{{\em
  Eur. Phys. J. C} {\bfseries 81} no.~4, (2021) 341},
  \href{http://arxiv.org/abs/2012.04684}{{\ttfamily arXiv:2012.04684
  [hep-ph]}}.

\bibitem{AbdulKhalek:2018rok}
R.~Abdul~Khalek, S.~Bailey, J.~Gao, L.~Harland-Lang, and J.~Rojo, ``{Towards
  Ultimate Parton Distributions at the High-Luminosity LHC},''
  \href{http://dx.doi.org/10.1140/epjc/s10052-018-6448-y}{{\em Eur. Phys. J. C}
  {\bfseries 78} no.~11, (2018) 962},
  \href{http://arxiv.org/abs/1810.03639}{{\ttfamily arXiv:1810.03639
  [hep-ph]}}.

\bibitem{Cipriano:2013ooa}
P.~Cipriano, S.~Dooling, A.~Grebenyuk, P.~Gunnellini, F.~Hautmann, H.~Jung, and
  P.~Katsas, ``{Higgs boson as a gluon trigger},''
  \href{http://dx.doi.org/10.1103/PhysRevD.88.097501}{{\em Phys. Rev. D}
  {\bfseries 88} no.~9, (2013) 097501},
  \href{http://arxiv.org/abs/1308.1655}{{\ttfamily arXiv:1308.1655 [hep-ph]}}.

\bibitem{Hautmann:2002tu}
F.~Hautmann, ``{Heavy top limit and double logarithmic contributions to Higgs
  production at m(H)**2 / s much less than 1},''
  \href{http://dx.doi.org/10.1016/S0370-2693(02)01761-6}{{\em Phys. Lett. B}
  {\bfseries 535} (2002) 159--162},
  \href{http://arxiv.org/abs/hep-ph/0203140}{{\ttfamily arXiv:hep-ph/0203140}}.

\bibitem{Contino:2016spe}
R.~Contino {\em et~al.}, ``{Physics at a 100 TeV pp collider: Higgs and EW
  symmetry breaking studies},''
  \href{http://arxiv.org/abs/1606.09408}{{\ttfamily arXiv:1606.09408
  [hep-ph]}}.

\bibitem{ATLAS:2019erb}
{\bfseries ATLAS} Collaboration, G.~Aad {\em et~al.}, ``{Search for high-mass
  dilepton resonances using 139 fb$^{-1}$ of $pp$ collision data collected at
  $\sqrt{s}=$13 TeV with the ATLAS detector},''
  \href{http://dx.doi.org/10.1016/j.physletb.2019.07.016}{{\em Phys. Lett. B}
  {\bfseries 796} (2019) 68--87},
  \href{http://arxiv.org/abs/1903.06248}{{\ttfamily arXiv:1903.06248
  [hep-ex]}}.

\bibitem{ATLAS:2019lsy}
{\bfseries ATLAS} Collaboration, G.~Aad {\em et~al.}, ``{Search for a heavy
  charged boson in events with a charged lepton and missing transverse momentum
  from $pp$ collisions at $\sqrt{s} = 13$ TeV with the ATLAS detector},''
  \href{http://dx.doi.org/10.1103/PhysRevD.100.052013}{{\em Phys. Rev. D}
  {\bfseries 100} no.~5, (2019) 052013},
  \href{http://arxiv.org/abs/1906.05609}{{\ttfamily arXiv:1906.05609
  [hep-ex]}}.

\bibitem{CMS:2018ipm}
{\bfseries CMS} Collaboration, A.~M. Sirunyan {\em et~al.}, ``{Search for
  high-mass resonances in dilepton final states in proton-proton collisions at
  $\sqrt{s}=$ 13 TeV},'' \href{http://dx.doi.org/10.1007/JHEP06(2018)120}{{\em
  JHEP} {\bfseries 06} (2018) 120},
  \href{http://arxiv.org/abs/1803.06292}{{\ttfamily arXiv:1803.06292
  [hep-ex]}}.

\bibitem{CMS:2021ctt}
{\bfseries CMS} Collaboration, A.~M. Sirunyan {\em et~al.}, ``{Search for
  resonant and nonresonant new phenomena in high-mass dilepton final states at
  $ \sqrt{s} $ = 13 TeV},''
  \href{http://dx.doi.org/10.1007/JHEP07(2021)208}{{\em JHEP} {\bfseries 07}
  (2021) 208}, \href{http://arxiv.org/abs/2103.02708}{{\ttfamily
  arXiv:2103.02708 [hep-ex]}}.

\bibitem{CidVidal:2018eel}
X.~Cid~Vidal {\em et~al.}, ``{Report from Working Group 3}: {Beyond the
  Standard Model physics at the HL-LHC and HE-LHC},''
  \href{http://dx.doi.org/10.23731/CYRM-2019-007.585}{{\em CERN Yellow Rep.
  Monogr.} {\bfseries 7} (2019) 585--865},
  \href{http://arxiv.org/abs/1812.07831}{{\ttfamily arXiv:1812.07831
  [hep-ph]}}.

\bibitem{Accomando:2019ahs}
E.~Accomando, F.~Coradeschi, T.~Cridge, J.~Fiaschi, F.~Hautmann, S.~Moretti,
  C.~Shepherd-Themistocleous, and C.~Voisey, ``{Production of Z'-boson
  resonances with large width at the LHC},''
  \href{http://dx.doi.org/10.1016/j.physletb.2020.135293}{{\em Phys. Lett. B}
  {\bfseries 803} (2020) 135293},
  \href{http://arxiv.org/abs/1910.13759}{{\ttfamily arXiv:1910.13759
  [hep-ph]}}.

\bibitem{Fiaschi:2021okg}
J.~Fiaschi, F.~Giuli, F.~Hautmann, and S.~Moretti, ``{Lepton-Charge and
  Forward-Backward Asymmetries in Drell-Yan Processes for Precision Electroweak
  Measurements and New Physics Searches},''
  \href{http://dx.doi.org/10.1016/j.nuclphysb.2021.115444}{{\em Nucl. Phys. B}
  {\bfseries 968} (2021) 115444},
  \href{http://arxiv.org/abs/2103.10224}{{\ttfamily arXiv:2103.10224
  [hep-ph]}}.

\bibitem{Liu:2019bua}
D.~Liu, L.-T. Wang, and K.-P. Xie, ``{Broad composite resonances and their
  signals at the LHC},''
  \href{http://dx.doi.org/10.1103/PhysRevD.100.075021}{{\em Phys. Rev. D}
  {\bfseries 100} no.~7, (2019) 075021},
  \href{http://arxiv.org/abs/1901.01674}{{\ttfamily arXiv:1901.01674
  [hep-ph]}}.

\bibitem{Panico:2015jxa}
G.~Panico and A.~Wulzer,
  \href{http://dx.doi.org/10.1007/978-3-319-22617-0}{{\em {The Composite
  Nambu-Goldstone Higgs}}}, vol.~913.
\newblock Springer, 2016.
\newblock \href{http://arxiv.org/abs/1506.01961}{{\ttfamily arXiv:1506.01961
  [hep-ph]}}.

\bibitem{Kaplan:1983fs}
D.~B. Kaplan and H.~Georgi, ``{SU(2) x U(1) Breaking by Vacuum Misalignment},''
  \href{http://dx.doi.org/10.1016/0370-2693(84)91177-8}{{\em Phys. Lett. B}
  {\bfseries 136} (1984) 183--186}.

\bibitem{Kaplan:1983sm}
D.~B. Kaplan, H.~Georgi, and S.~Dimopoulos, ``{Composite Higgs Scalars},''
  \href{http://dx.doi.org/10.1016/0370-2693(84)91178-X}{{\em Phys. Lett. B}
  {\bfseries 136} (1984) 187--190}.

\bibitem{DeCurtis:2011yx}
S.~De~Curtis, M.~Redi, and A.~Tesi, ``{The 4D Composite Higgs},''
  \href{http://dx.doi.org/10.1007/JHEP04(2012)042}{{\em JHEP} {\bfseries 04}
  (2012) 042}, \href{http://arxiv.org/abs/1110.1613}{{\ttfamily arXiv:1110.1613
  [hep-ph]}}.

\bibitem{Agashe:2004rs}
K.~Agashe, R.~Contino, and A.~Pomarol, ``{The Minimal composite Higgs model},''
  \href{http://dx.doi.org/10.1016/j.nuclphysb.2005.04.035}{{\em Nucl. Phys. B}
  {\bfseries 719} (2005) 165--187},
  \href{http://arxiv.org/abs/hep-ph/0412089}{{\ttfamily arXiv:hep-ph/0412089}}.

\bibitem{Giudice:2007fh}
G.~F. Giudice, C.~Grojean, A.~Pomarol, and R.~Rattazzi, ``{The
  Strongly-Interacting Light Higgs},''
  \href{http://dx.doi.org/10.1088/1126-6708/2007/06/045}{{\em JHEP} {\bfseries
  06} (2007) 045}, \href{http://arxiv.org/abs/hep-ph/0703164}{{\ttfamily
  arXiv:hep-ph/0703164}}.

\bibitem{Fiaschi:2021sin}
J.~Fiaschi, F.~Giuli, F.~Hautmann, and S.~Moretti, ``{Enhancing the Large
  Hadron Collider sensitivity to charged and neutral broad resonances of new
  gauge sectors},'' \href{http://dx.doi.org/10.1007/JHEP02(2022)179}{{\em JHEP}
  {\bfseries 02} (2022) 179}, \href{http://arxiv.org/abs/2111.09698}{{\ttfamily
  arXiv:2111.09698 [hep-ph]}}.

\bibitem{Greljo:2021kvv}
A.~Greljo, S.~Iranipour, Z.~Kassabov, M.~Madigan, J.~Moore, J.~Rojo, M.~Ubiali,
  and C.~Voisey, ``{Parton distributions in the SMEFT from high-energy
  Drell-Yan tails},'' \href{http://dx.doi.org/10.1007/JHEP07(2021)122}{{\em
  JHEP} {\bfseries 07} (2021) 122},
  \href{http://arxiv.org/abs/2104.02723}{{\ttfamily arXiv:2104.02723
  [hep-ph]}}.

\bibitem{Gao:2012qpa}
J.~Gao, C.~S. Li, and C.~P. Yuan, ``{NLO QCD Corrections to dijet Production
  via Quark Contact Interactions},''
  \href{http://dx.doi.org/10.1007/JHEP07(2012)037}{{\em JHEP} {\bfseries 07}
  (2012) 037}, \href{http://arxiv.org/abs/1204.4773}{{\ttfamily arXiv:1204.4773
  [hep-ph]}}.

\bibitem{Gao:2013kp}
J.~Gao, ``{CIJET: A program for computation of jet cross sections induced by
  quark contact interactions at hadron colliders},''
  \href{http://dx.doi.org/10.1016/j.cpc.2013.05.019}{{\em Comput. Phys.
  Commun.} {\bfseries 184} (2013) 2362--2366},
  \href{http://arxiv.org/abs/1301.7263}{{\ttfamily arXiv:1301.7263 [hep-ph]}}.

\end{thebibliography}\endgroup

\end{document}